\documentclass[11pt,A4,headsepline,headinclude,footexclude,DIV14,BCOR9mm,noappendixprefix]{scrbook}
\usepackage{amsthm}                  
\usepackage{amsmath}
\usepackage{amssymb}

\usepackage[all]{xy}
\usepackage[ps]{xypic}

\usepackage[german]{babel}
\usepackage[latin1]{inputenc}
\usepackage[T1]{fontenc}

\newenvironment{enumerateR}{%
  \begin{enumerate}}{\end{enumerate}}

\newtheorem{Satz}{Satz}[section]
\newtheorem{Lemma}[Satz]{Lemma}
\newtheorem{prop}[Satz]{Proposition}

\newtheorem{ASatz}{Satz}[chapter]
\newtheorem{ALemma}[ASatz]{Lemma}

\theoremstyle{definition}
\newtheorem{Definition}[Satz]{Definition}
\newtheorem{ADefinition}[ASatz]{Definition}
\theoremstyle{remark}
\newtheorem{Bemerkung}[Satz]{Bemerkung}
\newtheorem{Bemerkungen}[Satz]{Bemerkungen}
\newtheorem{Beispiel}[Satz]{Beispiel}
\newtheorem{Beispiele}[Satz]{Beispiele}

\newtheorem{Folgerung}[Satz]{Folgerung}

\newtheorem{ABemerkung}[ASatz]{Bemerkung}

\newtheorem{ABeispiele}[ASatz]{Beispiele}

\newcommand{\ra}{\rightarrow}

\newcommand{\lra}{\longrightarrow}

\newcommand{\hra}{\hookrightarrow}

\newcommand{\RA}{\Rightarrow}

\newcommand{\R}{\mathbb{R}}

\newcommand{\op}{\operatorname}

\newcommand{\F}{\mathcal{F}}
\newcommand{\B}{\mathcal{B}}

\newcommand{\g}{\mathfrak{g}}
\renewcommand{\(}{\left(}
\renewcommand{\)}{\right)}

\newcommand{\Lie}{\mathcal{L}}
\newcommand{\dif}{\textup{d}}

\newcommand{\mulV}[1]{\mathfrak{X}^{#1}}
\newcommand{\homM}[2]{\mathfrak{X}^{#1,#2}}
\newcommand{\Wedge}{\textstyle{\bigwedge}}

\newcommand{\elide}[1]{\stackrel{#1}{\Hat{\ldots}}\:\!}
\newcommand{\elidetwo}[2]{\stackrel{#1}{\Hat{\ldots}}\,\stackrel{#2}{\Hat{\ldots}}\:\!}

\newcommand{\dcd}{\,\cdot\;\!,\:\!\cdot\,}

\newcommand{\gra}{\op{graph}}
\newcommand{\stCA}[1][M]{T #1 \oplus T^\ast\! #1}
\newcommand{\sect}{\Gamma^\infty}
\newcommand{\scal}[1]{\langle{#1}\rangle}
\newcommand{\D}{\mathcal{D}}
\newcommand{\dL}{\dif_L}
\newcommand{\prL}{\op{pr}_L}
\newcommand{\prLs}{\op{pr}_{L^\ast}}
\newcommand{\dnd}[1][t]{\frac{\dif}{\dif #1}\Big|}
\newcommand{\FL}{\mathbb{F}L}
\newcommand{\partdif}[1]{\frac{\partial}{\partial #1}}
\newcommand{\roth}[1]{\{#1\}_{{}_{\!\mathcal{R}}}}
\newcommand{\te}{\tilde{e}}
\newcommand{\ta}{\tilde{a}}

\newcommand{\teta}{\tilde{\eta}}

\newcommand{\G}{{{}_\mathcal{G}}}
\newcommand{\NR}{{{}_\mathcal{NR}}}
\newcommand{\CM}{{{}_{CM}}}
\newcommand{\E}{{{}_E}}
\newcommand{\dE}{{\dif_{{E}}}}
\newcommand{\dF}{{\dif_{{F}}}}
\newcommand{\ad}{\op{ad}}
\newcommand{\C}{{{}_\mathcal{C}}}
\newcommand{\dirac}[1]{\{#1\}_{{}_L}}
\newcommand{\invJ}{\mathcal{I}}
\newcommand{\der}{\mathsf{D}}
\newcommand{\invL}{\op{R}}
\newcommand{\Pol}{\mathcal{P}}
\usepackage{scrpage2}

\ofoot[]{}
\ifoot[]{}
\ihead{\headmark}
\ohead[]{\pagemark}

\addtokomafont{pagehead}{\scshape} 
\addtokomafont{sectioning}{\normalfont\bf} 

\begin{document}
\pagestyle{scrheadings} 
\pagestyle{empty}
                              
\begin{center}
    \thispagestyle{empty}
    {\ \vspace{10mm}}\\
    {\huge Deformation von Lie-Algebroiden
      \vspace{3mm} und Dirac-Strukturen}\\
    \vspace{45mm}
    {\Large  Diplomarbeit}\\
    \vspace{3mm}
    \large vorgelegt von\\
    \vspace{4mm}
    {\Large Frank Keller}\\
    \vspace{10mm}
    {\large Dezember 2004}\\
    \vspace{30mm}
    {\large Wissenschaftliche Betreuung: \\
      \vspace{4mm}
      Prof. Dr. Hartmann Römer\\[2mm]
      PD Dr. Stefan Waldmann}\\
    {\Large \vspace{35mm}
      \sc  Albert-Ludwigs-Universität Freiburg\\
      \vspace{4mm} Fakultät für Mathematik und Physik}
\end{center}

\cleardoublepage 
 
\pagestyle{plain}
\pagenumbering{roman}

\ \vspace{10mm}

\centerline{\LARGE \bf Abstract}

\vspace{4mm}

\begin{quote}
    In this diploma thesis we discuss the deformation theory of Lie
    algebroids and Dirac structures. The first chapter gives a short
    introduction to Dirac structures on manifolds as introduced by
    Courant in 1990. We also give some physical applications of Dirac
    structures. In the second chapter we present the deformation
    theory of Lie algebroids following a recent work of Crainic and
    Moerdijk. We discuss the subject from three different points of
    view and show the equivalence of these different interpretations.
    In the third chapter we give definitions for smooth and formal
    deformations of Dirac structures on Courant algebroids. To
    investigate the formal theory, we write the Courant bracket with
    the use of the Rothstein super-Poisson bracket as a derived
    bracket. As the main result we show that the obstruction for
    extending a given formal deformation of a certain order lies in
    the third Lie algebroid cohomology of the Lie algebroid given by
    the undeformed Dirac structure.
\end{quote}

\cleardoublepage


\chapter*{Einleitung}
\pagestyle{plain}

Die Behandlung der klassischen Mechanik mit Hilfe der symplektischen
Geometrie hat sich in den letzten 40 Jahren als äußerst gewinnbringend
herausgestellt. Zum einen wurde durch eine geometrische Beschreibung
erreicht, dass die Theorie unabhängig von Koordinatensystemen
formuliert werden kann. Zum anderen zeigt die
Phasenraumreduktion für Systeme mit Symmetrien, die sogenannte
Marsden-Weinstein-Reduktion der gleichnamigen Autoren, dass auch
durchaus Phasenräume mit komplizierten Geometrien in der Physik eine
Rolle spielen.

Für die Formulierung der Hamiltonschen Gleichungen muss der Phasenraum
jedoch nicht zwingend mit einer symplektischen Struktur ausgestattet
sein. Notwendig ist lediglich eine Poisson-Klammer für die Funktionen
auf dem Phasenraum. Dies führt auf den Begriff der
Poisson-Mannigfaltigkeiten, als Beispiel seien hier die
Euler-Gleichungen des starren Körpers auf dem Phasenraum
$\mathfrak{so}(3)$ genannt, siehe z.B. \cite{marsden.ratiu:1999a}.

Dirac-Strukturen können als eine Verallgemeinerung von symplektischen
Mannigfaltigkeiten und Poisson-Mannigfaltigkeiten aufgefasst werden.
Darüberhinaus können aber auch Systeme, die durch Zwangsbedingungen
eingeschränkt sind, mit Hilfe einer geeigneten Dirac-Struktur auf dem
Phasenraum beschrieben werden.  Dirac-Strukturen wurden 1990 von
Courant \cite{courant:1990a} in Zusammenarbeit mit Weinstein einführt.
Weitere Arbeiten sind unter anderem von Bursztyn, Crainic, Liu,
Radko, Roytenberg, Weinstein und Xu
\cite{liu.weinstein.xu:1997a,roytenberg:1999a,roytenberg:2002a,
  roytenberg:2002b,severa.weinstein:2001a, 
  bursztyn.radko:2003a,bursztyn.crainic.weinstein.zhu:2004a}
erschienen. Dabei wurde auch eine Axiomatisierung der Theorie
vorgenommen, was zu Dirac-Strukturen in Courant-Algebroiden führte.

Bei der Betrachtung von Systemen mit Eichfreiheitsgraden, die durch
eine Lagrangefunk\-tion auf dem Konfigurationsraum $TQ$ gegeben sind,
ist es nicht ohne weiteres möglich, zu einer Hamiltonschen
Formulierung zu gelangen. Bezeichnen wir mit $M\subseteq T^\ast Q$ das
Bild von $TQ$ unter der Legendretransformation, so ist auf $M$ immer
noch eine Dirac-Struktur gegeben. Die Bewegungsgleichungen auf $M$
können dann mit Hilfe dieser Dirac-Struktur formuliert werden. Die
konkrete Lösung dieser Gleichungen entspricht gerade dem
Dirac-Algorithmus \cite{dirac:1964a} zur Behandlung solcher Systeme,
was auch den Namen Dirac-Struktur erklärt.  In Anwendungen zur
Feldtheorie, in der solche Systeme mit Eichfreiheitsgraden vorkommen,
werden die zu betrachtenden Mannigfaltigkeiten in der Regel natürlich
nicht endlichdimensional sein. Man hofft jedoch, durch die Betrachtung
endlichdimensionaler Problemen Kenntnisse zu erlangen, die
anschließend auch zu einem bessere Verständnis von Eichtheorien
beitragen.

Behandelt man Systeme mit Symmetrien, so treten Dirac-Strukturen auf
natürliche Weise auf. Setzt man eine Observable wie beispielsweise den
Drehimpuls auf einen bestimmten Wert fest, so wird das System auf eine
Untermannigfaltigkeit $N$ des gesamten Phasenraums eingeschränkt, auf
der im Allgemeinen keine symplektische Struktur mehr gegeben ist.
Jedoch haben wir auf $N$ immer noch eine Dirac-Struktur, mit deren
Hilfe wir die Mechanik beschreiben können.

In der Kontrolltheorie finden Dirac-Strukturen aufgrund ihrer
Eigenschaft, Systeme mit Zwangsbedingungen beschreiben zu können,
vielfach Verwendung. Dabei werden auch verallgemeinerte oder nicht
integrable Dirac-Strukturen zur Behandlung von Systemen mit
nichtholono\-men Zwangsbedingungen diskutiert. Hier sind insbesondere
Arbeiten von Blankenstein, Maschke und van der Schaft
\cite{blankenstein.ratiu:2004,blankenstein.schaft:2001a,
  dalsmo.schaft:1999a,dalsmo.schaft:1999a} zu nennen. Für die
Behandlung von nichtlinearen Differentialgleichungen, die in der
Kontrolltheorie auftreten, wurde von Dorfman eine algebraische
Version von Dirac-Strukturen entwickelt \cite{dorfman:1993a}, die auch
für solche unendlichdimensionalen Probleme geeignet ist.

In dieser Arbeit soll die Deformationstheorie für Dirac-Strukturen
betrachtet werden. Eine algebraische Deformationstheorie für
assoziative Algebren wurde in den 60er Jahren von Gerstenhaber
\cite{gerstenhaber:1963a,gerstenhaber:1964a} 
entwickelt. Um dabei keine (im Allgemeinen sehr komplizierten)
analytischen Probleme behandeln zu müssen, erfolgt die Deformation in
einem formalen Rahmen, d.h., man fasst das zu deformierende Objekt als
eine formale Potenzreihe in einem formalen Deformationsparameter
auf. Falls man sogar eine Deformation findet, die glatt
von diesem Deformationsparameter abhängt, entspricht die formale
Reihe der Taylorentwicklung dieser glatten Deformation an der Stelle
Null. 

Für die Physik sind Deformationstheorien unter dem Gesichtspunkt
kinematischer oder dynamischer Stabilitätsanalysen wichtig. Damit ist
gemeint, dass das Verhalten der Kinematik bzw. Dynamik eines Systems
unter kleinen Störungen analysiert werden soll. Jede Störungstheorie
kann damit als Beispiel für eine Deformationstheorie dienen. Aber auch
die spezielle Relativitätstheorie kann als eine Deformation der
Newtonschen Mechanik aufgefasst werden, wobei der Deformationsparameter
das Inverse der Lichtgeschwindigkeit ist.

Als wichtige Anwendung einer Deformationstheorie ist weiter die
Deformationsquantisierung zu nennen. Deformiert wird dabei das
assoziative, kommutative Produkt der klassischen Observablen zu einem
zwar weiterhin assoziativen, aber nicht mehr kommutativen Produkt,
einem sogenannten Sternprodukt. Die Rolle des Deformationsparameters
übernimmt dabei das Planck\-sche Wirkungsquantum $\hbar$. Die
Deformationsquantisierung wurde 1978 von Bayen, Flato, Fr{\o}nsdal,
Lichnerowicz und Sternheimer \cite{bayen.et.al:1978a,fronsdal:1978a,
  lichnerowicz:1988a} eingeführt.

Die Deformation von Dirac-Strukturen ist nun in zweierlei Hinsicht
interessant. Zum einen, um wie oben beschrieben, eine
Stabilitätsanalyse durchzuführen. Zum anderen erhofft man sich, durch
die klassische Deformationstheorie wichtige Informationen zu sammeln,
die für eine spätere Quantisierung hilfreich sein können. Dabei spielt
insbesondere das Formalitätstheorem von Kontsevich eine wichtige Rolle
\cite{kontsevich:1997a,kontsevich:1997b}. Demnach sind die
Äquivalenzklassen von Sternprodukten auf Poisson-Mannigfaltigkeiten in
Bijektion zu den Äquivalenzklassen von Deformationen des
Poisson-Tensors. Es lässt sich deshalb zumindest vermuten, dass für
ein noch zu definierendes Sternprodukt auf einer
Dirac-Mannigfaltigkeit die klassische Deformationstheorie ebenfalls
wichtig sein wird.

Das Ziel dieser Arbeit ist, eine Definition der formalen
Deformationstheorie für Dirac-Struk\-turen zu geben und deren
Eigenschaften zu untersuchen. Dabei soll insbesondere gezeigt werden,
dass die Obstruktionen für die Fortsetzbarkeit einer gegebenen
Deformation  einer bestimmten Ordnung durch eine Kohomologieklasse
gegeben sind.

Im ersten Kapitel werden die grundlegenden Definitionen zu
Dirac-Strukturen sowie einige wichtige Resultate und Anwendungen
vorgestellt. Wir betrachten dazu zunächst lineare Dirac-Strukturen,
d.h. Dirac-Strukturen auf Vektorräumen, um die Fragen, die im Bereich
der linearen Algebra liegen, zu diskutieren. Anschließend werden wir
dann zu Dirac-Mannigfaltigkeiten übergehen, wobei eine zusätzliche
Integrabilitätsbedingung an die Dirac-Strukturen gestellt wird.  Wir
folgen dabei, soweit nichts anderes erwähnt wird, hauptsächlich der
Arbeit Courants \cite{courant:1990a}. Zum Abschluss dieses Kapitels
geben wir mit den Impliziten Hamiltonschen Systemen und der Diracschen
Theorie von Zwangsbedingungen noch zwei wichtige physikalische
Anwendungsbeispiele für Dirac-Strukturen.

Bevor wir uns der Deformation von Dirac-Strukturen zuwenden, studieren
wir im zweiten Kapitel zunächst die Deformationstheorie von
Lie-Algebroiden. Dabei betrachten wir, aufbauend auf einer Arbeit von
Crainic und Moerdijk \cite{crainic.moerdijk:2004a}, drei verschiedene Formulierungen
von Lie-Algebroiden sowie den entsprechenden Deformationstheorien und
zeigen deren Äquivalenz. Dies dient einerseits als Einführung in die
formale Deformationstheorie, ist aber andererseits auch unter dem
Gesichtspunkt interessant, dass Dirac-Strukturen insbesondere auch
Lie-Algebroide sind. Dieses Kapitel ist jedoch weitgehend unabhängig
von den anderen.

Im dritten Kapitel soll schließlich die Deformation von
Dirac-Strukturen untersucht werden. Zuvor werden wir jedoch die
Objekte, mit denen wir uns befassen, noch weiter verallgemeinern, d.h.
im Folgenden werden wir Dirac-Strukturen in einem Courant-Algebroid
betrachten.  Anschließend wollen wir eine Definition für glatte
Deformationen von Dirac-Strukturen sowie einen Äquivalenzbegriff dafür
angeben.

Um zur formalen Deformationstheorie übergehen zu können, müssen wir
unser Problem zunächst auf geeignete Weise umformulieren.  Dies ist
notwendig, weil die formale Deformationstheorie verlangt, dass von den
zu deformierenden Objekten formale Reihen gebildet werden können. Da
Dirac-Strukturen aber Untervektorbündel sind, ist dies auf direktem
Weg nicht möglich. Wir betrachten eine Deformation $L_t$ einer
Dirac-Struktur $L \subseteq E$ in einem Courant-Algebroid $E$ deshalb,
zumindest lokal, als den Graphen einer Abbildung $\omega_t:L\ra L'$ in
einer geeigneten Aufspaltung $E=L\oplus L'$, wobei $L = L_0$ der Graph
von $\omega_0 = 0$ ist.  Damit können wir jetzt die formale
Deformationstheorie formulieren.  Wir werden eine Gleichung für
$\omega_t$ herleiten, die die Bedingung an $L_t$, eine Dirac-Struktur
zu sein, kodiert und diese Gleichung untersuchen. Um dies systematisch
durchführen zu können, geben wir an, wie die Courant-Klammer in dem
Courant-Algebroid $E$ mit Hilfe der Rothstein-Klammer als abgeleitete
Klammer im Sinne von \cite{roytenberg:1999a} geschrieben werden kann,
ein Resultat, dass sicher auch für sich allein interessant ist.  Damit
können wir zeigen, dass die Deformationsbedingung ein rekursives,
kohomologisches System von Gleichungen für eine Deformation $\omega_t
= t \omega_1 + t^2 \omega_2 +\ldots$ liefert. Wir erhalten damit auch
für die Deformation von Dirac-Strukturen das Ergebnis, dass die
Obstruktionen für die Fortsetzbarkeit von Deformationen in einer
dritten Kohomologie, in diesem Fall der Lie-Algebroid-Kohomologie von
$L$, liegen.

Im Anhang A wird eine kurze Einführung zu Lie-Algebroiden,
Lie-Bialgebroiden und damit verwandten Themen gegeben.
Schließlich befindet sich  im Anhang B noch eine sehr kurze Einführung
in die Theorie der formalen Potenzreihen, die für uns im Zusammenhang
mit den formalen Deformationen gebraucht werden.



\tableofcontents
\pagestyle{scrheadings}

\chapter{Dirac-Strukturen auf Mannigfaltigkeiten}
\pagenumbering{arabic} 

Dirac-Mannigfaltigkeiten, d.h. Mannigfaltigkeiten mit einer
Dirac-Struktur, stellen eine Verallgemeinerung von symplektischen
Mannigfaltigkeiten und Poisson-Mannigfaltigkeiten dar. Denn ist eine
symplektische Form $\omega \in \Omega^2(M)$ auf einer Mannigfaltigkeit
$M$ gegeben, so definiert der Graph von $\omega$ eine Dirac-Struktur
auf $M$. Dabei wird $\omega$ durch die Definition $X \mapsto
i_X\omega$ für $X \in TM$ als eine Abbildung $\omega:TM \ra T^\ast M$
aufgefasst, es gilt also $\op{graph}(\omega) \subseteq TM \oplus
T^\ast M$.  Genauso ist der Graph eines Poisson-Tensors $\pi \in
\mathfrak{X}^2(M)$, interpretiert als Abbildung $\pi:T^\ast M \ra TM$,
eine Dirac-Struktur. Jedoch braucht nicht jede Dirac-Struktur von der
Form eines dieser beiden Beispiele zu sein. In dem allgemeineren
Rahmen der Dirac-Mannigfaltigkeiten lassen sich dadurch unter anderem
auch physikalische Systeme behandeln, die durch Zwangsbedingungen
einschränkt sind. Ebenso ein Beispiel für Dirac-Mannigfaltigkeiten
sind präsymplektische Mannigfaltigkeiten, also Mannigfaltigkeiten mit
einer geschlossenen, eventuell jedoch ausgearteten Zweiform. Systeme
dieser Art treten auf, wenn man für eine nichtreguläre
Lagrangefunktion zum Hamiltonschen Formalismus übergeht.

In diesem Kapitel sollen die grundlegenden Definitionen sowie einige
Aussagen zu Dirac-Strukturen auf Mannigfaltigkeiten gegeben werden.
Dabei folgen wir weitgehend der Arbeit Courants \cite{courant:1990a},
in der Dirac-Strukturen zum ersten mal vorgestellt wurden. Im Anschluss
an den  mathematischen Teil, den wir zunächst bearbeiten müssen,
werden wir am Ende diese Kapitels schließlich noch zwei für die Physik
wichtige Anwendungen vorstellen.

\section{Lineare Dirac-Strukturen}
Bevor wir uns der Betrachtung von Dirac-Strukturen auf
Mannigfaltigkeiten zuwenden, wollen wir zunächst lineare
Dirac-Strukturen, d.h. Dirac-Strukturen auf Vektorräumen, betrachten.
Diese Ergebnisse können anschließend punktweise auf Mannigfaltigkeiten
übertragen werden.

\subsection{Definition und Beispiele}
Sei V ein Vektorraum mit einer Bilinearform $(\cdot\,,\cdot)$. Ein
Untervektorraum $W\subseteq V$ heißt genau dann isotrop, wenn
$W\subseteq W^{\perp}$ gilt, wobei $W^{\perp} := \{v\in V | (v,w)=0\;
\forall \; w\in W\}$ den Orthogonalraum zu $W$ bezeichnet. Ist
$(\cdot\,,\cdot)$ nicht ausgeartet, dann gilt $\dim W + \dim W^\perp =
\dim V$ und für $W$ isotrop folgt damit wegen $\dim W \leqslant \dim
W^\perp$, dass $\dim W \leqslant \frac{\dim V}{2}$.  Ein isotroper
Unterraum $W$ heißt maximal isotrop, wenn es keinen isotropen
Unterraum $W'$ gibt, so dass $W$ ein echter Unterraum von $W'$ ist.

\begin{Lemma}\label{bili}
    Sei $V$ ein endlichdimensionaler Vektorraum mit einer nicht
    ausgearteten symmetrischen Bilinearform $(\cdot\,,\cdot)$, $W$ ein
    isotroper Unterraum. Sei $U$ ein Komplement von $W$ in $W^\perp$,
    d.h. wir haben eine orthogonale Zerlegung
    $$W^\perp = W \perp U.$$
    Schließlich sei $w_1,\ldots w_k$ eine Basis von $W$.
    Dann gibt es Vektoren $v_1,\ldots v_k$ in $U^\perp$, so dass
    $$(v_i,v_j)=0, \quad (w_i,v_j)= \delta_{ij}$$
    für $i,j =1,\ldots,k$ gilt, und $V$ die orthogonale Summe
    $$
    V = \op{span} (w_1,v_1) \perp \ldots \perp \op{span} (w_k,v_k)
    \perp U$$
    ist.
\end{Lemma}

\begin{proof}
    Sei $U_1 = \op{span}(w_2,\ldots,w_k)\oplus U$. Dann gilt $U_1
    \varsubsetneq W^\perp \; \RA W^{\perp \perp} = W \varsubsetneq
    U_1^\perp$. Wir wählen $u_1 \in U_1^\perp \backslash W$, also ist
    $(u_1,w_1) \ne 0$ und wir können annehmen, dass $(u_1,w_1) = 1$.
    Sei $P_1 = \op{span}(w_1,u_1)$. Wir finden ein $\alpha \in \R$, so
    dass
    $$
    (\alpha w_1 + u_1,\alpha w_1 + u_1) = 2 \alpha (w_1,u_1) +
    (u_1,u_1) = 0,$$
    und setzen $v_1 = \alpha w_1 +u_1$. Damit gilt
    also
    $$
    (w_1,w_1) = 0, \qquad (v_1,v_1) = 0, \qquad (w_1,v_1) = 1.$$
    Sei
    nun $W_1 = \op{span}(w_2,\ldots,w_k)$. $W_1$ ist isotrop und es
    gilt $W_1^\perp = W_1 \perp P_1 \perp U.$ Für $k = \dim W =1$ sind
    wir fertig, sonst folgt die Behauptung durch Induktion.
\end{proof}

\begin{Folgerung}\label{maxiso}
    Sei $V$ ein $n$-dimensionaler Vektorraum mit einer nicht
    ausgearteten symmetrischen Bilinearform $(\cdot\,,\cdot)$, deren
    Matrix in Normalform $q$ mal $1$ und $p$ mal $-1$ enthält. Sei $W$
    ein isotroper Unterraum. Dann gilt $\dim W \leq \min(q,p)$ und es
    gibt einen isotropen Unterraum $W'$ mit $\dim W' = \min(q,p)$, so
    dass $W \subseteq W'$. Maximal isotrope Unterräume haben also alle
    die Dimension $\min(q,p)$.
\end{Folgerung}
\begin{proof}
    Sei $\dim W = k$. Wir wissen schon, dass $k \leq \frac{\dim
      V}{2}$. Sei weiter $w_1,\ldots,w_k$ eine Basis von $W$. Nach
    Lemma \ref{bili} gibt es $v_1,\ldots,v_k$ und
    $u_1,\ldots,u_{n-2k}$, so dass die Matrixdarstellung von
    $(\cdot\,,\cdot)$ in der Basis $w_1,\ldots,w_k$, $v_1,
    \ldots,v_k,u_1$, $\ldots,u_{n-2k}$ die folgende Gestalt hat:
    $$\left ( \begin{array}{c|c}
            \begin{array}{cc}
                0 & \mathbb{E}_k \\
                \mathbb{E}_k & 0 \\
            \end{array}
            & \; 0 \; \\
            \hline 
            \rule[-3mm]{0mm}{8mm}
            0 & A \\
        \end{array} \right )_{\displaystyle .}
    $$
    Wählen wir als neue Basisvektoren $a_i = \frac{1}{2}(w_i +
    v_i)$, $b_i = \frac{1}{2}(w_i - v_i),\; i=1,\ldots,k$ und
    geeignete $u_i$'s, in denen die Matrix $A$ Normalform annimmt, so
    wird die Bilinearform dargestellt durch
    $$\left ( \begin{array}{c|c}
            \begin{array}{cc}
                \mathbb{E}_k & 0\\
                0 & -\mathbb{E}_k\\
            \end{array}
            & \; 0 \; \\
            \hline 
            0 &
            \begin{array}{cc}
                \mathbb{E}_{q-k} & 0\\
                0 & -\mathbb{E}_{p-k}\\
            \end{array}
        \end{array} \right )_{\displaystyle .}
    $$
    An dieser Form kann man zum einen sehen, dass $k = \dim W \leq
    \min(q,p)$ gelten muss. Andererseits ist jetzt auch klar, wie $W$
    zu einem isotropen Unterraum $W'$ der Dimension $\min(q,p)$
    erweitert werden kann.
\end{proof}

\begin{Beispiel}
    Betrachte $V = \R^4$ mit der Minkowski-Metrik $(v,w)=v_1 w_1 -
    v_2 w_2-v_3 w_3 -v_4 w_4$. Abgesehen vom Nullraum sind die isotropen
    Unterräume alle eindimensional. Die Elemente der isotropen
    Unterräume werden in diesem Fall lichtartig genannt. 
\end{Beispiel}

Sei $V$ ein endlichdimensionaler reeller Vektorraum, $\dim V =
n$. Durch die Definition
$$
\langle (x,\eta),(y,\mu)\rangle = \eta(y)+\mu(x)$$
ist auf $V\oplus
V^\ast$ auf kanonische Weise eine symmetrische Bilinearform
$\scal{\dcd}$ gegeben.
\begin{Definition}
    Eine Dirac-Struktur auf einem endlichdimensionalen reellen
    Vektorraum V ist ein Untervektorraum $L\subseteq V\oplus V^\ast$,
    der bezüglich der kanonischen Bilinearform $\langle \dcd \rangle$
    maximal isotrop ist.  Die Menge aller Dirac-Strukturen auf $V$
    bezeichnen wir mit $\op{Dir}(V)$.
\end{Definition}
\begin{Bemerkung}
    Nach Folgerung \ref{maxiso} sind Dirac-Strukturen also genau die
    $n$-dimensionalen, isotropen Unterräume von $V \oplus V^\ast$.
    (Offensichtlich gibt es $n$-dimensionale isotrope Unterräume, z.B.
    $V$ selbst, und diese müssen bereits maximal sein.)
\end{Bemerkung}
\begin{Bemerkung}
    Für $(x,\eta)\; \mbox{und} \; (y,\mu) \in L$ gilt
    $$
    \langle (x,\eta),(y,\mu)\rangle = \eta(y)+\mu(x) = 0 \ \RA \ 
    \eta(y)=-\mu(x),$$
    $$
    \langle (x,\eta),(x,\eta)\rangle = 2\eta(x) = 0.$$
\end{Bemerkung}
Im weiteren bezeichnen wir mit $\rho: V \oplus V^\ast \lra V$ bzw.
$\rho^\ast: V \oplus V^\ast \lra V^\ast$ die kanonischen Projektionen
auf $V$ bzw. $V^\ast$.

\begin{Lemma} Sei $L \subseteq V \oplus V^\ast$ eine lineare Dirac-Struktur
    auf V und $W\subseteq V$ ein Untervektorraum. Sei ferner $W^\circ
    := \{\lambda \in V^\ast |\lambda(w)=0\; \forall w \in W\}$ der
    Annihilatorraum von $W$.  Dann gilt
    \begin{eqnarray*}
        \rho(L)^\circ &=& L \cap V^\ast,\\
        \rho^\ast (L)^\circ &=& L \cap V.
    \end{eqnarray*}
\end{Lemma}

\begin{Bemerkung}
    Wir schreiben für $L \cap (V \oplus \{0\})$ einfach $L \cap V$ und
    fassen $L \cap V$ je nach Situation als Unterraum von $V$ oder von
    $L$ auf. Entsprechendes gilt für $L \cap V^\ast.$
\end{Bemerkung}

\begin{proof}
    Sei $\lambda \in L \cap V^\ast$, und sei $x\in \rho(L)$. Dann gibt
    es ein $\mu \in V^\ast$ mit $(x,\mu) \in L$, und es folgt $$0=\langle
    (x,\mu),(0,\lambda)\rangle = \lambda(x) \;\RA\; \lambda \in
    \rho(L)^\circ.$$
    Sei nun umgekehrt $\lambda \in \rho(L)^\circ$
    vorausgesetzt. Angenommen, es gilt $\lambda \notin L\cap V^\ast$.
    Wir bilden dann den Unterraum $L'=L\oplus \mbox{span}(\lambda)$,
    wobei $L$  echt in $L'$ enthalten ist. Weiter ist $L'$ 
    isotrop, denn seien $(x,\mu), (y,\eta) \in L$ und $\lambda_1,
    \lambda_2 \in \mbox{span}(\lambda)$, dann folgt
    $$
    \langle (x,\mu)+\lambda_1,(y,\eta)+\lambda_2 \rangle
    =\lambda_1(y)+\lambda_2(x) = 0.$$
    $L$ war aber bereits maximal isotrop,
    d.h. unsere Annahme führt zu einem Widerspruch. Es folgt also
    $\lambda \in L\cap V^\ast$, womit die erste
    Gleichung gezeigt ist. Die zweite Gleichung lässt sich analog
    nachweisen.
\end{proof}

\begin{Beispiel}
    Sei $(V,\omega)$ ein präsym\-plek\-tischer Vektorraum, d.h ein
    Vektorraum $V$ mit einer antisymmetrischen Bilinearform $\omega$.
    Wir fassen die präsym\-plek\-tische Form als eine Abbildung
    $\omega:V\lra V^\ast$, $v \mapsto \omega(V,\cdot)$ auf. Dann ist
    graph$(\omega)=\{(v,\omega(v))|v \in V\}\subseteq V\oplus V^\ast$
    eine Dirac-Struktur auf $V$.
\end{Beispiel}
\begin{Beispiel}
    Sei $(V,\pi)$ ein Vektorraum zusammen mit einer linearen
    antisymmetrischen Abbildung $\pi:V^\ast \lra V$. Dann ist
    graph$(\pi)=\{(\pi(\lambda),\lambda)|\lambda \in V^\ast\}$ eine
    Dirac-Struktur auf $V$.
\end{Beispiel}
Diese beiden Beispiele können gewissermaßen als Spezialfälle des
folgenden Satzes angesehen werden.
\begin{Satz}\label{repres}
    Eine Dirac-Struktur $L$ auf $V$ ist äquivalent zu:
    \begin{enumerate}
    \item Einem Untervektorraum $R$ von $V$ zusammen mit einer
        antisymmetrischen Bilinearform $\Omega_L$ auf $R$.
    \item Einem Untervektorraum $K$ von $V$ zusammen mit einer
        antisymmetrischen Bilinearform $\pi_L$ auf $(V/K)^\ast$, d.h.
        einem Bivektor auf $V/K$.
    \end{enumerate}
    Dabei gilt $R=\rho(L)$ und $K=V\cap L = \ker\Omega$.
\end{Satz}
\begin{Bemerkung}
    Wenn Verwechslungen ausgeschlossen sind, lassen wir im folgenden
    den Index $L$ an $\Omega_L$ und $\pi_L$ weg.
\end{Bemerkung}
\begin{proof}
    Sei $L$ eine Dirac-Struktur. Wir setzen $R=\rho(L)$ und definieren
    eine Bilinearform $\Omega$ auf $K$ durch
    $\Omega(\rho(v))=\rho^\ast(v)|_R$, also $\Omega(x) = \eta|_R$ für
    ein $\eta \in V^\ast$ mit $(x,\eta)\in L$. $\Omega$ ist
    wohldefiniert, denn für $x\in L$ mit $\rho(x) = 0$ folgt
    $\rho^\ast(x) \in \rho(L)^\circ = R^\circ$. Weiter ist $\Omega$
    antisymmetrisch, denn mit $v_1 = (x,\eta), v_2 =(y,\mu) \in L$
    gilt $\eta(y)=- \mu(x)$ und es folgt
    $$\Omega(x,y)=\eta(y) = -\mu(x) = -\Omega(y,x).$$
    Außerdem ist
    $\ker \Omega = L\cap V$.  Genauso können wir eine antisymmetrische
    Bilinearform $\pi$ auf $\rho^\ast(L)$ definieren, und mit dem
    kanonischen Isomorphismus $\rho^\ast(L)^\ast =
    V/\rho^\ast(L)^\circ = V/V\cap L$ erhalten wir schließlich einen
    Bivektor $\pi$ auf $V/K$, wobei $K=V \cap L$.
    
    Ist umgekehrt ein Unterraum $R$ mit einer Bilinearform $\Omega$
    wie in 1. gegeben, so definieren wir einen isotropen Unterraum
    durch
    $$L = \{(x,\eta)|x\in R, \eta \in V^\ast\; \mbox{mit}\; \eta|_R =
    \Omega(x)\},$$
    und da $L$ die komplementären Unterräume $(R,\ast)$
    und $(0,R^\circ)$ enthält, ist $\dim L = \dim V$ und damit eine
    Dirac-Struktur.  Im zweiten Fall definieren wir
    $$L = \{(x,\eta)|x\in V, \eta \in K^\circ\; \mbox{mit}\;
    [x]=\pi(\eta)\},$$
    wobei wir wieder $(V/K)^\ast$ mit $K^\circ$
    identifizieren, und $[x]$ die Äquivalenzklasse von $x$ in $V/K$
    bezeichnet. $L$ ist isotrop und enthält die Unterräume
    $(\ast,K^\circ)$ und $(K,0)$.
\end{proof}

\begin{Bemerkung}
    Definieren wir auf $V \oplus V^\ast$ eine antisymmetrische
    Bilinearform durch
    $$\scal{(x,\eta),(y,\mu)}_{\_} = \eta(x) - \mu(y),$$
    dann ist
    $\Omega$ durch die Gleichung
    $$
    \rho_L^\ast \Omega =\frac{1}{2} i^\ast \scal{\dcd}_\_$$
    bestimmt, wobei $\rho_L = \rho|_L$ die Einschränkung von $\rho$
    auf die Dirac-Struktur  und $i:L \ra V \oplus V^\ast$ die
    Einbettung ist.
\end{Bemerkung}

Ist die Dirac-Struktur als Graph einer Zweiform auf $V$ oder $V^\ast$
gegeben, dann entspricht die Definition bereits einer der beiden
Charakterisierungen aus Satz \ref{repres}. Es ist interessant, auch
die jeweils andere Beschreibung zu betrachten.
\begin{Beispiel}
    Ist die Dirac-Struktur durch eine antisymmetrische Bilinearform
    $\Omega$ auf $V$ gegeben, so ist $\pi$ wegen $\ker \pi = L \cap
    V^\ast = \{0\}$ eine nicht ausgeartete antisymmetrische
    Bilinearform auf $(V/\ker \Omega)^\ast$ und es gilt $\pi^{-1} =
    \Omega_\mathrm{red}$, wobei $\Omega_\mathrm{red}$ die induzierte
    symplektische Form auf $V/\ker \Omega $ bezeichnet.
\end{Beispiel}
\begin{Beispiel}
    Ist die Dirac-Struktur durch eine antisymmetrische Bilinearform
    $\pi$ auf $V^\ast$ gegeben, so ist $K=\{0\}$, $R = \op{im} \pi$
    und $\Omega$ ist die durch $\pi$ induzierte symplektische Form auf
    $R$. (Es gilt $\ker \Omega = K = \{0\}$.) Genauer ist $\Omega$
    gegeben durch $\Omega(x) = \eta|_{\op{im} \pi}$ für ein $\eta \in
    V^\ast$ mit $\pi(\eta) = x$.
\end{Beispiel}

\subsection{Dirac-Abbildungen}

Seien $V$ und $W$ Vektorräume. Wir wollen untersuchen, wie eine
lineare Abbildung $\phi:V \lra W$ Abbildungen $\F\phi:\op{Dir}(V) \lra
\op{Dir}(W)$ und $\B\phi:\op{Dir}(W)\lra \op{Dir}(V)$ induziert. Falls
$\phi$ ein Isomorphismus ist, so können wir $\phi^{-1}$ und
$(\phi^{-1})^\ast$ bilden und es ist klar, wie $\F \phi$ und $\B \phi$
zu definieren sind. Durch die zwei verschiedenen Charakterisierungen
von Dirac-Strukturen nach Satz \ref{repres} können wir aber auch im
Allgemeinen beide induzierten Abbildungen definieren. Ist $L_W$ eine
Dirac-Struktur auf $W$, so definieren wir eine Dirac-Struktur $L_V$
auf $V$ durch $\rho(L_V)=\phi^{-1}(\rho(L_W))$ und $\Omega_V =
\phi^\ast \Omega_W$. Umgekehrt ist zu einer gegebenen Dirac-Struktur
auf $V$ durch die Vorgaben $K_W = \phi(K_V)$ und $\pi_W = \phi_\ast
\pi_V$ eine Dirac-Struktur auf $W$ gegeben. Dies kann zur Definition
der Abbildungen $\B\phi$ und $\F\phi$ dienen, etwas mehr Übersicht
bringt aber der folgende Weg nach \cite{bursztyn.radko:2003a}.

Wir übertragen die grundlegenden Definitionen aus der Theorie der
kanonischen Relationen \cite{weinstein:1980b} vom symplektischen Fall
auf Vektorräume mit einer nicht ausgearteten symmetrischen
Bilinearform der Signatur Null. Alles was wir hier brauchen ist in
\cite[Abschnitt 5.3] {abraham.marsden:1985a} zu finden, die Beweise
lassen sich direkt übertragen.

Seien $(E_i,g_i),\; i=1,2,3$ Vektorräume mit symmetrischen nicht
ausgearteten Bilinearformen der Signatur Null. Mit $\overline{E}_i$
bezeichnen wir $E_i$ zusammen mit der Form $-g_i$.  Eine kanonische
Relation $L\subseteq E_1 \times \overline{E}_2$ ist ein maximal
isotroper Unterraum von $E_1\times \overline{E}_2$. Die Menge der
kanonischen Relationen auf $E_1 \times \overline{E}_2$ bezeichnen wir
in Analogie zu Lagrangeschen Unterräumen von symplektischen
Vektorräumen mit $\op{Lag}( E_1 \times \overline{E}_2)$.

Für zwei kanonische Relationen
$L_1 \subseteq E_1 \times \overline{E}_2$ und $L_2 \subseteq E_2
\times \overline{E}_3$ definieren wir eine Verknüpfung $\circ$ durch
$$L_1 \circ L_2 = \{(e_1,e_3)\in E_1 \times E_3\;|\; \exists\; e_2\in
E_2\; \op{mit}\; (e_1,e_2)\in L_1 \;\op{und}\; (e_2,e_3) \in L_2\}.$$
Man sieht leicht, dass $L_1 \circ L_2$ ein isotroper Unterraum von
$E_1\times \overline{E}_3$ ist. Tatsächlich ist $L_1 \circ L_2$ sogar
maximal isotrop \cite[Prop. 5.3.12]{abraham.marsden:1985a}.

Seien nun $V$ und $W$ Vektorräume, und sei $\phi:V\ra W$ eine lineare
Abbildung. Wir setzen $E =(V\oplus
V^\ast,\langle\cdot\,,\cdot\rangle)$ und $F =(W\oplus
W^\ast,\langle\cdot\,,\cdot\rangle)$. Dann sind
\begin{eqnarray*}
    \mathcal{F}\phi &:=&\{(\phi(x),\eta,x,\phi^\ast \eta)\;|\;x\in V,
    \eta\in W^\ast\}\subseteq F\times \overline{E}\\ 
    \mathcal{B}\phi &:=&\{(x,\phi^\ast \eta,\phi(x), \eta)\;|\;x\in V,
    \eta\in W^\ast\}\subseteq E\times \overline{F}\\ 
\end{eqnarray*}
kanonische Relationen.
\begin{Bemerkung}\label{circ}
    Sei $U$ ein weiterer Vektorraum und $\psi: W \ra U$ linear sowie
    $H = (U \oplus U^\ast,\langle\cdot\,,\cdot\rangle)$. Dann zeigt
    eine kurze Rechnung
    \begin{eqnarray*}
        \F \psi \circ \F \phi  &=& \F(\psi \circ \phi) \; \in \;
        \op{Lag}(H\times \overline{E})\\ 
        \B \phi \circ \B \psi    &=& \B(\psi \circ \phi) \;\, \in \;
        \op{Lag}(E\times \overline{H}). 
    \end{eqnarray*}
\end{Bemerkung}
Durch die Identifikation von Dirac-Strukturen auf $V$ bzw. $W$ mit
kanonischen Relationen auf $E \times \{0\}$ bzw. $F\times \{0\}$
erhalten wir jetzt die gesuchten Abbildungen
$$\begin{array}{cccc}
    \F\phi:&\op{Dir}(V) &\lra& \op{Dir}(W)\\
    &L_V &\longmapsto& \F\phi \circ L_V\\
\end{array}
$$
und
$$\begin{array}{cccc}
    \B\phi:&\op{Dir}(W) &\lra& \op{Dir}(V)\\
    &L_W &\longmapsto& \B\phi \circ L_W.\\
\end{array}
$$
Die Symbole $\F\phi$ und $\B\phi$ stehen jetzt einerseits für
kanonische Relationen und andererseits für Abbildungen auf Mengen von
Dirac-Strukturen. Aus dem Kontext geht aber hervor, was jeweils
gemeint ist. Weiter übertragen sich die Gleichungen aus Bemerkung
\ref{circ} offensichtlich auch auf die zweite Interpretation von
$\F\phi$ bzw. $\B\phi$.  Eine kleine Rechnung führt auf die folgenden
expliziten Formeln:
$$\F\phi(L_V)=\{(\phi(x),\eta)|x\in V,\eta \in
W^\ast,(x,\phi^\ast\eta)\in L_V\},$$
$$\B\phi(L_W)=\{(x,\phi^\ast\eta)|x\in V,\eta\in W^\ast,(\phi(x),\eta)
\in L_W\}.$$
Man beachte, dass die Abbildungen $\F\phi$ und $\B\phi$
im allgemeinen nicht invers zueinander sind. Ist aber $\phi$ injektiv,
so ist $\B\phi\circ \F\phi=\op{id}$, und ist $\phi$ surjektiv, dann
gilt $\F\phi\circ\B\phi=\op{id}$.  Der nächste Satz stellt die
Verbindung zu der anfangs erwähnten alternativen Definition her.
\begin{Satz}
    Sei $\phi:V\lra W$ eine lineare Abbildung, und seien $L_V \in
    \op{Dir}(V)$, $L_W \in \op{Dir}(W)$.
    \begin{enumerate}
    \item Ist $L_W = \F\phi(L_V)$, dann gilt $\ker \Omega_{L_W} =
        \phi(\ker\Omega_{L_V})$ und $ \pi_{L_W} =
        \phi_\ast(\pi_{L_V}).$
    \item Ist $L_V = \B\phi(L_W)$, dann gilt
        $\rho({L_V)=\phi^{-1}(\rho(L_W}))$ und $\Omega_{L_V} =
        \phi^\ast(\Omega_{L_W}).$
    \end{enumerate}
\end{Satz}
\begin{proof}
    Sei $L_W = \F\phi (L_V)$. An der expliziten Form für $\F\phi
    (L_V)$ sehen wir, dass 
    $$\ker(\Omega_{L_W})=W\cap L_W =
    \{\phi(x)\,|\,x\in V,\, (x,0)\in L_V\},$$
    und da
    $\ker(\Omega_{L_V})=V\cap L_V=\{x\,|\,x\in V,\, (x,0)\in L_V\}$
    folgt $\ker(\Omega_{L_W})=\phi(\ker(\Omega_{L_V}))$.
    
    Die Abbildung $\phi:V\ra W$ induziert eine Abbildung $V/(V\cap
    L_V) \ra W/\phi(V\cap L_V)=W/(W\cap L_W)$ auf den Quotienten und
    $\phi_\ast \pi_{L_V}$ ist für $\eta \in W/(W \cap L_W)$ definiert
    durch die Gleichung $\phi_\ast \pi_{L_V}(\eta) =
    \phi(\pi_{L_V}(\phi^\ast\eta))$. Nach der Definition von
    $\pi_{L_V}$ gilt weiter $\phi(\pi_{L_V}(\phi^\ast\eta)) =\phi(x)$
    für ein $x\in V$ mit $(x,\phi^\ast \eta) \in L_V$. Ebenso ist
    $\pi_{L_W}(\eta) =y$ für ein $y\in W$ mit $(y,\eta)\in L_W$. Ist
    nun $L_W = \F\phi(L_V)$ so ist $(y,\eta)\in L_W$ genau dann, wenn
    $y=\phi(x)$ und $(x,\phi^\ast\eta) \in L_V$. Es folgt also
    $\pi_{L_W}=\phi_\ast \pi_{L_V}$.
    
    Der zweite Teil des Satzes folgt genauso wie das eben gezeigte.
\end{proof}

\begin{Bemerkung}
    Falls $L_V = \op{graph}(\pi)$ für einen Bivektor $\pi$ auf $V$, so
    gilt also $\F\phi(L_V)=\op{graph}(\phi_\ast \pi)$. Ebenso folgt
    für $L_W = \op{graph}(\Omega)$ mit einer $2$-Form $\Omega$, dass
    $\B\phi(L_W)=\op{graph}(\phi^\ast \Omega)$.
\end{Bemerkung}

\begin{Definition}
    Seien $V$ und $W$ Vektorräume mit Dirac-Strukturen $L_V$ und $L_W$,
    und sei $\phi:V \ra W$ eine lineare Abbildung.
    \begin{enumerate}
    \item $\phi$ heißt forward-Dirac-Abbildung, wenn $L_W =
        \F\phi(L_V)$.
    \item $\phi$ heißt backward-Dirac-Abbildung, wenn $L_V =
        \B\phi(L_W)$.
    \end{enumerate}
\end{Definition}
Da $\F\phi$ und $\B\phi$ im Allgemeinen nicht invers zueinander sind,
sind die beiden Definitionen nicht äquivalent.

\begin{Beispiel}
    Ist $i:W\ra V$ ein Untervektorraum und $L_V = \op{graph}(\Omega)$
    mit einer $2$-Form $\Omega$ sowie $L_W = \op{graph}(i^\ast
    \Omega)$, dann ist $L_W=\B i(L_V)$ und $i$ damit eine
    backward-Dirac-Abbildung.
\end{Beispiel}

\begin{Beispiel}\label{linRed}
    Sei $L\in \op{Dir}(V)$. Auf $V/\ker(\Omega)$ definieren wir die
    Dirac-Struktur $\tilde{L}= \op{graph}(\pi_L)$. Dann ist die
    Projektion $V \ra V/\ker(\Omega_L)$ eine forward-Dirac-Abbildung.
    Ist insbesondere $L=\op{graph}(\Omega)$, dann folgt
    $\tilde{L}=\op{graph}(\Omega_\mathrm{red})$, wobei
    $\Omega_\mathrm{red}$ durch $\Omega = \pi^\ast
    \Omega_{\mathrm{red}}$ gegeben ist.
\end{Beispiel}


\section{Dirac-Mannigfaltigkeiten}

Wir erweitern  jetzt die Definition von Dirac-Strukturen auf
Mannigfaltigkeiten, indem wir Unterbündel  in der (faserweisen)
direkten Summe  von Tangentialbündel und Kotangentialbündel
betrachten, die punktweise Dirac-Strukturen sind. Zusätzlich werden
wir aber noch eine Integrabilitätsbedingung fordern, die im linearen Fall
nicht auftaucht.

Sei also $M$ eine Mannigfaltigkeit. Auf dem Vektorbündel $TM\oplus T^\ast
M$ haben wir durch
$$\langle (X,\eta),(Y,\mu)\rangle =\eta(Y)+\mu(X) $$
kanonisch eine Bilinearform sowie durch
$$\scal{(X,\eta),(Y,\mu)}_\_ =\eta(Y)-\mu(X) $$
eine antisymmetrisch Form gegeben, wobei  $X,Y \in
\Gamma^\infty (TM)$ und $\eta, \mu \in \Omega^1(M)$.

\begin{Definition}
    Eine verallgemeinerte Dirac-Struktur auf $M$ ist ein bezüglich
    $\langle\dcd \rangle$ maximal isotropes Untervektorbündel $L$ von
    $TM\oplus T^\ast M$.
\end{Definition}

Weiter definieren wir die Vektorbündelhomomorphismen
$$\begin{array}{ccccccccc}
    \rho:& \stCA &\ra& TM &\qquad&\rho^\ast:& \stCA &\ra& TM
    \\[2mm] 
    & (X,\alpha) &\mapsto& X &\qquad&  & (X,\alpha) &\mapsto& \alpha\\
\end{array}$$
und erhalten wie im linearen Fall für eine
Dirac-Struktur $L$ punktweise die Gleichungen 
\begin{eqnarray*}
    \rho(L)^\circ &=& L \cap T^\ast M,\\
    \rho^\ast (L)^\circ &=& L \cap TM.
\end{eqnarray*}

\begin{Beispiele}
    Die Beispiel, die wir im Fall linearer Dirac-Strukturen gegeben
    haben, übertragen sich auf offensichtliche Weise:
\begin{enumerate}
\item Sei $\omega \in \Omega^2(M)$. Dann ist $\gra(\omega) \subseteq
    TM \oplus T^\ast M$ eine verallgemeinerte Dirac-Struktur.
\item Sei $\pi \in \mulV{2}(M) = \Gamma^\infty(\Wedge^2 TM)$. Dann ist
    $\gra(\pi) \subseteq TM \oplus T^\ast M$ eine verallgemeinerte
    Dirac-Struktur.
\end{enumerate}
\end{Beispiele}

\begin{Definition}
    Seien $M,N$ Mannigfaltigkeiten mit verallgemeinerten
    Dirac-Strukturen $L_M$ und $L_N$. Sei $\phi:M \ra N$ eine glatte
    Abbildung. Dann heißt $\phi$ forward-Dirac-Abbildung, wenn $T_m
    \phi: T_m M \ra T_{\phi(m)}N$ für alle $m \in M$ eine
    forward-Dirac-Abbildung ist. Entsprechend heißt $\phi$
    backward-Dirac-Abbildung, wenn $T_m\phi$ für alle $m \in M$ eine
    backward-Dirac-Abbildung ist.
\end{Definition}

\subsection{Courant-Klammer und Integrabilität}
Wir haben gesehen, dass der Graph einer Zweiformen $\omega$ auf $M$ eine
verallgemeinerte Dirac-Struktur definiert. Man ist aber 
natürlich besonders an solchen Formen $\omega$ interessiert, die
geschlossen sind. Ebenso wird man vor allem solche Graphen von
Bivektorfeldern $\pi$ betrachten wollen, für die die Bedingung
$[\pi,\pi] = 0$ erfüllt ist. Dabei bezeichnet $[\dcd]$ die
Schouten-Nijenhuis-Klammer auf $M$, siehe \ref{defSchouten}. Wir
werden deshalb jetzt auf $\sect(\stCA)$ eine Verknüpfung einführen,
mit deren Hilfe wir eine Bedingung an die Dirac-Strukturen stellen
können, die eine Verallgemeinerung der beiden oben genannten
Forderungen darstellt. 

\begin{Definition}
    Die Courant-Klammer $$[\dcd]:\sect(\stCA)\times \sect(\stCA) \ra
    \sect(\stCA)$$
    ist für $(X,\eta),(Y,\mu) \in \sect(\stCA)$ durch
    $$[(X,\eta),(Y,\mu)]=([X,Y],\Lie_X \mu - i_Y \dif \eta)$$
    gegeben.
\end{Definition}

\begin{Bemerkung}
    Abweichend von \cite{courant:1990a} definieren wir die
    Courant-Klammer in der nicht-anti"-symmetrischen Form, da dies die
    heute gebräuchlichere Definition ist. Zur Äqui\-valenz
    der verschiedenen Definitionen siehe \cite[Prop
    2.6.5]{roytenberg:1999a} sowie \cite{liu.weinstein.xu:1997a}.
\end{Bemerkung} 

Wir verwenden für die Courant-Klammer auf $\stCA$ die
gleiche Bezeichnung wie für die Lieklammer von Vektorfeldern bzw. wie
für die Schouten-Nijenhuis-Klammer (siehe \ref{defSchouten})  von Multivektorfeldern. Dies ist
unproblematisch, da die Einschränkung aller drei Klammern auf
$\mathfrak{X}(M) = \sect(TM)$ übereinstimmt.

\begin{Definition}
    Eine verallgemeinerte Dirac-Struktur $L$ heißt integrabel oder kurz
    Dirac-Struktur, wenn $L$ bezüglich der Courant-Klammer
    abgeschlossen ist,
    $$[\sect(L),\sect(L)] \subseteq \sect(L).$$
\end{Definition}

\begin{Lemma}
\begin{enumerate}
\item Sei $\omega \in \Omega^2(M)$. Dann ist $\gra(\omega)$ genau dann
    eine Dirac-Struktur, wenn $\omega$ geschlossen ist,
    $$\dif \omega =0.$$
\item Sei $\pi \in \mulV{2}(M)$. Dann ist $\gra( \pi ) $ genau dann
    eine Dirac-Struktur, wenn gilt
    $$[\pi,\pi] = 0,$$
    wobei $[\dcd]$ die Schouten-Nijenhuis-Klammer auf $M$ bezeichnet.
\end{enumerate}
\end{Lemma}
\begin{proof}
    Der erste Teil folgt aus der Gleichung (vgl. \ref{defSchouten})
    $$i_{[X,Y]}\omega = \Lie_X i_Y \omega- i_Y \Lie_X \omega = \Lie_X
    i_Y \omega- i_Y \dif i_X \omega -i_Y i_X\dif\omega,$$
    der zweite ergibt sich mit (siehe
    \cite{kosmann-schwarzbach.magri:1990a}) 
    $$\frac{1}{2}i_\beta i_\alpha[\pi,\pi] = \pi(\Lie_{\pi(\alpha)}\beta -
    i_{\pi(\beta)} \dif \alpha) - [\pi(\alpha),\pi(\beta)].$$
    Wir werden diese beiden Aussagen später (Abschnitt \ref{graphPreSymp}
    und \ref{graphPoisson}) aber auch aus allgemeineren Betrachtungen 
    wiedergewinnen.
\end{proof}

Ist $\phi: M\ra M$ ein Diffeomorphismus, dann haben wir durch
$(T\phi,T_\ast\phi):\stCA \ra \stCA$ mit $T_\ast \phi = (T
\phi^{-1})^\ast$ einen kanonischen Lift von $\phi$ auf $\stCA$
gegeben. Ist $L$ eine Dirac-Struktur, dann gilt $\F\phi(L) =
(T\phi,T_\ast\phi)(L)$. Man sieht leicht, dass $(T\phi,T_\ast\phi)$
eine Isometrie von $\scal{\dcd}$ sowie natürlich bezüglich der
Courant-Klammer ist, mit anderen Worten ein Automorphismus von $\stCA$
ist. Somit sind $\F\phi(L)$ und $\B\phi(L) = \F\phi^{-1}(L)$ ebenfalls
Dirac-Strukturen.

\begin{Definition}[Eichtransformationen]
    Sei $B$ eine Zweiform auf M. Betrachten wir $B$ als eine Abbildung
    $B:TM \ra T^\ast M$, dann ist durch $\tau_B(X,\alpha) = (X,B(X)
    +\alpha)$ eine Abbildung $\tau_B:\stCA \ra \stCA$ definiert. Wir
    nennen $\tau_B$ eine Eichtransformation.
\end{Definition}

\begin{Lemma}\label{eichtrafo} 
    $\tau_B$ erhält genau dann die Courant-Klammer auf $\stCA$, wenn
    $B$ geschlossen ist.
\end{Lemma}

\begin{proof}
    Dass $\tau_B$ den Anker erhält, ist klar. Weiter folgt aus der
    Antisymmetrie von $B$, dass $\tau_B$ eine Isometrie von
    $\scal{\dcd}$ ist. Damit bleibt noch, das Verhalten der
    Courant-Klammer unter $\tau_B$ zu untersuchen,
    \begin{eqnarray*}
        {[\tau_B(X,\alpha),\tau_B(Y,\beta)]}
        &=&[(X,B(X)+\alpha),(Y,B(Y)+\beta)] \\ 
        &=& ([X,Y],\Lie_X \beta - i_Y
        \dif \alpha +\Lie_X i_Y B - i_Y \dif i_X B)\\
        &=& [(X,\alpha),(Y,\beta)] + \Lie_X i_Y B - i_Y \Lie_X B + i_Y
        i_X \dif B\\
        &=& [(X,\alpha),(Y,\beta)] + i_{[X,Y]}B + i_Y i_X \dif B\\
        &=& \tau_B([(X,\alpha),(Y,\beta)]) + i_Y i_X \dif B.
    \end{eqnarray*}
    Die Courant-Klammer bleibt also genau dann erhalten, wenn $B$
    geschlossen ist.
\end{proof}

Für spätere Anwendungen formulieren wir hier noch folgendes Lemma.
\begin{Lemma} \label{eichequi}
    Sei $\phi$ ein Diffeomorphismus von $M$ und $B$ eine Zweiform. Dann
    gilt
    $$\tau_B \circ (T\phi,T_\ast \phi) = (T\phi,T_\ast \phi) \circ
    \tau_{\phi^\ast B}.
    $$
\end{Lemma}
\begin{proof}
    Für $X \in \mathfrak{X}(M)$ und $\alpha\in\Omega^1(M)$ gilt
    \begin{eqnarray*}
        \tau_B\circ(T\phi,T_\ast\phi)(X,\alpha) &=&
        (T\phi(X),B(T\phi(X))+T_\ast(\alpha))\\
        &=& (T\phi(X), (\phi^\ast B)(X)\circ T\phi^{-1} +
        T_\ast\phi(\alpha))\\
        &=& (T\phi(X),T_\ast\phi ((\phi^\ast B)(X) + \alpha))\\
        &=& (T\phi,T_\ast\phi)\circ\tau_{\phi^\ast B} (X,\alpha).
    \end{eqnarray*}

\end{proof}

\begin{Lemma}\label{eigenCWK}
    Seien $e_1 =(X_1,\alpha_1)$, $e_2 =(X_2,\alpha_2)$, $e_3 =
    (X_3,\alpha_3) \in \sect{(\stCA)}$. Für die Courant-Klammer gelten
    folgende Identitäten:
    \begin{enumerate}
    \item Jacobi-Identität in der Form
        $$[e_1,[e_2,e_3]] = [[e_1,e_2],e_3]+[e_2,[e_1,e_3]].$$
        
    \item Leibniz-Identität:
        \begin{eqnarray*}
            [e_1,f e_2] &=& f [e_1,e_2] + (\rho(e_1) f) e_2\\
            &=&  f[e_1,e_2] + \Lie_{X_1}f\,e_2.
        \end{eqnarray*}

    \item Defekt in der Antisymmetrie:
        $$[e_1,e_2] + [e_2,e_1] = (0,\dif\scal{e_1,e_2}),$$
        insbesondere also  $[e_1,e_1] = \frac{1}{2}
        (0,\dif\scal{e_1,e_1}).$ 
        
    \item $\Lie_{X_1}\scal{e_2,e_3}=
        \scal{[e_1,e_2],e_3}+\scal{e_2,[e_1,e_3]}$. 
    \end{enumerate} 
\end{Lemma}

\begin{proof}
    Der Beweis erfolgt durch direktes Nachrechnen.
\end{proof}

Aus den ersten beiden Punkten folgt nun unmittelbar:
\begin{Lemma}
    Sei $L\subseteq \stCA$ eine verallgemeinerte Dirac-Struktur. Dann
    ist $L$ zusammen mit der auf $L$ eingeschränkten Courant-Klammer
    und dem auf $L$ eingeschränkten Anker genau dann ein Lie-Algebroid
    (siehe \ref{defLieAlg}), wenn $L$ integrabel ist.
\end{Lemma}

Aus der Theorie der Lie-Algebroide erhalten wir folgendes Lemma.
\begin{Lemma}
    Sei $L \subseteq TM\oplus T^\ast M$ eine Dirac-Struktur (also
    integrabel). Dann ist $\rho(L) \subseteq TM$ eine integrable
    singuläre Distribution (siehe \cite{sussmann:1973a} oder
    \cite[3.21]{michor:2004a}). D.h. es gibt eine (singuläre)
    Blät\-ter\-ung von $M$, so dass die Tangentialräume der Blätter
    punktweise mit $\rho(L)$ übereinstimmen. Singulär bedeutet dabei,
    dass die Dimensionen der einzelnen Blätter nicht übereinstimmen müssen.
\end{Lemma} 
\begin{proof}
    Seien $s_1,\ldots,s_n$ lokale Basisschnitte der Dirac-Struktur $L$, und
    seien durch
    $$[s_i,s_j]=c_{ij}^k s_k $$
    lokale Funktionen $c_{ij}^k$ definiert. Dann spannen die
    Vektorfelder $\rho(e_1),\ldots,\rho(s_n)$ lokal die Distribution $\rho(L)
    \subseteq TM$ auf, und es gilt
    $$[\rho(e_i),\rho(e_j)] = c_{ij}^k \rho(e_k),$$
    womit die
    Integrabilität folgt\footnote{Man beachte, dass diese Bedingung
      stärker ist als nur die Involutivität, welche bei singulären
      Distributionen für die Integrabilität bekanntlich noch nicht
      hinreichend ist.}, vgl. \cite[Abschnitt 8]{sussmann:1973a}.
\end{proof}

\begin{Lemma}
    Sei $L$ eine verallgemeinerte Dirac-Struktur und seien
    $e_1,e_2,e_3 \in \sect(L)$. Wir definieren eine Abbildung
    $$T: \sect(L)\times \sect(L) \times \sect(L) \lra \R\quad$$
    durch
    $$T(e_1,e_2,e_3)=\scal{[e_1,e_2],e_3}.$$
    Dann ist $T$
    antisymmetrisch und $C^\infty(M)$-linear, also $T\in
    \sect(\Wedge^3 L^\ast)$, und $L$ ist genau dann integrabel, wenn
    $T=0$ gilt.
\end{Lemma}
\begin{proof}
    Die Funktionenlinearität in den ersten beiden Argumenten zeigt man
    mit Lemma \ref{eigenCWK} unter Beachtung der Isotropie von $L$,
    die Funktionenlinearität im letzten Argument ist klar.  Unter
    Ausnutzung der Isotropie vom $L$ lässt sich $T$ umformen zu
    $$\begin{array}{ccl} T((X,\eta),(Y,\mu),(Z,\nu)) &=& i_X \dif
        i_Y \nu - i_Y \dif i_X
        \nu + i_Z \dif i_X \mu \\[2mm]
        &&\quad -\dif \omega(Y,Z) - \dif \mu(Z,X) -\dif \nu(X,Y)\\[3mm]
        &=& -\frac{1}{2}\bigl( i_X \dif \scal{(Y,\mu),(Z,\nu)}_\_ +
        i_Y \dif \scal{(Z,\nu),(X,\omega)}_\_\\[2mm]
        &&\qquad\quad+ i_Z \dif \scal{(X,\omega),(Y,\mu)}_\_\bigr) \\[2mm]
        &&\quad- \dif \omega(Y,Z) - \dif \mu(Z,X) -\dif \nu(X,Y),
    \end{array}
    $$
    womit die Antisymmetrie von $T$ folgt.  Ist $L$ integrabel, so
    ist offensichtlich $T=0$. Da aber $L$ maximal isotrop ist, gilt
    auch die Umkehrung.
\end{proof}

\subsection{Präsymplektische Blätterung}
Wenn wir Lemma \ref{repres} punktweise auf eine Dirac-Struktur
anwenden, erhalten wir eine antisymmetrische Abbildung
$$\Omega:\rho(L)\times \rho(L) \ra \R,$$
wobei für $X,Y \in \rho(L)$ mit $(X,\alpha),(Y,\beta) \in L$ gilt, dass
$$\Omega(X,Y) = \alpha(Y) = - \beta(X).$$

Sei $i:N \hra M$ ein Blatt zu der Distribution $\rho(L)$, und sei
$\Omega_N$ die auf $N$ eingeschränkte Zweiform $\Omega$. Weiter sei
$\rho_N$ die auf $L|_N$ eingeschränkte Projektion, $\rho_N=
\rho|_{L_N}$. Dann haben wir folgendes Diagramm
$$\xymatrix{ TN=\rho(L)|_N \ar[ddr] & L|_N \ar[dd]
  \ar@{->>}[l]_-<<{\rho_N} \ar@{^{(}->} [r]^-{\overline{i}} &\, \stCA
  \ar[dd]\\
  &&\\
  &N \ar[r]^-i & M }
$$
wobei jetzt alle Abbildungen Vektorbündelhomomorphismen sind.  Wie
im linearen Fall gilt jetzt
$$\rho_N^\ast \Omega_N = \frac{1}{2}\,i^\ast \scal{\dcd}_\_\,,$$
womit folgt, dass $\Omega_N$ eine glatte Zweiform auf $N$ ist.

Seien jetzt $X_1=\rho(e_1)$, $X_2 =\rho(e_2)$, $X_3 = \rho(e_3) \in
\sect(TN) \subset \sect(\rho(L))$. Es lässt sich zeigen
\cite{courant:1990a}, dass die Gleichung
$$\dif \Omega_N(X_1,X_2,X_3)=-T(e_1,e_2,e_3)$$
gilt, und wir erhalten die
\begin{Folgerung}
    Eine integrable Dirac-Struktur hat eine Blätterung mit
    präsymplektischen Blät\-tern, d.h. auf jedem Blatt $N$ ist eine
    geschlossene Zweiform $\Omega_N$ gegeben.
\end{Folgerung}
\begin{Beispiel}
    War $L$ der Graph einer präsymplektischen Form, so ist $\rho(L) =
    TM$ und das einzige Blatt ist die Mannigfaltigkeit $M$ selbst.  Im
    Falle, dass $L$ der Graph einer Poisson-Struktur $\pi$ ist, haben
    wir $\rho(L) = \pi(T^\ast M)$. Ist $N$ ein Blatt dieser
    Distribution, so folgt mit $\ker \Omega_N = L\cap TM|_N = \{0\}$,
    dass $(N,\Omega_N)$ eine symplektische Mannigfaltigkeit ist. Wir
    erhalten also die bekannte Blätterung einer
    Poisson-Mannigfaltigkeit durch symplektische Blätter
    \cite[Abschnitt 5.1]{cannasdasilva.weinstein:1999a}.
\end{Beispiel}

\subsection{Die Poisson-Algebra der zulässigen Funktionen}

\begin{Definition}
    Eine Funktion $f \in C^\infty(M)$ heißt zulässig, wenn $\dif f|_m
    \in \rho^\ast(L)$ für alle $m \in M$.
\end{Definition}
\begin{Bemerkung}
    In der Physik spielen die zulässigen Funktionen die Rolle der
    eichinvarianten Funktionen, vgl. Satz \ref{PhasenRed}.
\end{Bemerkung}
Ist $f$ eine zulässige Funktion, dann gibt es ein Vektorfeld $X_f$, so
dass $(X_f,\dif f) \in \sect(L)$. Wir nennen $X_f$ ein Hamiltonsches
Vektorfeld zu $f$. Im Allgemeinen ist $X_f$ jedoch nicht eindeutig.
Trotzdem ist für zwei zulässige Funktionen $f,g$ durch $$\dirac{f,g}=
X_f(g) = \Omega_L(X_f,X_g)$$
eine Klammer definiert. Denn sei $X'_f$
ein zweites Vektorfeld mit $(X'_f,\dif f) \in \sect(L)$, dann folgt
$X_f - X'_f \in \sect(L\cap TM) = \rho^\ast(L)^\circ$, und damit $(X_f
-X'_f)(g) = 0$.
\begin{Bemerkung}
    Ist $X_f$ ein Hamiltonsches Vektorfeld zu einer zulässigen
    Funktion $f$, so verläuft der Fluss zu $X_f$ innerhalb eines
    Blattes der verallgemeinerten Distribution $\rho(L)$, denn es gilt
    ja  $(X_f,\dif f) \in \sect(L)$ und damit $X_f \in
    \sect(\rho(L))$.
    Die Blätterung einer Dirac-Struktur bildet also eine Art
    Superauswahlregel.
\end{Bemerkung}
\begin{Satz}[{\cite[Prop. 2.5.1]{courant:1990a}}]
    Die Menge der zulässigen Funktionen zusammen mit der eben
    definierten Klammer $\dirac{\dcd}$ ist eine Poisson-Algebra.
\end{Satz}
\begin{proof}
    Es ist klar, dass die zulässigen Funktionen einen $\R$-Vektorraum
    bilden. Seien $f,g$ und $h$ zulässige Funktionen und $X_f, X_g$
    und $X_h$ zugehörige Hamiltonsche Vektorfelder. Dann ist durch
    $$g (X_f,\dif f) + f (X_g,\dif g) = (g X_f + f X_g,\dif(f g))$$
    ein Schnitt in $L$ definiert, womit die Abgeschlossenheit der
    zulässigen Funktionen unter der Multiplikation folgt. Weiter gilt
    $$\dirac{f g,h} = g X_f( h) + f X_g(h)  = g\dirac{f,h} +
    f\dirac{g,h},$$
    und damit die Leibnizregel für $\dirac{\dcd}$.
    Für die Abgeschlossenheit unter $\dirac{\dcd}$ berechnen wir 
    $$[(X_f,\dif f),(X_g,\dif g)] = ([X_f,X_g],\Lie_{X_f} \dif g) =
    ([X_f,X_g],\dif\dirac{f,g})$$
    und schließlich ist die
    Jacobi-Identität äquivalent zu der Integrabilität von $L$, denn es
    gilt
   \begin{eqnarray*}
       T\big((X_f,\dif f),(X_g,\dif g),(X_h,\dif h)\big) &=&
       \scal{([X_f,X_g],\dif\dirac{f,g}),(X_h,\dif h)}\\
       &=& [X_f,X_g](h) + X_h(\dirac{f,g}) \\
       &=& X_f(\dirac{g,h})-X_g(\dirac{f,h}) + X_h(\dirac{f,g})\\
       &=& \dirac{f,\dirac{g,h}}+ \dirac{g,\dirac{h,f}} +
       \dirac{h,\dirac{f,g}}.
       \end{eqnarray*}
\end{proof}

\subsection{Dirac-Mannigfaltigkeiten und Reduktion}\label{diracRedu}

Sei $L \subseteq \stCA$ eine Dirac-Struktur. Wir betrachten die
Distribution 
$\ker \Omega = L \cap TM = \ker \rho^\ast|_L$ auf $M$.  Mit der
Integrabilität von $L$ folgt sofort, dass $L \cap TM$ involutiv ist.
\begin{Folgerung}
    Hat die Distribution $L \cap TM$ konstante Dimension, so ist sie
    integrabel.
\end{Folgerung}
\begin{Bemerkung}
    Im Fall, dass $L$ der Graph einer präsymplektische Form $\omega$
    mit konstantem Rang ist, folgt wegen $L\cap TM = \ker\omega$ die
    Integrabilität der charakteristischen Distribution \cite[Abschnitt
    4.3]{abraham.marsden:1985a}.
\end{Bemerkung}

\begin{Satz}[{\cite[Corollary 2.6.3]{courant:1990a}}]\label{PhasenRed}
    Sei die Dimension von $L \cap TM$ konstant, und bezeichne $\F$ die
    induzierte Blätterung. Ist $M/\F$ eine Mannigfaltigkeit und die
    Projektion $\pi:M\ra M/\F$ eine glatte Submersion, dann existiert
    auf $M/\F$ eine Poisson-Struktur, so dass die Projektion eine
    forward-Dirac-Abbildung ist.
\end{Satz}
\begin{proof}
    Man überlegt sich, dass Funktionen auf $M/\F$ mit den zulässigen
    Funktionen auf $M$ identifiziert werden können. Dadurch wird
    $C^\infty(M/\F)$ zu einer Poisson-Algebra. Die letzte Aussage
    folgt wie in Beispiel \ref{linRed}.
\end{proof}

Sei $L \in \stCA$ eine verallgemeinerte Dirac-Struktur und sei $i:N
\hra M$ eine Untermannigfaltigkeit. Wir definieren 
$$L_N = \frac{L \cap (TN \oplus T^\ast M)}{L \cap TN^\circ}$$
und
identifizieren $L_N$ punktweise mit einem Unterraum von $\stCA[N]$,
wobei $L_N$ punktweise maximal isotrop ist. Ist $L_N$ sogar ein
glattes Unterbündel, dann folgt, dass $L_N$ eine verallgemeinerte
Dirac-Struktur auf $N$ ist.  Es gilt nun der folgende Satz, für einen
Beweis siehe \cite[Abschnitt 3.1]{courant:1990a}.
\begin{Satz}\label{DiracRed}
    Gelten die oben genannten Voraussetzungen, dann sind äquivalent:
    \begin{enumerateR}
    \item $L \cap (TN \oplus T^\ast M)$ hat konstante Dimension.
    \item $L \cap TN^\circ$ hat konstante Dimension.
    \end{enumerateR}
    Falls eine (und damit beide) dieser Bedingungen gelten, so ist
    $L_N$ eine verallgemeinerte Dirac-Struktur auf $N$ und die
    Inklusion $i:N \hra M$ ist eine backward-Dirac-Abbildung. War
    außerdem $L$ integrabel, so ist auch $L_N$ integrabel.
\end{Satz}

\begin{Beispiel}[Marsden-Weinstein-Reduktion,
    {\cite[Theo. 4.3.1]{abraham.marsden:1985a}}] 
    Sei $(M,\omega)$ eine
    symplektische Mannigfaltigkeit und $J:M\ra \g^\ast$ eine
    äquivariante Impulsabbildung zu einer symplektischen Gruppenaktion
    auf $M$. Weiter sei $i:J^{-1}(\mu)\hra M$ für $\mu \in \g^\ast$
    eine Untermannigfaltigkeit sowie $L= \op{graph}(\omega)$.  Man
    findet, dass die Bedingungen von Satz \ref{DiracRed} erfüllt sind
    und dass $L_{J^{-1}(\mu)} = \op{graph}(i^\ast\omega)$. Weiter gilt
    aufgrund der Voraussetzungen an Gruppenaktion und Impulsabbildung,
    dass $\op{ker}(i^\ast \omega) = L_{J^{-1}(\mu)} \cap
    T\(J^{-1}(\mu)\)$ konstante Dimension hat.  Bezeichnet
    $\mathcal{F}$ die durch $\op{ker}(i^\ast\omega)$ definierte
    Foliation von $J^{-1}(\mu)$, dann erhalten wir unter den in Satz
    \ref{PhasenRed} genannten Voraussetzungen insgesamt folgendes
    Diagramm:
    $$
    \xymatrix{
      & M \\
      J^{-1}(\mu) \ar@{^{(}->}[ru]^-i \ar@{->>}[rd]^-\pi & \\
      & J^{-1}(\mu)/\F }
    $$
    In diesem Fall gehört die Poisson-Struktur auf $J^{-1}(\mu)/\F$
    sogar zu einer symplektischen Form $\omega_{red}$. Die Aussage, dass
    $i$ eine backward- und $\pi$ eine forward-Dirac-Abbildung ist,
    überträgt sich zu
    $$\pi^\ast \omega_{red} = i^\ast \omega.$$
    War $M$ eine
    Poisson-Mannigfaltigkeit, so erhalten wir unter den entsprechenden
    Voraussetzungen, dass auch $J^{-1}(\mu)/\F$ eine
    Poisson-Mannigfaltigkeit ist, siehe \cite[Abschnitt
    3.3]{courant:1990a}.
\end{Beispiel}

\section{Anwendungen}

\subsection{Implizite Hamiltonsche Systeme}
Als Anwendungsbeispiel für Dirac-Strukturen geben wir eine kurze
Darstellung der Grundlagen Impliziter Hamiltonscher Systeme. Für
weiterführende Informationen siehe z.B. \cite{ dalsmo.schaft:1999a,
  blankenstein.schaft:2001a, blankenstein.ratiu:2004, schaft:1998a}.
\begin{Definition}
Sei $M$ eine Mannigfaltigkeit mit (verallgemeinerter) Dirac-Struktur
$L$ und Hamiltonfunktion $H \in C^\infty(M)$. Das (verallgemeinerte)
implizite Hamiltonsche System ist die Menge aller glatten Kurven
$\gamma: \R \supset I \ra M$ mit 
$$\(\dot{\gamma}(t),dH|_{\gamma(t)}\) \in L.$$
\end{Definition}
Die Isotropie der Dirac-Struktur stellt sicher, dass die
Hamiltonfunktion eine Erhaltungsgröße ist, dass heißt es gilt
$$\frac{\dif}{\dif t} H(\gamma(t)) = \scal{\dif H,
  \dot{\gamma(t)}}=0.$$
Die Bewegungsgleichung für ein implizites
Hamiltonsches System hat im Allgemeinen nicht durch jeden Punkt von
$M$ eine Lösung. Dies ist schon deshalb klar, da die Gleichung nur für
Punkte $m\in M$ mit $\dif H|_m \in \rho^\ast(L)$ überhaupt erfüllt
sein kann. Tatsächlich müssen wir aber im Allgemeinen zu einer noch
kleineren Teilmenge von $M$ übergehen, um die Bewegungsgleichung
überall lösen zu können. Weiter muss die Lösung, falls eine existiert,
auch nicht eindeutig sein. 

Die Lösungskurven $\gamma(t)$ zu einem impliziten Hamiltonschen System
verlaufen immer in einem bestimmten Blatt der Distribution $\rho(L)$.
Schränken wir uns auf ein Blatt $i:N \hra M$ ein, so entspricht die
Lösung des implititen Hamiltonschen Systems der Lösung der
Hamiltonschen Gleichungen auf der präsymplektischen Mannigfaltigkeit
$(N,\Omega_N)$ zur Hamiltonfunktion $H|_N$. Siehe dazu auch den
nächsten Abschnitt über Diracsche Zwangsbedingungen.

Wir wollen sehen, wie die Bewegungsgleichungen lokal aussehen. Sei
deshalb jetzt $M = \R^n$. Nehmen wir weiter an, dass $\rho^\ast(L)$
konstante Dimension $n-k$ hat. Wir definieren dann wie in Satz
\ref{repres} beschrieben eine Abbildung $J:\rho^\ast(L) \ra
(\rho^\ast(L))^\ast = TM/(L\cap TM)$, die wir beliebig zu einer
Abbildung $J:T^\ast M \ra TM/(L\cap TM)$ fortsetzen. Seien nun
$g_1,\ldots,g_k$ Basisschnitte von $L \cap TM = (\rho^\ast(L))^\circ$
und $g:\R^k \ra L\cap TM$ die dadurch gegebene Abbildung. Dann folgt
$$\dif H \in \rho^\ast(L)\quad \Leftrightarrow \quad g^\ast \dif H =
\dif H \circ g = 0.$$
Identifizieren wir nun noch $TM/(L\cap TM)$ mit
einem zu $L\cap TM$ komplementären Unterbündel von $TM$, so 
übersetzt sich die Bedingung $(\dot{x},\dif H)\in L$ in die folgenden
Gleichungen:
\begin{eqnarray*}
    \dot{x} &=& J(\dif H|_x) +  \lambda_i g_i|_x\\
    0  &=&  g^\ast \dif H
\end{eqnarray*}
Dabei sind die Lagrangeschen Multiplikatoren
$\lambda_1,\ldots,\lambda_k$ Funktionen von $t$ und durch die
Forderung der zweiten Gleichung eingeschränkt.  Im Allgemeinen werden
dadurch nicht alle $\lambda_i$'s bestimmt sein.  Für den Fall, dass
die Dirac-Struktur der Graph eines Poisson-Tensors $J:T^\ast M \ra TM$
ist, gilt $L \cap TM = (\rho^\ast(L))^\circ = \{0\}$ und damit $g=0$,
und wir erhalten die gewöhnlichen Hamiltonschen Gleichungen.

Ebenso können wir, sofern $\rho(L)$ konstante Dimension $n-k$ hat,
eine Abbildung $\omega:\rho(L) \ra \rho(L)^\ast$ definieren und diese
zu einer Abbildung $\omega:TM \ra \rho(L)^\ast$ erweitern. Sind jetzt
$p_1,\ldots,p_k$ Basisschnitte von $L\cap T^\ast M = (\rho(L))^\circ$
und somit $p:\R^k \ra L\cap T^\ast M$, erhalten wir die Hamiltonschen
Gleichungen
\begin{eqnarray*}
    \dif H &=& i_{\dot{x}}\omega + \lambda_i p_i\\
    0 &=& p^\ast(\dot{x})
\end{eqnarray*}
mit Lagrangeschen Multiplikatoren $\lambda_i$, wobei wir noch
$\rho(L)^\ast$ mit einem zu $\rho(L)^\circ$ komplementären Unterbündel
identifiziert haben.

\begin{Beispiel}[Kinematische Zwangsbedingungen]
Sei $Q$ der Konfigurationsraum eines physikalischen Systems. Seien
$(q_1,\ldots,q_n)$ lokale Koordinaten und seien
$r$ Zwangsbedingungen lokal durch 
$$\sum_{j=1}^n \alpha^i_j(q) \dot{q}^j = 0,\qquad \text{für}\;
i=1,\ldots,r$$
als Bedingung an die Lösungskurven $q(t)$ gegeben. Die
$a_j^i$ sind dabei lokal definierte Funktionen auf $Q$. Wir wollen
annehmen, dass es $k$ global definierte Einsformen $\alpha_i$ gibt, so
dass die Zwangsbedingungen in der Form $\alpha_i(\dot{\gamma})=0$
geschrieben werden können, wobei $\gamma$ eine Kurve in $Q$
bezeichnet. Sei $\tau: T^\ast Q \ra Q$ die Projektion, dann definieren
wir eine Dirac-Struktur $L$ auf $M=T^\ast Q$ durch die Forderung
$$L \cap T^\ast M =
\op{span}\{\tau^\ast\alpha_1,\ldots,\tau^\ast\alpha_r\}$$
sowie durch
die Einschränkung der kanonischen symplektischen Zweiform $\omega_0
\in \Omega^2(T^\ast Q)$ auf $\rho(L) = (L\cap T^\ast M)^\circ$,
vergleiche Satz $\ref{repres}$. Man kann zeigen, dass die so
definierte verallgemeinerte Dirac-Struktur genau dann integrabel ist,
wenn die Zwangsbedingungen holonom sind \cite{dalsmo.schaft:1999a}.
Sei nun $H:M\ra\R$ eine Hamiltonfunktion. Die Gleichungen des durch
$(M,L,H)$ gegebenen impliziten Hamiltonschen Systems können wir damit
schreiben als
\begin{eqnarray*}
    \dif H &=& i_{\dot{x}}\omega + \lambda_r \tau^\ast\alpha_r \\
    0 &=& \alpha(T\tau(\dot{x}))
\end{eqnarray*}
oder, wenn wir zu der oben zuerst genannten Formulierung
übergehen und außerdem kanonische Koordinaten verwenden,
\begin{eqnarray*}
    \dot{q}_i &=& \frac{\partial H}{\partial p_i} \\
    \dot{p}_j &=& -\frac{\partial H}{\partial q_j}  + \lambda_r
    \alpha^r_i \\
    0&=& \alpha^r_i \frac{\partial H}{\partial p_i}.
\end{eqnarray*}
Dies sind die Hamiltonschen Gleichungen für Systeme mit
nicht-holonomen Zwangsbedingungen, siehe z.B. \cite[Abschnitt
5.8]{bloch:2003a}. Ist die kinetische Energie durch eine positiv
definite Metrik auf $Q$ gegeben, so kann man zeigen
\cite{dalsmo.schaft:1999a}, dass die Hamiltonschen Gleichungen durch
jeden Punkt der Untermannigfaltigkeit
$$M_C = \{x\in T^\ast Q\,|\, \dif H|_x \in\rho^\ast(L)\} \subseteq
T^\ast Q$$
eine eindeutige Lösung haben.

\end{Beispiel}

\subsection{Diracsche Theorie der Zwangsbedingungen}

Sei $Q$ der Konfigurationsraum eines physikalischen Systems mit
einer Lagrangefunktion $L \in C^\infty(TQ)$. Wir definieren durch
$$\FL(v)(w)= \dnd_{t=0}L(v+t\, w)$$
die Faserableitung $\FL:TQ\ra T^\ast Q$. Ist $\FL$ ein Diffeomorphismus,
so können wir die Hamiltonfunktion $H\in C^\infty(T^\ast Q)$ als die
Legendretransformierte von $L$ definieren,
$$H\circ \FL(v) = \FL(v)v-L(v).$$
Ist $\FL$ kein Diffeomorphismus, so
ist die Hamiltonfunktion im Allgemeinen nicht definiert. In bestimmten
Situationen können wir jedoch eine Hamiltonfunktion auf dem Bild der
Faserableitung definieren. Nehmen wir an, dass $M:=\FL(TQ)
\stackrel{i}{\hra} T^\ast Q$ eine Mannigfaltigkeit ist, sowie dass die
Urbilder von Punkten in $M$ unter $\FL$ zusammenhängend sind, dann
können wir zeigen, dass die Funktion $v \mapsto \FL(v)v - L(v)$ auf
dem Urbild eines Punktes $\alpha \in M$ konstant ist. Sei dazu $v_1,v_2
\in \FL^{-1}(\alpha)$ und $\gamma: [0,1] \ra \FL^{-1}(\alpha)$ eine
differenzierbare Kurve\footnote{Wir nehmen hier also zusätzlich an,
  dass je zwei Punkte von $\FL^{-1}(\alpha)$ durch eine
  differenzierbare Kurve verbunden werden können.} von $v_1$ nach
$v_2$. Dann folgt\footnote{Man beachte beim Differenzieren von
  $L(\gamma(t))$, dass die Kurve $\gamma$ immer im Tangentialraum über
  dem gleichen Punkt verläuft.}
$$\frac{\dif}{\dif t}\Big( \FL(\gamma(t))\gamma(t) -
L(\gamma(t))\Big)= \alpha(\dot{\gamma}(t)) -
\FL(\gamma(t))\dot{\gamma}(t) = 0.$$
Damit ist $H$ durch die obige
Gleichung auf $M$ wohldefiniert. Indem wir die kanonische
symplektische Form $\omega_0$ von $T^\ast Q$ auf $M$ zurückziehen,
$\omega:=i^\ast \omega_0$, wird $(M,\omega)$ zu einer
präsymplektischen Mannigfaltigkeit. Lösungen der Hamiltonschen
Gleichungen sind jetzt Kurven $\gamma: I \ra M$ mit $I \subseteq \R$ ,
die die Gleichung $i_{\dot{\gamma}(t)}\omega = \dif H$ erfüllen. Dies
entspricht der Gleichung
$$(\dot{\gamma}(t),\dif H) \in L$$
zu dem impliziten Hamiltonschen
System $(M,L,H)$ mit $L = \op{graph}(\omega)$. Wir wollen jetzt
zusätzlich annehmen, dass $\op{ker}(\omega)$ konstante Dimension hat.
Da $\omega$ im Allgemeinen ausgeartet ist, können wir nicht mehr damit
rechnen, dass durch jeden Punkt von $M$ Lösungen dieser Gleichung
führen. Dirac gab in \cite{dirac:1964a} ein Verfahren an, wie
zusätzliche Zwangsbedingungen eingeführt werden können, so dass auf
einem dadurch weiter eingeschränkten Phasenraum das Hamiltonsche
Vektorfeld $i_{X_H}\omega = \dif H$ an jedem Punkt existiert. (Siehe
auch \cite{henneaux.teitelboim:1992a,sudarshan.mukundu:1974a} für eine
Darstellung dieses Verfahrens sowie \cite{gotay.nester.hinds:1978a}
für eine geometrische Beschreibung.) 

Die Menge der zulässigen
Funktionen auf $M$ bzgl. der Dirac-Struktur $L =\op{graph}(\omega)$
ist gegeben durch
$$\{f \in C^\infty(M)\,|\;\dif f \in \rho^\ast(L) = (L \cap TM)^\circ
= (\op{ker} \omega)^\circ\}.$$
Die zulässigen Funktionen sind also die
Funktionen, die entlang der charakteristischen Distribution
$\op{ker}(\omega)$ konstant sind und auf diesen ist damit eine
Poisson-Klammer definiert.

Sei nun $N \stackrel{j}{\hra} M$ die durch
den Dirac-Algorithmus gewonnene Teilmenge von $M$, die wir als glatte
Untermannigfaltigkeit annehmen wollen. Nach Konstruktion von $N$
können wir jetzt die Gleichung
$$i_{X_H} \omega = \dif H$$
eingeschränkt auf $N$ mit $X_H \in TN$
lösen. Man beachte, dass dies nicht damit übereinstimmt, die Gleichung
$i_{X_H} j^\ast \omega = j^\ast \dif H$ zu lösen. $X_H$ ist im
Allgemeinen jedoch nicht eindeutig bestimmt. Lokal können wir die
Bestimmung von $X_H$ auch auf das Lösen der Gleichung
$$i_{X_H}\omega_0 = \dif \tilde{H}$$
zurückführen, wobei $\tilde{H}$
eine lokale Fortsetzung von $H$ auf $T^\ast Q$ ist. Um dies
einzusehen, wählen wir eine offene Umgebung $U \subseteq T^\ast Q$ von
$p \in N$ und Funktionen $\phi_1,\ldots,\phi_k \in C^\infty(U)$ 
derart, dass
$$M \cap U = \bigcap_{i=1}^k\phi_i^{-1}(0)$$
und die $\dif \phi_i$ auf
$U$ linear unabhängig sind. Sei nun $H'$ eine beliebige Fortsetzung
von $H|_{U\cap M}$ auf $U$. Man kann zeigen \cite[Appendix
1.A]{henneaux.teitelboim:1992a}, dass jede lokale Fortsetzung von $H$
durch
$$\tilde{H} = H' + \lambda_i \phi_i$$
mit $\lambda_i \in C^\infty(T^\ast Q)$ gegeben ist. 
Betrachten wir also die Gleichung 
$$i_{X_H} \omega_0 = \dif \tilde{H} = \dif {H'} + \lambda_i \dif
\phi_i,$$
die jetzt für jedes $\lambda = (\lambda_1,\ldots,\lambda_k)$
eine eindeutige Lösung hat, so ist durch die Forderung $X_H \in TN$,
die nach Konstruktion von $N$ auch erfüllt werden kann, die Freiheit
bei der Wahl der $\lambda_i$'s auf $\op{dim}\ker(\omega)$ Parameter
eingeschränkt. Ist $M$ koisotrop und damit $\op{ker}(\omega)=TM \cap
TM^\perp = TM^\perp$, so bleiben alle $\lambda_i$'s frei wählbar. Ist
dagegen $M$ symplektisch, also $\op{ker}(\omega)=0$, so sind alle
$\lambda_i$'s eindeutig bestimmt.

\newpage

\chapter{Lie-Algebroid-Deformation}

Die formale Deformationstheorie von kommutativen Algebren wurde in
\cite{gerstenhaber:1963a} von Gerstenhaber vorgestellt und später auf
Lie-Algebren und weitere Strukturen wie symplektische
Formen und Poisson-Tensoren erweitert. Die Betrachtung im Rahmen der
formalen Potenzreihen (siehe Anhang \ref{formalSeries}) ermöglicht
dabei eine rein algebraische Beschreibung. Zwar ist in physikalischen
Anwendungen im Allgemeinen eine glatte Deformationstheorie zu fordern,
also beispielsweise eine Deformation $\pi_t$ einer Poisson-Struktur
$\pi = \pi_0$, die glatt von dem Parameter $t$ abhängt. Eine
systematische Untersuchung glatter Deformationen ist jedoch sehr
kompliziert und nur in einigen besonders einfachen Fällen wirklich
konkret durchführbar.  Man betrachtet deshalb zunächst die formale
Deformationstheorie, um zumindest etwaige Obstruktionen für die
Deformierbarkeit zu bestimmen. In einem zweiten Schritt wäre dann der
Übergang zur glatten Deformationstheorie zu vollziehen, was sich aber,
wie gesagt, in den meisten Fällen als außerordentlich schwierig wenn
nicht gar unmöglich herausstellt.

In der Physik spielt die Deformationstheorie die Rolle einer
Störungstheorie zur dynamischen oder kinematischen Stabilitätsanalyse.
Bei der dynamischen Stabilitätsanalyse betrachtet man eine Deformation
der die Dynamik generierenden Hamiltonfunktion bzw. des Hamiltonschen
Vektorfeldes. Als Beispiel sei hier das
Kolmogorov-Arnold-Moser-Theorem (KAM-Theorem) zur Störung integrabler
Systeme genannt, siehe z.B. \cite{tabor:1989a}. Die Deformationen, die
wir betrachten, dienen dagegen zur kinematischen Stabilitätsanalyse.
Im Gegensatz zu dem eben erwähnten dynamischen Fall betrachten wir
dabei nicht Störungen der Hamiltonfunktion, sondern Deformationen der
Bewegungsgleichungen selbst. Ein derartiges Vorhaben mag zunächst
verwundern, da ja die Bewegungsgleichungen Teil der physikalischen
Theorie sind und normalerweise nicht als mit Messfehlern behaftet
angesehen werden. Doch  kann beispielsweise der Übergang von
der nichtrelativistischen Mechanik zur speziellen Relativitätstheorie
als Anwendung einer Deformationstheorie betrachtet werden. Das Objekt,
dass dabei deformiert wird, ist die Galileigruppe, und es ist ja
durchaus ein Ergebnis von Messungen, dass diese Gruppe als
Transformationsgruppe in der Mechanik Verwendung findet. Und
tatsächlich stellt sich bei noch genaueren Messungen heraus, dass die
Galileigruppe die Transformationen nur näherungsweise beschreibt. Die
Deformationstheorie der Galileigruppe zeigt andererseits, dass diese
Gruppe nicht stabil unter Deformationen ist, d.h. wir können die
Galileigruppe auf nichttriviale Weise deformieren, wobei der
Deformationsparameter die Rolle von $\frac{1}{c}$ mit der
Lichtgeschwindigkeit $c$ spielt. Damit erhalten wir zumindest einen
Hinweis darauf, dass die Galileigruppe einer genügend genauen
experimentellen Prüfung möglicherweise nicht standhält.  Andererseits
stellt sich die Lorentzgruppe als stabil unter (physikalisch
sinnvollen) Deformationen heraus und wir erhalten einen guten
Anhaltspunkt dafür, dass die spezielle Relativitätstheorie auch
weiteren Prüfungen standhält.

Ein zweites Beispiel für eine Deformationstheorie ist die
Deformationsquantisierung. Dabei wird das kommutative Produkt der
klassischen Observablen auf dem Phasenraum unter Beachtung gewisser
physikalischer Nebenbedingungen zu einem nichtkommutativen
Produkt de\-for\-miert. Auch hier liefert die Instabilität der klassischen
Observablenalgebra einen Hinweis darauf, dass sich die klassische Theorie
 bei ausreichend genauer Messung als nicht haltbar erweist.  Als
deformierte Theorie erhält man dann eine mögliche Quantisierung und
die Rolle des Deformationsparameters  wird von $\hbar$ übernommen.

Diese Beispiele zeigen, dass die Beschäftigung mit
Deformationstheorien auch für die Physik fruchtbar ist. Im folgenden
werden wir deshalb die Deformationstheorie von Lie-Algebroiden
studieren. Lie-Algebroide treten in der Physik an verschiedenen
Stellen auf. Sei beispielsweise ein Phasenraum $M$ gegeben, auf dem
eine Symmetriegruppe $G$ operiert. Indem wir zu den infinitesimalen
Erzeugern dieser Gruppe, also ihrer Lie-Algebra $\g$, übergehen, lässt
sich diese Operation  als eine Abbildung $\phi:\g \ra
\sect(TM)$ beschreiben.  Das Wirkungs-Lie-Algebroid zu der Wirkung $\phi$ ist dann
durch $E = M\times \g$ gegeben. Zur Definition der Lieklammer auf
$\sect(E)$ sowie des Ankers siehe z.B. \cite[Abschnitt
16.2]{cannasdasilva.weinstein:1999a}. Damit ist die gesamte
Information, die in Phasenraum, Symmetriegruppe und in der Art der
Wirkung enthalten sind, in einem einzigen Objekt, dem
Wirkungs-Lie-Algebroid, enthalten.  Da wir am Beispiel der
Galileigruppe bereits gesehen haben, dass sich die Untersuchung der
Deformation von Liegruppen physikalisch auszahlen kann, sollte es
nicht verwundern, wenn selbiges auch auf die Deformation von
Lie-Algebroiden zutrifft.

Lie-Algebroide treten aber auch in der Form von integrablen
Unterbündeln $E\subseteq TM$ des Tangentialbündels auf, wobei $E^\circ
\subseteq T^\ast M$ die Zwangsfläche zu in den Impulsen linearen
Zwangsbedingungen ist, siehe \cite{eilks:2004a}.

Die Deformationstheorie von Lie-Algebroiden sollte damit hinreichend
motiviert sein. Zu\-nächst wollen wir aber zur Einführung die
Deformationstheorie von Liealgebren nach Gerstenhaber untersuchen.

\section{Deformationen von Lie-Algebren}

Sei $V$ ein endlichdimensionaler $\R$-Vektorraum. Wir bezeichnen für
$n > 0$ mit $\mathcal{M}^n(V) = \bigotimes^n V^\ast \otimes V$ die
multilinearen und mit $\mathcal{A}^n(V) = \bigwedge^n V^\ast \otimes
V$ die antisymmetrischen multilinearen Abbildungen $f:V\times \ldots
\times V \lra V$. Weiter sei $\mathcal{M}^0(V)=\mathcal{A}^0(V)=V$
sowie $\mathcal{M}^n(V)=\mathcal{A}^n(V) = \{0\}$ für $n<0$.
Schließlich definieren wir
$$\mathcal{M}^\bullet(V)=\bigoplus_{n\in \mathbb{Z}}\mathcal{M}^n(V)$$
sowie
$$\mathcal{A}^\bullet(V)=\bigoplus_{n\in
  \mathbb{Z}}\mathcal{A}^n(V).$$
Eine Algebra-Struktur auf $V$
definiert damit ein Element in $\mathcal{M}^2(V)$. Ist $V=\g$ eine
Liealgebra, dann ist durch die Lieklammer gemäß $\mu(x,y)=[x,y]$ ein
Element $\mu \in \mathcal{A}^2(\g)$ gegeben.

Es soll nun auf $\mathcal{M}^\bullet(V)$ bzw. auf
$\mathcal{A}^\bullet(V)$ eine Algebra-Struktur erklärt werden, mit
der im ersten Fall die Assoziativität und im zweiten Fall die
Jacobi-Identität für eine auf $V$ vorhandene Algebra-Struktur
kontrolliert werden kann. Die folgende Konstruktion geht zurück auf
Gerstenhaber \cite{gerstenhaber:1963a,gerstenhaber:1964a}. Man
beachte, dass dafür nur die Vektorraumstruktur von $V$ benutzt wird. Wir
definieren zunächst für $f \in \mathcal{M}^m(V)$ und $g \in
\mathcal{M}^n(V)$ sowie $ 1 \leq i \leq m$ eine Verknüpfung $\circ_i$
durch
$$(f \circ_i g) (x_1,\ldots,x_{m+n-1}) :=
f(x_1,\ldots,x_{i-1},g(x_i,\ldots,x_{i+n-1}),x_{i+n},\ldots,x_{m+n-1})$$
wobei $x_1,\ldots,x_{m+n-1} \in V$, 
und definieren dann das Gerstenhaber-Produkt als
$$f \circ g := \sum_{i=1}^m(-1)^{(i-1)(n-1)}f\circ_i g,$$
sowie für $f
\in \mathcal{A}^m(V)$, $g \in \mathcal{A}^n(V)$ das antisymmetrisierte
Gerstenhaber-Produkt durch
$$(f\diamond g)(x_1,\ldots,x_{m+n-1}) = \sum_{\pi}(-1)^\pi
f\(g(x_{\pi(1)},\ldots,x_{\pi(n)}),x_{\pi(n+1)},\ldots,x_{\pi(n+m-1)}\),$$
wobei die Summe über alle
$(n,m-1)$-Shuffles\footnote {Eine Permutation
  $$\pi = \(\begin{array}{ccccccc}
      1 & 2 & \ldots & p & p+1 & \ldots & p+q \\
      \pi(1) & \pi(2) & \ldots & \pi(p) & \pi(p+1) & \ldots & \pi(p+q)
  \end{array} \) \in \mathcal{S}_{p+q}
  $$
  ist ein $(p,q)$-Shuffle, wenn die Zahlen
  $\pi(1),\pi(2),\ldots,\pi(p)$ und $\pi(p+1),
  \pi(p+2),\ldots,\pi(p+q)$ aufsteigend geordnet sind.} geht. Man
überlegt sich leicht, dass $f\diamond g$ bis auf einen Vorfaktor
tatsächlich die Antisymmetrisierung von $f\circ g$ ist.

Die so erhaltenen Produkte sind zwar im allgemeinen nicht assoziativ,
dennoch lässt sich zeigen, dass durch die Gerstenhaberklammer
$$[f,g]_\G=f\circ g - (-1)^{(m-1)(n-1)}g \circ f$$
eine Super-Lieklammer
auf $\mathcal{M}^\bullet(V)$ bzw. durch die
Nijenhuis-Richardson-Klammer \cite{nijenhuis.richardson:1967a}
$$[f,g]_\NR =f\diamond g - (-1)^{(m-1)(n-1)}g \diamond f$$
eine
Super-Lieklammer auf $\mathcal{A}^\bullet(V)$ definiert ist.  Ist
$\mu\in \mathcal{M}^2(V)$, so ist $\mu$ assoziativ genau dann, wenn
$[\mu,\mu]_\G=0$.  Ist $\mu\in \mathcal{A}^2(V)$, so erfüllt $\mu$ die
Jacobi-Identität genau dann, wenn $[\mu,\mu]_\NR=0$.

Im weiteren beschränken wir uns auf den antisymmetrischen Fall. Sei
also durch $\mu\in \mathcal{A}^2(\g)$ eine Lieklammer $[\dcd]$ auf
$\g$ gegeben, d.h. $[\mu,\mu]_\NR=0$.  Wir definieren eine Abbildung
\begin{eqnarray*}
\delta_\mu: \mathcal{A}^\bullet(\g) &\lra& \mathcal{A}^{\bullet+1}(\g)\\
        f &\longmapsto& [\mu,f]_\NR,\\
\end{eqnarray*}
und mit der Super-Jacobi-Identität zeigt man unter Verwendung von
$[\mu,\mu]_\NR=0$ die Gleichung $\delta_\mu^2=0$. Mit einer kleinen
Rechnung folgt, dass für $f\in \mathcal{A}^n(\g)$ die Beziehung
$\delta_\mu f = (-1)^n\delta_{\op{CE}} f$ gilt, wobei
$\delta_{\op{CE}}$ das Chevalley-Eilenberg-Differential zu $\mu$ ist,
\begin{eqnarray*}
    \delta_{\op{CE}}f(x_1,\ldots,x_{n+1})&=&\sum_{i=1}^{n+1}
    (-1)^{i+1}\mu(x_i,f(x_1,\stackrel{i}{\Hat{\ldots}}\,,x_{n+1}))\\ &&\qquad  + \sum_{1\leq
      i < j \leq
      n+1}(-1)^{i+j}f(\mu(x_i,x_j),x_1,\stackrel{i}{\Hat{\ldots}}\,\stackrel{j}{\Hat{\ldots}}\,,x_{n+1}).
\end{eqnarray*} 

Wir betrachten jetzt für eine Folge $\mu_i \in \mathcal{A}^2(\g)$ eine
formale Deformation $\mu_t = \mu_0 + t \mu_1 +t^2 \mu_2 +\ldots \in
\mathcal{A}^2(\g)[[t]]$ von $\mu_0 = [\,\cdot\;,\,\cdot\;]$, der
Lieklammer auf $\g$. $\mu_t$ erfüllt die Jacobi-Identität genau dann,
wenn
$$[\mu_t,\mu_t]_\NR=0$$
gilt. Diese Bedingung in den einzelnen Ordnungen von $t$ ausgewertet
führt auf ein System von Gleichungen
\begin{eqnarray*}
    t^0: & [\mu_0,\mu_0]_\NR &= \:0 \\
    t^1: & [\mu_0,\mu_1]_\NR + [\mu_1,\mu_0]_\NR &= -2 \delta \mu_1 = 0\\
    t^k, k \geq 2: & [\mu_0,\mu_k]_\NR + [\mu_1,\mu_{k-1}]_\NR + \ldots +
    [\mu_k,\mu_0]_\NR &= -2 \delta \mu_k +
    \sum_{i=1}^{k-1}[\mu_i,\mu_{k-i}]_\NR = 
    0,
\end{eqnarray*}
wobei wir jetzt kurz  $\delta$ für $ \delta_{\mu_0}$ schreiben.

Zunächst folgt aus der ersten Gleichung $[\mu_0,\mu_0]_\NR=0$, dass
$\mu_0$ die Jacobi-Identität erfüllen muss, d.h. die undeformierte
Klammer muss bereits eine Lieklammer sein. Die zweite Gleichung
$\delta \mu_1 =0$ bedeutet, dass $\mu_1$ ein
Chevalley-Eilenberg-Kozyklus sein muss. Angenommen, man hat bereits
$\mu_0,\ldots,\mu_{k-1}$ gefunden, so dass die obigen Gleichungen bis
zur Ordnung $k-1$ erfüllt sind. Die Frage ist, ob die Deformation mit
einem $\mu_k$ bis zur Ordnung $k$ fortgesetzt werden kann. Es ist also
die Gleichung
$$2 \delta \mu_k = \sum_{i=1}^{k-1}[\mu_i,\mu_{k-i}]_\NR$$
zu lösen. Dies
ist nur möglich, wenn die rechte Seite der Gleichung ein Kozyklus ist.
Mit Hilfe der Super-Jacobi-Identität für die Klammer
$[\cdot\,,\cdot]_\NR$ sowie der Gültigkeit der Gleichungen für die
niedrigeren Ordnungen lässt sich jedoch tatsächlich zeigen, dass
$\sum_{i=1}^{k-1}[\mu_i,\mu_{k-i}]_\NR$ geschlossen ist. Für die
Lösbarkeit der Gleichung ist es aber notwendig, dass die auftretende
Summe exakt ist. Es ergibt sich also eine Obstruktion in der dritten
Chevalley-Eilenberg-Kohomologie $H^3_{\op{CE}}(\g)$ für unser
Deformationsproblem.

\subsection{Triviale Deformationen}

\begin{Definition}
    Sei $\mu_0 \in \mathcal{A}^2(\g)$ eine Liealgebrastruktur auf $\g$.
\begin{enumerate}
\item Zwei formale Deformationen $\mu_t = \mu_0 + t\, \mu_1 + \ldots$
    und $\mu'_t = \mu_0 + t\, \mu'_1 +\ldots$ von $\mu_0$ heißen
    äquivalent, wenn es eine $\R[[t]]$-lineare Abbildung
    $$\phi_t = \op{id} + t\, \phi_1 +\ldots \; : \g[[t]] \lra \g[[t]] $$
    gibt, so dass
    $$
    \mu'_t (x,y) = \phi_t^{-1}\big(\mu_t(\phi_t(x),\phi_t(y))\big)$$
    als
    Gleichung in formalen Potenzreihen.
\item Eine Deformation $\mu_t = \mu_0 + t\, \mu_1 + \ldots$ heißt
    trivial, wenn $\mu_t$ äquivalent zu $\mu_0$ ist.
\end{enumerate}
\end{Definition} 

Sind $\mu_t$ und $\mu'_t$ äquivalente Deformationen, so erhalten wir,
indem wir die Bedingung für Äquivalenz in erster Ordnung von $t$
betrachten, die Gleichung 
$$\mu'_1 - \mu_1 = [\mu_0,\phi_1]_\NR = \delta \phi_1,$$
d.h. $\mu'_1$ und
$\mu_1$ definieren das gleiche Element in $H_{\op{CE}}^2(\g)$.  Es
gilt nun das folgende
\begin{Lemma}[{\cite[Prop. 1]{gerstenhaber:1964a}}]
    Sei $\mu_t$ eine formale Deformation von $\mu_0$. Dann ist $\mu_t$
    äquivalent zu einer Deformation $\mu'_t = \mu_0 + t^n \mu'_n +
    \ldots$, wobei das erste nicht verschwindende $\mu'_n$
    geschlossen, jedoch nicht exakt ist.
\end{Lemma}
\begin{proof}
    Ist $\mu'_n = \delta \phi_n$ exakt, so liefert $\phi_t = \op{id} +
    t^n\, \phi_n$ eine Äquivalenztransformation auf eine Deformation der
    Form $\mu_0 + t^{n+1} \mu''_{n+1} + \ldots$
\end{proof}
Damit ergibt sich die
\begin{Folgerung}\label{starr}
    Ist $H^2_{\op{CE}}(\g) = 0$, so ist $\g$ starr, d.h. alle formalen
    Deformationen sind trivial.
\end{Folgerung}


\section{Lie-Algebroid-Strukturen als Multiderivationen}

Wir wollen jetzt versuchen, die Konstruktion von Gerstenhaber auf
Lie-Algebroide anzuwenden. Dabei folgen wir
\cite{crainic.moerdijk:2004a}. Die direkte Übertragung der obigen
Überlegungen würde nur zu einer Deformation der Lie-Algebra-Struktur
auf $\sect(E)$ führen, die Leibniz-Regel wäre im
Allgemeinen für die deformierte Klammer allerdings nicht mehr erfüllt. Es zeigt
sich jedoch, dass die Gerstenhaberkonstruktion, angewandt auf einen
Unterkomplex von $\mathcal{A}^\bullet(\sect(E))$, das gewünschte
liefert. Dieser Unterkomplex ist der Komplex der Multiderivationen
$Der^\bullet(E)$ von $E$, den wir jetzt vorstellen wollen.
\begin{Definition}
    Eine Multiderivation vom Grad $n \geq 0$ ist eine
    antisymmetrische, multilineare Abbildung
    $$D:\underbrace{\sect(E)\times \ldots \times \sect(E)}_{n+1\;
      \op{mal}} \lra \sect(E)$$
    derart, dass es eine Abbildung
    $$\sigma_D:\underbrace{\sect(E)\times \ldots \times \sect(E)}_{n\;
      \op{mal}} \lra \sect(TM)$$
    gibt, so dass für $X_0,\ldots,X_n \in
    \sect(E), f\in C^\infty(M)$ die Gleichung
    $$D(X_0,X_1,\ldots,X_{n-1},f X_n)=f
    D(X_0,\ldots,X_n)+\sigma_D(X_0,\ldots,X_{n-1})(f) X_n$$
    gilt. Wir nennen $\sigma_D$ das Symbol von $D$. Für
    $n\geq 0$ sei $Der^n(E)$ der Raum der Multiderivationen auf $E$.
    Weiter setzen wir $Der^{-1}(E)$ $= \sect(E)$ sowie $Der^n(E) =
    \{0\}$ für $n\leq -2$. Schließlich sei
    $$Der^\bullet(E) = \bigoplus_{n\in \mathbb{Z}} Der^n(E).$$
\end{Definition}
Die Abbildung $\sigma_D$ ist offensichtlich ebenfalls antisymmetrisch.
Man kann weiter zeigen, dass für Vektorbündel mit Faserdimension $\geq
2$ folgt, dass $\sigma_D$ sogar $C^\infty(M)$-linear ist, also
$$\sigma_D \in \sect(\Wedge^n E^\ast \otimes TM).$$
Für
Faserdimension $=1$ ist ohnehin $Der^k(E)=0$ für $k>0$, so dass nur
Symbole $\sigma \in \mathfrak{X}(M)=\sect(TM)$ auftauchen.

Wir übertragen jetzt die Gerstenhaberkonstruktion auf
$Der^\bullet(E)$. Dabei wählen wir gemäß \cite{crainic.moerdijk:2004a}
eine vom vorherigen Abschnitt leicht abweichende Normierung und
Vorzeichenkonvention. Für $D_1 \in Der^p(E)$ und $D_2 \in Der^q(E)$
setzen wir das Gerstenhaberprodukt fest als
$$D_1 \circ D_2 (s_0,\ldots,s_{p+q}) = \sum_\pi (-1)^\pi
D_1\bigl(D_2(s_{\pi(0)},\ldots,s_{\pi(q)}),s_{\pi(q+1)},
\ldots,s_{\pi(p+q)}\bigr),$$
wobei die Summe über alle
$(q+1,p)$-Shuffles läuft. Die Super-Lieklammer ist jetzt
$$[D_1,D_2]_\CM=(-1)^{pq} D_1\circ D_2 - D_2 \circ D_1,$$
was bis auf
ein Supervorzeichen der Nijenhuis-Richardson-Klammer, eingeschränkt

auf die Multiderivationen $Der^\bullet(E)$, entspricht.  Man kann
zeigen, dass mit $D_1 \in Der^p(E)$ und $D_2 \in Der^q(E)$ folgt, dass
$[D_1,D_2]_\CM \in Der^{p+q}(E)$ gilt, d.h. $Der^\bullet(E)$ ist unter
der Gerstenhaberklammer abgeschlossen. Genauer rechnet man nach, dass
$$\sigma_{[D_1,D_2]_\CM} = (-1)^{pq} \sigma_{D_1} \circ D_2
-\sigma_{D_2} \circ D_1 +[\sigma_{D_1},\sigma_{D_2}],$$
wobei
$$\sigma_{D_1} \circ D_2\,(s_1,\ldots,s_{p+q}) = \sum_{\pi} (-1)^\pi
\sigma_{D_1}(D(s_{\pi(1)},\ldots,s_{\pi(q+1)}),
s_{\pi(q+2)},\ldots,s_{\pi(q+p)})$$
mit der Summe über alle
$(q+1,p)$-Shuffles, sowie
$$[\sigma_{D_1},\sigma_{D_2}]\,(s_1,\ldots,s_{p+q})=\sum_{\pi}
(-1)^\pi [\sigma_{D_1}(s_{\pi(1)},\ldots,s_{\pi(p)}),
\sigma_{D_2}(s_{\pi(p+1)},\ldots,s_{\pi(p+q)})],$$
summiert über alle
$(p,q)$-Shuffles.

Wegen $Der^n(E) \subseteq \mathcal{A}^{n+1}(\sect(E))$ ist damit
klar, dass $(Der^\bullet(E),[\cdot\,,\cdot\,]_\CM)$ eine gradierte
Liealgebra ist, und dass ein Element $m \in Der^1(E)$ genau dann eine
Liealgebra-Struktur auf $\sect(E)$ definiert, wenn  $ [m,m]_\CM=0$ gilt.
Tatsächlich ist in diesem Fall durch $m$ sogar eine
Lie-Algebroid-Struktur auf $E$ gegeben, wobei der Anker $\rho =
\sigma_m$ das Symbol von $m$ ist. Die Leibniz-Identität ist dabei
bereits durch die Voraussetzung $m \in Der^1(E)$ erfüllt.

Gegeben ein Element $m \in Der^1(E)$ mit $[m,m]_\CM=0$, so definieren wir
eine Abbildung $\delta_m: Der^k(E) \ra Der^{k+1}(E)$ durch $D \mapsto
[m,D]_\CM$, wobei mit der Super-Jacobi-Identität für $[\dcd]_\CM$ wieder
$\delta_m^2=0$  folgt. Wir erhalten also einen Kokettenkomplex
$$
\xymatrix{
Der^{-1}(E) \ar[r]^-{\delta_m} & Der^0(E) \ar[r]^-{\delta_m} & Der^1(E)
\ar[r]^-{\delta_m} & Der^2(E) \ar[r]^-{\delta_m} &}\ldots
$$
sowie die dazugehörigen Kohomologieklassen, die wir hier mit
$H^k_{m,CM}(E)$ für Crainic-Moer\-dijk bezeichnen wollen. Durch eine
kleine Rechnung erhalten wir für $D\in Der^p(E)$ die Gleichung 
$$\sigma_{\delta_m(D)}=\sigma_{[m,D]_\CM}=\delta(\sigma_D)+(-1)^p \rho
\circ D,$$
wobei jetzt
\begin{eqnarray*}
    \delta(\sigma_D)(s_1,\ldots,s_{p+1})& =& \sum_{i=1}^{p+1}(-1)^{i+1}
    [\rho(s_i),\sigma_D(s_1,\elide{i},s_{p+1})]\\
    &&\qquad + \sum_{i < j}(-1)^{i+j}
    \sigma_D([s_j,s_j],s_1,\elidetwo{i}{j},s_{p+1}).
\end{eqnarray*}
gemeint ist.
\begin{Bemerkung}\label{H_TM}
    Ist $E=TM$ mit $m \in Der^1(TM)$ der Lieklammer auf Vektorfeldern
    und $\rho=\op{id}$, dann gilt $\sigma_D \in Der^{p-1}(TM)$ sowie
    $\delta(\sigma_D)= \delta_m(D) = [m,D]_\CM$, also
    $$\sigma_{\delta_m(D)} = \delta_m(\sigma_D)+(-1)^p D$$
    woraus
    sofort folgt
    $$H_{m,CM}^p(TM) = 0.$$
\end{Bemerkung}

Wie vorher bei der Deformation von Lie-Algebren betrachten wir eine
formale Reihe $m_t = m_0 + t m_1 + t^2 m_2 + \ldots \in Der^1(E)[[t]]$
und fordern $[m_t,m_t]_\CM=0$. Analog zu den Überlegungen im
vorherigen Abschnitt erhalten wir, dass die Obstruktionen für die
Deformationen in $H^2_{m_0,CM}(E)$ zu suchen sind.

\section{Deformationen linearer Poisson-Strukturen}

Ein Lie-Algebroid $(\pi: E\ra M,[\cdot\,,\cdot\,]_\E,\rho)$ definiert
bekanntlich eine Poisson-Struktur auf dem dualen Bündel $\tau:E^\ast\lra
M$. Bezeichnet man mit $\hat{s} = \mathcal{J}(s) \in C^\infty(E^\ast)$
die durch $s\in \sect(E)$ gegebene faserweise lineare Funktion auf
$E^\ast$, und sei $\invJ_0$ die Umkehrabbildung dazu, so ist die
Poisson-Struktur auf $E^\ast$ durch die Forderung der Gleichung
$$\invJ_0(\{\hat{s}_1,\hat{s}_2\}) = [s_1,s_2]_\E ,\qquad s_1,s_2 \in
\sect(E)$$
sowie der Leibnizregel für die Poisson-Klammer eindeutig
bestimmt. Es gilt dann mit $f,g \in C^\infty(M), s\in\sect(E)$
$$\{\hat{s},\tau^\ast f\} = \tau^\ast\(\rho(s)f\),\qquad \{\tau^\ast
f,\tau^\ast g\} = 0.$$
Poisson-Strukturen dieser Art können durch das
Eulervektorfeld charakterisiert werden. Dabei ist für ein Vektorbündel
$\pi:E\ra M$ das Eulervektorfeld $\xi \in \sect(TE)$ definiert als das
Vektorfeld zu dem Fluss $\Phi^\xi_t(v_m)= e^t v_m$, $v_m \in E_m =
\pi^{-1}(m)$. Ist nun $P\in \mathfrak{X}^2(E^\ast)=\sect(\Wedge^2
TE^\ast)$ der Poisson-Tensor zu der Poisson-Struktur auf $E^\ast$, so
lässt sich zeigen, dass $\Lie_\xi P = -P$ gilt (siehe z.B.  \cite[Prop.
5.3.3]{marle:2002} und Abschnitt \ref{linPoissonDef}).

\subsection{Homogene Multivektorfelder}
Wir wollen unsere Betrachtungen ein wenig verallgemeinern und machen
dazu folgende Definition.
\begin{Definition}
    Seien $p,k\in \mathbb{Z}$. Die Multivektorfelder $X \in
    \mathfrak{X}^p(E) = \sect(\Wedge^p TE^\ast)$ mit
    $$\Lie_\xi X = k X$$
    heißen homogen vom Grad $k$. Als Bezeichnung sei
    $$\mathfrak{X}^{p,k}(E)=\{X\in \mathfrak{X}^p(E)\,|\, \Lie_\xi X =
    k X\}$$
    vereinbart.
\end{Definition}
\begin{Bemerkung}
    Es gilt $\mathfrak{X}^{p,k}(E)={0}$ für $k<-p$. Zur Begründung
    überlegt man sich zunächst, dass für Funktionen $f \in C^\infty(E)
    = \mathfrak{X}^0(E)$ die Differentialgleichung $\Lie_\xi f = k
    \xi$ nur für $k\geq 0$ auf ganz $E$ glatte Lösungen hat. Seien
    $x^i,v^\alpha$ lokale Koordinaten von $E$, wobei die $x^i$ durch
    Koordinaten auf $M$ induziert und die $v^\alpha$ durch eine
    passende lokale Trivialisierung gegeben sind. Dann gilt
    $$\Lie_\xi \frac{\partial}{\partial x^i} = 0,\qquad \Lie_\xi
    \frac{\partial}{\partial v^\alpha} = - \frac{\partial}{\partial
      v^\alpha}.$$
    Damit ergibt sich
    \begin{eqnarray*}
        \lefteqn{\Lie_\xi \(f \frac{\partial}{\partial
          x^{i_1}}\wedge\ldots\wedge\frac{\partial}{\partial x^{i_r}}
        \wedge 
        \frac{\partial}{\partial
          v^{\alpha_1}}\wedge\ldots\wedge\frac{\partial}{\partial
          v^{\alpha_s}}\)}\hspace{15mm}\\ 
        &=&\(\Lie_\xi f\) \frac{\partial}{\partial
          x^{i_1}}\wedge\ldots\wedge\frac{\partial}{\partial x^{i_r}}
        \wedge 
        \frac{\partial}{\partial
          v^{\alpha_1}}\wedge\ldots\wedge\frac{\partial}{\partial
          v^{\alpha_s}}\\
        && -s f \frac{\partial}{\partial
          x^{i_1}}\wedge\ldots\wedge\frac{\partial}{\partial x^{i_r}}
        \wedge 
        \frac{\partial}{\partial
          v^{\alpha_1}}\wedge\ldots\wedge\frac{\partial}{\partial
          v^{\alpha_s}}\;.
    \end{eqnarray*}
    Da Funktionen mindestens den Homogenitätsgrad Null haben, folgt
    daraus die Behauptung.
\end{Bemerkung}
Man sieht leicht, dass $f \in C^\infty(E)$ genau dann faserweise
konstant ist, wenn $\Lie_\xi f = 0$ gilt. Es gibt dann eine eindeutig
bestimmte Funktion $g\in C^\infty(M)$ mit $f= \pi^\ast g$.

Lokal ist jedes $P \in \homM{p}{k}(E)$ Summe von Termen der Form
$$
f^{i_1\ldots i_{p-r}\alpha_1\ldots \alpha_r}
\frac{\partial}{\partial
  x^{i_1}}\wedge\ldots\wedge\frac{\partial}{\partial x^{i_{p-r}}}
\wedge \frac{\partial}{\partial
  v^{\alpha_1}}\wedge\ldots\wedge\frac{\partial}{\partial
  v^{\alpha_r}}
$$
mit $r \geq -k$ und $f^{i_1\ldots i_{p-r}\alpha_1\ldots \alpha_r}
\in \homM{0}{k+r}(E)$.
\begin{Lemma}\label{gradlemma}
    Sei $p,k,r \in \mathbb{Z}$, $p\geq 0$, $k \geq -p$ und $0\leq r
    \leq p$ sowie $X \in \mathfrak{X}^{p,k}(E)$. Seien weiter
    $f_1,\ldots,f_r \in C^\infty(M)$ Funktionen auf $M$ und
    $s_1,\ldots,s_{p-r} \in \mathfrak{X}^{0,1}(E)$ lineare Funktionen
    auf $E$.  Dann gilt
    $$
    i_X (\pi^\ast\dif f_1 \wedge \ldots \wedge \pi^\ast\dif f_r
    \wedge \dif s_1 \wedge \ldots \wedge \dif s_{p-r}) \in
    \mathfrak{X}^{0,p+k-r}(E)$$
    und damit für $r>p+k$
    $$i_X (\pi^\ast\dif f_1 \wedge \ldots \wedge \pi^\ast\dif f_r
    \wedge \dif s_1 \wedge \ldots \wedge \dif s_{p-r}) = 0$$
\end{Lemma}

\begin{proof}
    Man überlegt sich leicht, dass $\Lie_\xi \pi^\ast\dif f_i = 0$
    sowie $\Lie_\xi \dif s_i =\dif s_i$ gilt. Mit der Identität
    $[\Lie_\xi,i_X]=i_{[\xi,X]}$ rechnet man nach, dass
    \begin{eqnarray*}
        \lefteqn{\Lie_\xi i_X  (\pi^\ast\dif f_1 \wedge \ldots \wedge
          \pi^\ast\dif f_r \wedge 
          \dif s_1 \wedge \ldots \wedge \dif s_{p-r})}\\[2mm]
        & \qquad \qquad & = i_{[\xi,X]} (\pi^\ast\dif f_1 \wedge
        \ldots \wedge \pi^\ast\dif f_r \wedge 
        \dif s_1 \wedge \ldots \wedge \dif s_{p-r})\\
        & & \qquad\qquad +i_X \Lie_\xi (\pi^\ast\dif f_1 \wedge \ldots
        \wedge \pi^\ast\dif f_r \wedge 
        \dif s_1 \wedge \ldots \wedge \dif s_{p-r} )\\[2mm]
        & & = k i_X(\pi^\ast\dif f_1 \wedge \ldots \wedge \pi^\ast\dif
        f_r \wedge 
        \dif s_1 \wedge \ldots \wedge \dif s_{p-r}) \\
        & & \qquad\qquad + (p-r) i_X (\pi^\ast\dif f_1 \wedge \ldots
        \wedge \pi^\ast\dif f_r \wedge 
        \dif s_1 \wedge \ldots \wedge \dif s_{p-r})\\[2mm]
        & & = (p+k-r)\; i_X (\pi^\ast\dif f_1 \wedge \ldots \wedge
        \pi^\ast\dif f_r \wedge 
        \dif s_1 \wedge \ldots \wedge \dif s_{p-r}),
    \end{eqnarray*}
    d.h. es gilt
    $$
    i_X (\pi^\ast\dif f_1 \wedge \ldots \wedge \pi^\ast\dif f_r
    \wedge \dif s_1 \wedge \ldots \wedge \dif s_{p-r}) \;\in\,
    \mathfrak{X}^{0,p+k-r}(E)$$
    und der betrachtete Ausdruck ist für
    $r > p+k$ gleich Null.
\end{proof}
\begin{Bemerkung}\label{homMBestimmung}
    Sei $P \in \homM{p}{k}(M)$ mit $-p \leq k \leq 0$. Dann folgt mit
    dem eben gezeigten Lemma, dass $P$ lokal bereits durch Angabe von
    \begin{eqnarray*}
        &P(\dif v^{\alpha_1},\ldots,\dif v^{\alpha_p})&\\
        &\vdots&\\
        &P(\dif x^{i_1},\ldots,\dif x^{i_{p+k}}, \dif
        v^{\alpha_1},\ldots,\dif v^{\alpha_{-k}})&
    \end{eqnarray*}
    bestimmt ist, wobei wie oben die $x^i$ durch eine Karte auf $M$
    und die $v^\alpha \in \homM{0}{1}(E)$ durch eine Vektorbündelkarte
    gegeben sind.
\end{Bemerkung}

\begin{Lemma}
    Sei $X \in \homM{1}{0}(E)$ ein vollständiges lineares Vektorfeld
    und sei $\Phi_t$ der dazugehörige lokale Fluss. Dann ist $\Phi_t$
    ein Vektorbündelisomorphismus, d.h. $\Phi_t$ bildet Fasern wieder
    auf Fasern ab, und es gilt für $x \in M$, $v,w \in E_x$ sowie
    $\alpha \in \R$ die Gleichung
    $$\Phi_t(\alpha v+w)=\alpha \Phi_t(v) + \Phi_t(w).$$
    Ist umgekehrt $\Phi_t$ ein $t$-abhängiger
    Vektorbündelisomorphismus, so ist durch
    $$X_s(v) = \dnd_{t=s}\Phi_t(\Phi_s^{-1}(v))$$
    ein zeitabhängiges Vektorfeld $X_s\in \homM{1}{0}(E)$ definiert.
    
\end{Lemma}
\begin{proof}
    Sei zunächst $X \in \homM{1}{0}(E)$. Wegen $\Lie_\xi X =
    [\xi,X]=0$ vertauschen die zu den Vektorfeldern gehörenden Flüsse,
    $$
    \Phi_t \circ \Phi^\xi_s = \Phi^\xi_s \circ \Phi_t$$
    d.h. es
    gilt für $v \in E$ und alle $s\in \R$
    $$\Phi_t(e^s v) = e^s \Phi_t(v).$$
    Damit gilt also die Gleichung
    $$\pi(\Phi_t(v))=\pi(e^s \Phi_t(v)) = \pi(\Phi_t(e^s v)),$$
    und
    wenn wir den Limes $s \ra -\infty$ bilden, folgt mit der Stetigkeit
    der beteiligten Abbildungen
    
    $$\pi(\Phi_t(v)) = \pi(\Phi_t(0)),$$
    womit gezeigt ist, dass
    $\Phi_t$ Fasern auf Fasern abbildet. Außerdem folgt auch, dass für
    $\alpha \geq 0$
    $$\Phi_t(\alpha v) = \alpha \Phi_t(v).$$
    Da aber
    $$\frac{\dif}{\dif t}\Big|_{t=0}(-\Phi_t(v))=-X(\Phi_t(v)) =
    X(-\Phi_t(v))$$
    sowie $-\Phi_0(v) =-v$ und der Fluss eindeutig
    ist, gilt also
    $$-\Phi_t(v) = \Phi_t(-v),$$
    und die obige Gleichung gilt für alle
    $\alpha \in \R$.
    
    Jetzt betrachten wir die Kurve
    $$\gamma_t = \Phi_t(v+w)-\Phi_t(v) -\Phi_t(w),$$
    die aufgrund der
    Fasertreue von $\Phi_t$ wohldefiniert ist.  Wir wollen
    $\dot{\gamma}$ in einer Karte bestimmen, d.h. wir beschränken uns
    auf den Fall $E=U \times \R^k$ mit $U \subseteq \R^n$ offen. Seien
    $x^i$ Koordinaten in $U$ sowie $v^\alpha$ Koordinaten in $\R^k$.
    $X$ hat dann die Gestalt
    $$
    X(x,v)=a^i(x)\frac{\partial}{\partial x^i} + b^\alpha_\beta(x)
    v^\beta \frac{\partial}{\partial v^\alpha}$$
    mit $a^i,
    b^\alpha_\beta:U \ra \R$ Funktionen. Den Fluss zu $X$ schreiben
    wir als
    $$\Phi_t(x,v)=(\phi_t(x,v),\Psi_t(x,v)),$$
    so dass $$\gamma_t =
    (\phi_t(x,v),\Psi_t(x,v+w)-\Psi_t(v)-\Psi_t(w)).$$
    Zunächst folgt
    $\dot{\phi}_t(x,v) = a^i(x) \frac{\partial}{\partial x^i}$ und
    $\phi_t$ hängt damit nur von $x$ ab. Weiter ist
    $$\dot{\Psi}_t(x,v)=b^\alpha_\beta(\phi_t(x))\Psi_t^\beta(x,v)
    \frac{\partial}{\partial v^\alpha}$$
    und wir erhalten
    \begin{eqnarray*}
        \dot{\gamma}_t &=& a^i(\phi_t(x)) \frac{\partial}{\partial x^i}+
        b^\alpha_\beta\(\phi_t(x)\)
        \(\Psi_t^\beta(x,v+w)-\Psi^\beta_t(x,v)-\Psi^\beta_t(x,w)\)
        \frac{\partial}{\partial  v^\alpha}\\
        &=& X(\phi_t(x),\Psi_t(x,v+w)-\Psi_t(v)-\Psi_t(w))\\
        &=& X(\gamma_t),
    \end{eqnarray*}
    also ist $\gamma_t$ eine Integralkurve von $X$ mit Anfangswert
    $\gamma_0 =(x,0)$. Nun ist aber $X|_M =
    a^i(x)\frac{\partial}{\partial x^i} \in TM$, d.h. $M$ ist
    invariant unter dem Fluss $\Phi_t$ zu $X$. Somit ist
    $\gamma_t=(\phi_t(x),0)$ für alle $t$, für die $\gamma_t$
    definiert ist.  Die Kurve $\gamma_t$ verläuft also im Nullschnitt
    des Bündels $U\times  \R^k$. Das ist aber eine
    koordinatenunabhängige Aussage, so dass wir auch für nichttriviale
    Bündel erhalten, dass die Kurve $\gamma_t$ im Nullschnitt
    verläuft, womit die erste Aussage des Lemmas bewiesen wäre.
    Die umgekehrte Richtung ist aber klar.
\end{proof}

\subsection{Lineare Poisson-Deformation}\label{linPoissonDef}

Sei nun $\pi \in \mathfrak{X}^{2,-1}(E^\ast)$ ein linearer
Poisson-Tensor auf $\tau:E^\ast\ra M$. Durch die Vorschrift
$$\widehat{[s_1,s_2]} = \{\hat{s}_1,\hat{s}_2\}=i_\pi(\dif \hat{s}_1
\wedge \dif \hat{s}_2),\qquad s_1,s_2 \in \sect(E)$$
ist eine
Lieklammer auf den Schnitten von $E$ definiert. Weiter definieren wir
den Anker $\rho:E\ra TM$ durch
$$
\tau^\ast\(\rho(s)f\) = \{\hat{s},\tau^\ast f\}.
$$
Dass $\rho$ dadurch wohldefiniert ist, folgt aus Lemma \ref{gradlemma},
denn damit gilt
$$\Lie_\xi \{\hat{s},\tau^\ast f\} = \Lie_\xi i_\pi (\dif\hat{s}
\wedge \dif\tau^\ast f) = 0.$$
Damit ist die Funktion
$\{\hat{s},\tau^\ast f\}$ faserweise konstant. Weiter gilt aufgrund
von $\{\tau^\ast f,\tau^\ast g \} = i_\pi (\tau^\ast \dif f \wedge
\tau^\ast \dif g) = 0 $, dass $\rho$ sogar $C^\infty(M)$-linear und
damit  tatsächlich ein Vektorbündelhomomorphismus ist. Man sieht
leicht, dass $(E,[\cdot\,,\cdot\,],\rho)$ mit den obigen Definitionen
ein Lie-Algebroid wird. Wir haben also eine eindeutige Beziehung
zwischen Lie-Algebroid-Strukturen auf $E$ und linearen
Poisson-Strukturen auf $E^\ast$.  Dies kann verallgemeinert werden zu
der folgenden Aussage.

\begin{prop}
    Sei $\pi:E\ra M$ ein Vektorbündel, und sei $\tau:E^\ast \ra M$ das
    duale Bündel. Dann gibt es für jedes $k\geq 0$ einen Isomorphismus
    $$
    \invJ_k:\homM{k}{1-k}(E^\ast) \lra Der^{k-1}(E).$$
    Dabei ist die
    Abbildung $\invJ_0$ das Inverse zu der Abbildung $s \mapsto
    \hat{s}$, die einem Schnitt $s \in \sect(E)$ die entsprechende
    lineare Funktion $\hat{s} \in \homM{0}{1}(E^\ast)$ zuordnet. Für
    $k \geq 1$ ist $\invJ_k$ für ein $P\in\homM{k}{1-k}(E^\ast)$
    definiert durch
    $$\invJ_k(P)(s_1,\ldots,s_k) = \invJ_0(i_P\, \dif \hat{s}_1 \wedge
    \ldots \wedge \dif \hat{s}_k) \qquad s_1,\ldots,s_k \in
    \sect(E).$$
    Weiter ist das Symbol $\sigma_{\invJ_k(P)}$ für eine
    Funktion $f\in C^\infty(M)$ durch
    $$\sigma_{\invJ_k(P)}(s_1,\ldots,s_{k-1})(f) = \invJ_0(i_P\, \dif
    \hat{s}_1 \wedge \ldots \wedge \dif \hat{s}_{k-1} \wedge \dif\,
    \tau^\ast\! f)$$
    gegeben.
\end{prop}

\begin{proof}
    Zunächst überlegt man sich mit Hilfe von Lemma \ref{gradlemma},
    dass für $P \in \homM{k}{1-k}(E^\ast)$ folgt, dass
    $$i_P\, \dif \hat{s}_1 \wedge \ldots \wedge \dif \hat{s}_k \in
    \homM{0}{1}(E^\ast),$$
    und $\invJ_k(P)$ damit wohldefiniert ist.
    Weiter ist $\invJ_k(P)$ offensichtlich antisymmetrisch. Um die
    Derivationseigenschaft nachzurechnen, überlegt man sich zunächst
    für die Abbildungen $s \mapsto \hat{s}$ und ihr Inverses $\invJ_0$
    die Gleichungen
    $$\widehat{f s} = \tau^\ast\! f\, \hat{s} \qquad \mbox{bzw.}
    \qquad \invJ_0(\tau^\ast\! f\, \hat{s}) = f \invJ_0(\hat{s}).$$
    Damit folgt jetzt
    \begin{eqnarray*}
        \invJ_k(P)(s_1,\ldots,f s_k) &=& \invJ_0(i_P\, \dif \hat{s}_1
        \wedge 
        \ldots \wedge \dif (\tau^\ast \! f\, \hat{s}_k))\\
        &=& \invJ_0(\tau^\ast\! f \, i_P \dif \hat{s}_1 \wedge
        \ldots \wedge \dif \hat{s}_k + i_P\, \dif \hat{s}_1 \wedge
        \ldots 
        \wedge \dif\, \tau^\ast\!f \hat{s}_k )\\
        &=& f \invJ_k(P)(s_1,\ldots,s_k) +
        \sigma_P(s_1,\ldots,s_{k-1})(f) s_k,
    \end{eqnarray*}
    wobei
    $$\sigma_P(s_1,\ldots,s_{k-1})(f) = \invJ_0(i_P\, \dif \hat{s}_1
    \wedge\ldots \wedge \dif\, \tau^\ast\! f)$$
    mit der
    Kurzschreibweise $\sigma_P$ für $\sigma_{\invJ_k(P)}$ gilt.  Dabei
    folgt wieder mit Lemma \ref{gradlemma}, dass $\sigma_P$
    wohldefiniert ist. Weiter ist $\sigma_P(s_1,\ldots,s_{k-1})$ eine
    Derivation von $C^\infty(M)$ und definiert deshalb ein Vektorfeld
    auf $M$. Indem man nochmals Lemma \ref{gradlemma} zu Hilfe nimmt,
    sieht man, dass $\sigma_P$ in den Argumenten $s_1,\ldots,s_{k-1}$
    auch $C^\infty(M)$-linear ist. Umgekehrt sieht man mit Bemerkung
    \ref{homMBestimmung}, dass für ein $D \in Der^{k-1}(E)$ mit Symbol
    $\sigma_D$ durch die angegebenen Gleichungen eindeutig ein $P \in
    \homM{k}{1-k}(E^\ast)$ mit $\invJ(P)=D$ definiert ist, womit die
    Bijektivität von $\invJ$ folgt.
\end{proof}

Wollen wir jetzt die Deformation von Lie-Algebroiden untersuchen, so
können wir äquivalent dazu auch die Deformation von linearen
Poisson-Strukturen auf Vektorbündeln untersuchen. Aus der
Deformationstheorie für Poisson-Mannigfaltigkeiten wissen wir, dass
die zweite und dritte Poisson-Kohomologie mit den Deformationen
zusammenhängen, siehe z.B.
\cite{cannasdasilva.weinstein:1999a,waldmann:2004a}.  Wir geben jetzt
einen Unterkomplex an, der die Deformation von linearen
Poisson-Strukturen regelt. Zunächst ein
\begin{Lemma}\label{linSchout}
    Sei $\pi:E\ra M$ ein Vektorbündel und seien $P\in
    \mathfrak{X}^{p,k}(E)$, $Q\in \mathfrak{X}^{q,l}(E)$ homogene
    Multivektorfelder. Dann gilt 
    $$[P,Q] \in \mathfrak{X}^{p+q-1,k+l}(E),$$
    d.h. die homogenen Multivektorfelder sind unter der
    Schouten-Nijenhuis-Klammer abgeschlossen.
\end{Lemma}
\begin{proof}
    Mit Hilfe des Eulervektorfeldes $\xi$ rechnen wir nach, dass
    \begin{eqnarray*}
        \Lie_\xi [P,Q] &=& [\xi,[P,Q]] =[[\xi,P],Q]+[P,[\xi,Q]]\\
        &=& k [P,Q] + l [P,Q]\\
        &=& (k+l) [P,Q].
    \end{eqnarray*}
\end{proof}

Ist jetzt $\pi \in \homM{2}{-1}(E)$ ein linearer Poisson-Tensor, so
gilt nach Lemma \ref{linSchout} für das zugehörige $\dif_\pi =
[\pi,\cdot\,]$, dass
$$\dif_\pi (\homM{p}{k}(E)) \subseteq \homM{p+1}{k-1},$$
und wir
erhalten für $k\geq 0$ die Unterkomplexe

$$
\xymatrix{ \homM{0}{k}(E) \ar[r]^-{\dif_\pi} & \homM{1}{k-1}(E)
  \ar[r]^{\dif_\pi} & \homM{2}{k-2}(E) \ar[r]^-{\dif_\pi}&} \ldots
$$
Wir interessieren uns natürlich für den Komplex, der die linearen
Bivektorfelder enthält, also
$$
\xymatrix{ \homM{0}{1}(E) \ar[r]^-{\dif_\pi} & \homM{1}{0}(E)
  \ar[r]^{\dif_\pi} & \homM{2}{-1}(E) \ar[r]^-{\dif_\pi} &
  \homM{3}{-2}(E) \ar[r]^-{\dif_\pi} &} \ldots
$$
und die dazugehörige Kohomologieklassen
$$H_{\pi,lin}^k(E) = \frac{\ker(\dif_{\pi}:\homM{k}{1-k}(E)\lra
  \homM{k+1}{-k}(E))}{\op{im}(\dif_{\pi}:\homM{k-1}{2-k}(E)\lra
  \homM{k}{1-k}(E))}.$$

Sei nun $\pi_t = \pi_0 + t \pi_1 + t^2 \pi_2 +\ldots \in
\mathfrak{X}^2(M)[[t]]$ eine formale Deformation von $\pi_0$. Damit
$\pi_t$ ein linearer Poisson-Tensor ist, müssen die Gleichungen
$$[\pi_t,\pi_t]=0\qquad \mbox{und} \qquad L_\xi \pi_t = -\pi_t$$
gelten.  Die erste Bedingung liefert wie üblich die Gleichungen
$$\dif_{\pi_0}\pi_n = \frac{1}{2}\sum_{i=1}^{n-1}[\pi_i,\pi_{n-i}]$$
für die einzelnen $\pi_n$, die zweite Bedingung ist einfach
$$L_\xi \pi_n = -\pi_n.$$
Wenn wir aber die Lösungen der ersten
Gleichung nicht in dem ganzen Poisson-Komplex suchen, sondern unseren
Unterkomplex verwenden, ist die zweite Gleichung offenbar automatisch
erfüllt. Damit liegt die Obstruktion für die Deformation von linearen
Poisson-Strukturen in der dritten linearen Poisson-Kohomologie
$H^3_{\pi_0,lin}(E)$.

Damit haben wir jetzt zwei äquivalente Möglichkeiten, wie die
Bedingungen für die Existenz von Deformationen von Lie-Algebroiden
beschrieben werden können. In der Tat gilt der folgende Satz.

\begin{Satz}
    Seien $P\in \homM{p}{1-p}(E^\ast)$ und $Q\in
    \homM{q}{1-q}(E^\ast)$ Multivektorfelder. Dann gilt für die
    Abbildung
    $$\invJ_\bullet : \homM{\bullet}{1-\bullet}(E^\ast) \lra
    Der^{\bullet-1}(E)$$
    die Gleichung
    $$\invJ_{p+q-1}([P,Q]) =
    (-1)^{(p-1)(q-1)}[\invJ_p(P),\invJ_q(Q)]_\CM.$$

\end{Satz}

Für den Beweis brauchen wir ein Lemma.

\begin{Lemma}
    Sei $M$ eine Mannigfaltigkeit, $P \in \mathfrak{X}^p(M)$ und
    seien $\alpha_1\ldots\alpha_r \in \Omega^1(M)$, $r\geq p$
    Einsformen. Dann gilt
    $$i_P\; \alpha_1 \wedge \ldots \wedge \alpha_r = \sum_\pi (-1)^\pi
    (i_P\, \alpha_{\pi(1)}\wedge \ldots \wedge \alpha_{\pi(p)})\,
    \alpha_{\pi(p+1)}\wedge \ldots \wedge \alpha_{\pi(r)},$$
    wobei die
    Summe über alle $(p,r-p)$-Shuffles läuft.
\end{Lemma}
\begin{proof}
    Sei $q=r-p$ und $Q \in\mathfrak{X}^q(E)$ beliebig. Dann gilt
    einerseits (Summen jeweils über alle $(p,q)$-Shuffles)
    \begin{eqnarray*}
        i_{P \wedge Q}\, \alpha_1 \wedge\ldots\wedge\alpha_{p+q} &=&
        (-1)^{\frac{p+q}{2} (p+q-1)} (P \wedge
        Q)(\alpha_1,\ldots,\alpha_{p+q}) \\ 
        &=& (-1)^{\frac{p+q}{2} (p+q-1)} \sum_{\pi} (-1)^\pi
        P(\alpha_{\pi(1)},\ldots,\alpha_{\pi(p)})
        Q(\alpha_{\pi(p+1)},\ldots,\alpha_{\pi(p+q)}) \\
        &=& (-1)^{pq} \sum_{\pi} (i_P\,
        \alpha_{\pi(1)}\wedge\ldots\wedge \alpha_{\pi(p)})\, (i_Q\,
        \alpha_{\pi(p+1)} \wedge \ldots \wedge \alpha_{\pi(p+q)}) \\
        &=& (-1)^{pq}\, i_Q\; \Bigl(\sum_{\pi} (i_P\,
        \alpha_{\pi(1)}\wedge\ldots\wedge \alpha_{\pi(p)})\,
        \alpha_{\pi(p+1)} 
        \wedge \ldots \wedge \alpha_{\pi(p+q)}\Bigr).
    \end{eqnarray*}
    Andererseits ist aber
    \begin{eqnarray*}
        i_{P \wedge Q}\, \alpha_1 \wedge\ldots\wedge\alpha_{p+q} &=& 
        (-1)^{pq}\, i_{Q \wedge P}\, \alpha_1
        \wedge\ldots\wedge\alpha_{p+q}
        \\
        &=&(-1)^{pq} i_Q \bigl( i_P\, \alpha_1
        \wedge\ldots\wedge\alpha_{p+q}
        \bigr)
    \end{eqnarray*}
    womit die Behauptung folgt.
\end{proof}

Damit können wir jetzt den Satz beweisen.
\begin{proof}
    Seien also $P \in \homM{p}{1-p}(E^ \ast)$ und $Q \in
    \homM{q}{1-q}(E^\ast)$ Multivektorfelder sowie $D_P = \invJ(P) \in
    Der^{p-1}(E)$ und entsprechend $D_Q = \invJ(Q)$.  Dann ist
    \begin{eqnarray*}
        i_{[P,Q]} &=& [\Lie_P,i_Q] = \Lie_P i_Q - (-1)^{(p-1)q} i_Q
        \Lie_P \\ 
        &=& i_P \dif i_Q - (-1)^p \dif i_P i_Q - (-1)^{(p-1)q} i_Q i_P
        \dif +(-1)^{(p-1)q+p} i_Q \dif i_P
    \end{eqnarray*}
    Seien weiter $s_i \in \sect(E)$ Schnitte, die wir mit linearen
    Funktionen auf $E^\ast$ identifizieren. Beachtet man noch, dass
    $$(-1)^{(p-1)q+p}=-(-1)^{(p-1)(q-1)},$$
    dann folgt (Summen über alle $(p,q-1)$-Shuffles)
    \begin{eqnarray*}
        \lefteqn{i_{[P,Q]} \dif s_1 \wedge \ldots\wedge \dif s_{p+q} =
        \bigl(i_P \dif  i_Q   -
        (-1)^{(p-1)(q-1)} i_Q \dif i_P\bigr) \dif s_1\wedge \ldots
        \wedge\dif s_{p+q}}\hspace{7mm} \\[2mm]
        &=&  \sum_{\pi} (-1)^\pi i_P\,
        \dif(D_Q( s_{\pi(1)},\ldots, s_{\pi(q)})) \wedge \dif
        s_{\pi(q+1)} \wedge 
        \ldots \wedge \dif s_{\pi(p+q-1)} \\
        && -(-1)^{(p-1)(q-1)}
        \sum_{\pi} (-1)^\pi i_Q\,
        \dif(D_P(s_{\pi(1)},\ldots,s_{\pi(p)})) \wedge \dif
        s_{\pi(p+1)} \wedge 
        \ldots \wedge \dif s_{\pi(p+q-1)} \\[2mm]
        \qquad&=&\sum_{\pi} (-1)^\pi
        D_P(D_Q(s_{\pi(1)},\ldots,s_{\pi(q)}),s_{\pi(q+1)},\ldots,
        s_{\pi(p+q-1)})   
        \\
        &&\quad -(-1)^{(p-1)(q-1)}
        \sum_{\pi} (-1)^\pi
        D_Q(D_P(s_{\pi(1)},\ldots,s_{\pi(p)}),s_{\pi(p+1)},
        \ldots,s_{\pi(p+q-1)})\\[2mm]
        &=& \bigl(\, D_P \circ D_Q - (-1)^{(p-1)(q-1)} 
        D_Q \circ  D_P\bigr)(s_1,\ldots,s_{p+q-1})\\[2mm]
        &=& (-1)^{(p-1)(q-1)}[D_P,D_Q]_\CM(s_q,\ldots,s_{p+q-1})
    \end{eqnarray*}
\end{proof}

\begin{Satz}\label{kohomIso}
    Sei $\pi \in \homM{2}{-1}(E^\ast)$ mit $[\pi,\pi]=0$ und
    $m=\invJ_2(\pi) \in Der^1(E)$. Dann gilt $[m,m]_\CM=0$ und die
    Abbildung $\invJ_\bullet$ hat die Eigenschaft
    $$\delta_m \circ \invJ\,(P) =
    (-1)^{p-1}\invJ\circ\delta_\pi(P)\qquad \text{für}\qquad
    P\in\homM{p}{1-p}(E^\ast),$$ 
    d.h $\invJ_\bullet$ induziert einen Isomorphismus der Kohomologie
    $$
    H^\bullet_{\pi,lin}(E^\ast) \cong H^{\bullet-1}_{m,CM}(E).$$
\end{Satz}
\begin{proof}
    Zunächst ist $[m,m]= [\invJ(\pi),\invJ(\pi)]=-\invJ([\pi,\pi])=0$,
    also ist durch $\delta_m = [m,\,\cdot\;]$ ein Differential
    definiert.  Weiter ist für $P \in \homM{p}{1-p}(E^\ast)$
    $$\delta_m \circ \invJ\,(P) = [\invJ(\pi),\invJ(P)]=
    (-1)^{p-1}\invJ([\pi,P])= (-1)^{p-1}\invJ\circ\delta_\pi(P)$$
    womit die Behauptung gezeigt ist.
\end{proof}

\subsection{Triviale Deformationen}

\begin{Definition}
    \begin{enumerate}
    \item Zwei formale Deformationen $\pi_t$ und $\pi'_t$ von $\pi_0 \in
        \homM{2}{-1}(E^\ast)$ heißen äquivalent, wenn es eine formale
        Reihe von Automorphismen der Form
        $$\phi_t = \exp(\Lie_{X_t}) = \sum_{k=0}^\infty
        \frac{t^k}{k!}(\Lie_{X_t})^k$$
        mit einer formalen Reihe von
        Vektorfeldern
        $$X_t = t\, X_1 + t^2\, X_2 +\ldots$$
        gibt, so dass $\pi'_t =
        \phi_t(\pi_t)$. Dabei ist $(\Lie_{X_t})^k \pi =
        [X_t,[\ldots[X_t,\pi]\ldots]]$ gemeint.
    \item Eine Deformation $\pi_t$ von $\pi_0$ heißt trivial, wenn
        $\pi_t$ äquivalent zu $\pi_0$ ist.
    \end{enumerate}
\end{Definition}

Da wir verlangen, dass $\phi_t$ lineare Poisson-Strukturen auf lineare
Poisson-Strukturen abbildet, muss gelten
$$\exp(\Lie_{X_t})\Lie_\xi \pi_t = -\exp(\Lie_{X_t})\pi_t = -\pi'_t =
\Lie_\xi \pi'_t = \Lie_\xi \exp(\Lie_{X_t}) \pi_t,$$
woraus folgt,
dass $[\Lie_\xi, \Lie_t]= \Lie_{[\xi,X_t]}= 0$, d.h. $X_t \in
\homM{1}{0}(E^\ast)[[t]]$.

\begin{Bemerkung}
    Im Rahmen der allgemeinen Deformationstheorie einer
    Poisson-Mannig\-faltig\-keit $M$ kann gezeigt werden, dass die
    Menge der formalen Diffeomorphismen $\exp(\Lie_{X_t})$ von $M$ mit
    der Gruppe der homogenen Automorphismen der Gerstenhaberalgebra
    $(\mathfrak{X}^\bullet(M)[[t]],$ $ [\dcd],\wedge)$ vom Grad Null,
    welche in unterster Ordnung mit der Identität beginnen,
    übereinstimmt, siehe z.B. \cite[Prop. 4.2.39]{waldmann:2004a}. Da
    $\homM{\bullet}{1-\bullet}(E^\ast)$ unter dem Dachprodukt nicht
    abgeschlossen ist, gilt diese Aussage hier nicht. Wegen der
    Abgeschlossenheit unter der Lieklammer bilden die formalen
    Diffeomorphismen $\exp(\Lie_{X_t})$ mit $X_i \in
    \homM{1}{0}(E^\ast)$ aufgrund der Baker-Campbell-Hausdorff-Formel
    aber trotzdem eine Gruppe.
\end{Bemerkung}

Sei $\pi_t = \pi_0 + t^n\pi_n + \ldots$ eine formale Deformation von
$\mu_0$ und sei $\pi_n = \delta_{\pi_0}(X) = [\pi_0,X]$ mit $X \in
\homM{1}{0}(E^\ast)$. Wir setzen $\phi_t = \exp(t^n \Lie_X)$ und
bilden
$$\phi_t (\pi_t) = (\op{id}+t^n \Lie_X +\ldots)(\pi_0+t^n
\pi_n+\ldots) = \pi_0 + t^n( \pi_n + [X,\pi_0])+t^{n+1}(\ldots) =
\pi_0 +t^{n+1}(\ldots)$$
Es gilt also
\begin{Folgerung}\label{starrAlg}
    Ist $H_{\pi_0,lin}^2(E^\ast)=0$, so ist die
    Poisson-Mannigfaltigkeit $(E^\ast,\pi_0)$ starr bezüglich der
    Deformation linearer Poisson-Strukturen.
\end{Folgerung}

Ist $\pi_0$ die kanonische symplektische Poisson-Struktur auf $T^\ast
M$, dann gilt $\Lie_\xi \pi_0 = - \pi_0$ \cite[Satz
3.2.6]{waldmann:2004a}, also $\pi \in \homM{2}{-1}(T^\ast M)$. Ist
weiter $m_0 = \invJ(\pi_0)$ die Lieklammer auf $TM$, so ergibt
Bemerkung \ref{H_TM} zusammen mit Satz \ref{kohomIso}
$$H_{\pi_0,lin}^\bullet(T^\ast M) = H_{m_0,CM}^{\bullet-1}(TM)=\{0\},$$
womit gilt:
\begin{Folgerung}\label{TMdef}
    Das Lie-Algebroid $TM$ ist starr.
\end{Folgerung}
Haben wir eine Deformation der linearen Poisson-Struktur $\pi_0$
gegeben, so erhalten wir mit dem Isomorphismus $\invJ$ eine Deformation
der Lie-Algebroid-Struktur $m_0 = \invJ(\pi_0)$. Sind weiter $\pi_t$
und $\pi'_t$ äquivalente Deformationen mit $\pi'_t=\exp(\Lie_{X_t})
\pi_t$ für $X_t = t\, X_1 + t^2 X_2+\ldots$, so erhalten wir durch
Konjugation von $\exp(\Lie_{X_t})$ mit $\invJ$ eine
Äquivalenztransformation zwischen $m_t = \invJ(\pi_t)$ und $m'_t =
\invJ(\pi'_t)$, genauer gilt
$$m'_t = \exp(\ad_{D_t}) m_t$$
wobei $D_t = \invJ(X_t)$ sowie $\ad_{D_t} D'= [D_t,D']$
für $D' \in Der^k(E)$ gesetzt wurde.
Man kann zeigen, dass dies auch äquivalent dazu ist, dass die Gleichung
$$T_t m_t(s_1,s_2) = m'_t(T_t s_1, T_t s_2)\qquad \mbox{mit}\qquad T_t
= \exp(D_t)$$
gilt, siehe \cite[Prop. 6.2.19]{waldmann:2004a}.

\section{Lie-Algebroid-Strukturen als Derivationen der
  Grassmannalgebra}

Wir haben jetzt bereits zwei mögliche Betrachtungsweisen für
Lie-Algebroide kennengelernt, sowie die entsprechenden
Deformationstheorien untersucht und auch formal gezeigt, dass diese
äquivalent sind. In diesen Abschnitt werden wir noch eine dritte
Sichtweise auf Lie-Algebroide kennenlernen. Weiter werden wir die
dadurch gegebene Deformationstheorie untersuchen sowie die Äquivalenz
zu dem bisherigen zeigen.

Ist $(E,[\cdot\,,\cdot\,],\rho)$ ein
Lie-Algebroid, dann ist auf $\Omega^\bullet(E)=\sect(\Wedge^\bullet
E^\ast)$ durch die Gleichung
$$\dE \alpha(s_1,\ldots,s_{k+1}) = \sum_{i=1}^{k+1} \rho(s_i)
\alpha(s_1,\elide{i}, s_{k+1}) + \sum_{i<j} \alpha([s_i,s_j],s_1,
\elidetwo{i}{j}, s_{k+1})$$
eine Derivation
$$\dE: \Omega^k(E) \lra \Omega^{k+1}(E)$$
bezüglich des
$\wedge$-Produktes definiert, und man rechnet nach, dass $\dE^2 = 0$
gilt.  Ist umgekehrt für ein Vektorbündel $E \ra M$ eine solche
Derivation von $\Omega^\bullet(E)$ gegeben, so lässt sich mit der
obigen Formel eine Lie-Algebroid-Struktur auf $E$ definieren, indem
man für $s_1,s_2 \in \sect(E)$, $f\in C^\infty(M)$ die Gleichungen
$$\alpha([s_1,s_2]) =
\dE(\alpha(s_2))(s_1)-\dE(\alpha(s_1))(s_2)-\dE \alpha(s_1,s_2),$$
$$\rho(s_1)(f) = \dE f (s_1)$$
fordert.  Wir wollen die Deformation
von Lie-Algebroiden nun als Deformation von Differentialen auf
$\Omega^\bullet(E)$ auffassen.  Das folgende ist bis auf eine leichte
Verallgemeinerung auf Lie-Algebroide aus \cite[Abschnitt
8]{kolar.michor.slovak:1993a} übernommen.

\begin{Definition}
    Sei $F \ra M$ ein Vektorbündel. Dann bezeichnen wir die
    $\wedge$-Superderiva\-tion\-en von $\Omega^\bullet(F)$, die vom Grad
    $k$ sind, mit
    $$Der_k\Omega(F),$$
    d.h.  $\der \in Der_k\Omega(F)$ ist eine
    Abbildung
    $$\der : \Omega^\bullet(F) \lra \Omega^{\bullet+k}(F)$$
    und mit
    $\alpha \in \Omega^p(F)$, $\beta \in \Omega^q(F)$ gilt
    $$\der(\alpha \wedge \beta) = \der \alpha \wedge \beta + (-1)^{p k}
    \alpha \wedge \der \beta. $$
    Weiter sei auf
    $$Der_\bullet\Omega(F) = \bigoplus_{k\in \mathbb{Z}}
    Der_k\Omega(F)$$
    für $\der_1 \in Der_k\Omega(F)$, $\der_2 \in
    Der_l\Omega(F)$ der Superkommutator durch
    $$[\der_1,\der_2] = \der_1 \circ \der_2 - (-1)^{kl} \der_2 \circ
    \der_1$$
    definiert.

\end{Definition}
Ist insbesondere $\dF \in Der_1\Omega(F)$, so folgt
$$[\dF,\dF]= \dF \dF + \dF \dF = 2 \dif_F^2,$$
d.h. $\dF$
definiert genau dann eine Lie-Algebroid-Struktur auf $F$, wenn
$[\dF,\dF]=0$ gilt. In diesem Fall erhalten wir mit $\delta_\dF =
[\dF, \cdot\,]$ einen Kokettenkomplex
$$\xymatrix{ Der_{-1}\Omega(F) \ar[r]^{\delta_\dF} & Der_0\Omega(F)
  \ar[r]^{\delta_\dF} & Der_1\Omega(F) \ar[r]^-{\delta_\dF} & }
\ldots
$$
und die zugehörigen Kohomologieklassen bezeichen wir mit 
$$H_{\dif_F}^k(\Omega(F)).$$

\subsection{Algebraische Derivationen}
\begin{Definition}
    Eine Derivation $\der \in Der_k \Omega(F)$ heißt algebraisch, wenn
    $\der|_{C^\infty(M)} = 0$.
\end{Definition}

Sei nun $\der\in Der_k \Omega(F)$ eine algebraische Derivation. Dann
gilt für $\omega \in \Omega(F)$ und $f\in C^\infty(M)$, dass $\der(f\,
\omega)= f\, \der(\omega)$, d.h. $D$ ist tensoriell und wir können für
$x \in M$ die Einschränkung $\der_x \in Der_k(\Wedge^\bullet F_x)$
betrachten. Als Derivation ist $\der_x$ durch die Einschränkung auf
1-Formen,
$$\der_x|_{F_x^\ast}:F_x^\ast \ra \Wedge^{k+1} F_x^\ast $$
eindeutig
bestimmt.  Bilden wir die duale Abbildung,
$$K_x \in \Wedge^{k+1} F_x^\ast \otimes F_x,$$
so erhalten wir einen
glatten Schnitt
$$K \in \sect(\Wedge^{k+1} F^\ast \otimes F^\ast) =:
\Omega^{k+1}(F,F),$$
d.h. eine $k+1$-Form auf $F$ mit Werten in $F$.
Wir bezeichnen $D$ jetzt mit $i_K$ und für Einsformen $\omega \in
\Omega^k(F)$ gilt $$\der(\omega) = i_K \omega= \omega \circ K .$$
Wie
üblich setzen wir
$$\Omega(F,F) = \bigoplus_{k\in \mathbb{Z}} \Omega^k(F,F).$$

Wir sehen auch, dass $Der_k\Omega(F) = 0$ für $k<-1$ sowie
$Der_{-1}\Omega(F) = \{i_s|\, s \in \sect(F)\}$, denn eine Derivation
$D$ vom Grad $<0$ ist notwendigerweise algebraisch und daher von der
Form $\der = i_s$ für $s \in \sect(F)$.

\subsection{Die Lieableitung}

Sei jetzt wieder $E$ ein Lie-Algebroid
und $\dE$ das zugehörige Differential. 

\begin{Definition}
    Für $K\in \Omega^k(E,E)$ ist die Lieableitung $\Lie_K:\Omega^l(E)
    \lra \Omega^{k+l}(E)$ durch $$\Lie_K = [i_K,\dif] = (-1)^{k-1}
    \delta_\dE (i_K)$$
    definiert. Man beachte, dass hiermit aufgrund $\mathfrak{X}(M) =
    \Omega^0(TM,TM)$ die gewöhnliche Lie\-ab\-lei\-tung von
    Differentialformen verallgemeinert wird.
\end{Definition}
Es gilt damit also $[\dE, \Lie_K] = (-1)^{k-1}\delta_\dE^2(i_K) =
0$. Ist der Anker $\rho:E \ra TM$ injektiv, so auch die Abbildung
$\Lie: \Omega^\bullet(E,E) \lra Der_\bullet\Omega(E)$. Im Allgemeinen
gilt mit $\Lie_K = 0$, also $i_K$ geschlossen, dass $K \in
 \Omega(E,\ker \rho)$, d.h. $K$ ist eine Form auf $E$ mit Werten in
$\ker \rho$. Denn ist $\Lie_K$ eine algebraische Derivation (also z.B.
$\Lie_K = 0$), dann folgt für alle $f\in C^\infty(M)$, dass
$$\Lie_K f = \rho(K)(f) = 0$$
und deshalb $\rho(K) = 0$. Nur wenn
$\rho$ injektiv ist, folgt daraus $K = 0$. Ist weiter $i_K$ exakt, so
ist $i_K$ insbesondere geschlossen und wir haben ebenfalls $K \in
\Omega(E,\ker \rho)$.

Wir betrachten jetzt die Abbildung
$$i|_{\Omega(E,\ker \rho)}: \Omega(E,\ker\ \rho) \lra i(\Omega(E,\ker
\rho)) \subset \{\mbox{algebraische Derivationen}\}.$$
\begin{Lemma}
    $i(\Omega(E,\ker \rho))$ ist abgeschlossen unter $\delta_\dE =
    [\dE,\,\cdot\;]$.
\end{Lemma}
\begin{proof}
    Sei $K \in \Omega^k(E,\ker \rho)$ und $f \in C^\infty(M)$. Dann ist
    $$\delta_\dE(i_K)f = (-1)^{k-1}[d,i_K]f = (-1)^{k-1} \rho(K)f =
    0,$$ also $\delta_\dE(K)$ eine algebraische Derivation. Wir finden
    deshalb ein eindeutig bestimmtes $M\in \Omega(E,E)$, so dass $i_M =
    \delta_\dE (i_K)$. Wir haben aber bereits gesehen, dass jetzt
    sogar $M \in \Omega(E,\ker \rho)$ folgt.
\end{proof}
Dieses Lemma liefert uns jetzt einen Unterkomplex von $Der_\bullet
\Omega(E)$:
$$\ldots \xymatrix{ \ar[r]^-{\delta_\dE} & i(\Omega^k(E,\ker \rho))
  \ar[r]^-{\delta_\dE} & i(\Omega^{k+1}(E,\ker \rho))
  \ar[r]^-{\delta_\dE} & } \ldots
$$
Seien $u$ und $s_1,\ldots,s_{k+1}$ Schnitte in $E$, wobei $u \in
\ker\rho$ sei. Es gilt $$\rho([u,s_1]_\E) = [\rho(u),\rho(s_1)] = 0,$$
d.h.  $\ker\rho$ ist ein Lie-Ideal. Insbesondere ist dann für $K \in
\Omega^k(E,\ker \rho)$ auch der Ausdruck
$[s_1,K(s_2,\ldots,s_{k+1})]_\E \in \ker \rho$ und wir können durch
\begin{eqnarray*}
\mathfrak{d} K(s_1,\ldots,s_{k+1}) &=& \sum_{i=1}^{k+1}(-1)^{i+1}
[s_i,K(s_1,\elide{i},s_{k+1})] \\
&&\qquad + \sum_{i<j}(-1)^{i+j}
K([s_i,s_j],s_1,\elidetwo{i}{j},s_{k+1})
\end{eqnarray*}
eine Abbildung $\mathfrak{d}: \Omega^k(E,\ker \rho) \ra
\Omega^{k+1}(E,\ker\rho)$ definieren. Mit einer kleinen Rechnung
erhält man, dass $\mathfrak{d}^2 = 0$ gilt.

\begin{Satz}
    Die Abbildung
    $$
    i:(\Omega^\bullet(E,\ker \rho),\mathfrak{d}) \lra
    (i(\Omega^\bullet(E,\ker \rho)),\delta_\dE)$$
    ist ein
    Isomorphismus von Kokettenkomplexen.
\end{Satz}
\begin{proof}
    Da für $K \in \Omega(E,\ker \rho)$ die Abbildung $i_K$ eine
    algebraische Derivation ist, genügt es für Einsformen $\alpha \in
    \Omega^1(E,\ker \rho)$ zu zeigen, dass
    $$
    i_{\mathfrak{d}(K)}\alpha = \delta_\dE i_K \alpha=
    [\dE,i_K]\alpha,$$
    was wir jetzt nachrechnen wollen.
    Zunächst gilt (man beachte $\rho \circ K = 0$)
    \begin{eqnarray*}
        i_K\dE \alpha(s_1,\ldots,s_{k+1}) &=&
        \frac{1}{k!}\sum_{\sigma\in \mathcal{S}_{k+1}}(-1)^\sigma \dE
        \alpha(K(s_{\sigma(1)},\ldots,s_{\sigma(k)}),s_{\sigma(k+1)})\\
        &=& \frac{1}{k!}\sum_{\sigma\in \mathcal{S}_{k+1}}(-1)^\sigma
        \Big( \rho(K(s_{\sigma(1)},\ldots,s_{\sigma(k)}))
        \alpha(s_{\sigma(k+1)}) \\
        &&\qquad\quad-
        \rho(s_{\sigma(k)})
        (\alpha(K(s_{\sigma(1)},\ldots,s_{\sigma(k)})))\\
        &&\qquad\qquad-\alpha([K(s_{\sigma(1)},\ldots,s_{\sigma(k)}),
        s_{\sigma(k+1)}])\Big)\\ 
        &=& (-1)^{k+1}\sum_{i=1}^{k+1}(-1)^{i+1}\rho(s_1)
        \alpha(K(s_1,\elide{i},s_{k+1})) \\
        &&\qquad\quad +(-1)^k
        \alpha\bigg(\sum_{i=1}^{k+1}(-1)^{i+1}
        [s_i,K(s_1,\elide{i},s_{k+1})]\bigg). 
    \end{eqnarray*}
    Weiter haben wir
    \begin{eqnarray*}
        \dE i_K \alpha(s_1,\ldots,s_{k+1}) &=&
        \sum_{i=1}^{k+1}(-1)^{i+1}
        \rho(s_i)i_K\alpha(s_1,\elide{i},s_{k+1})\\
        &&\qquad\quad + \sum_{ i <j}(-1)^{i+j}i_K
        \alpha([s_i,s_j],s_1,\elidetwo{i}{j},s_{k+i} ) \\
        &=&     \sum_{i=1}^{k+1}(-1)^{i+1}
        \rho(s_i)\alpha(K(s_1,\elide{i},s_{k+1}))\\
        &&\qquad\quad + \alpha\bigg(\sum_{i <j} (-1)^{i+j}
        K([s_i,s_j],s_1,\elidetwo{i}{j},s_{k+i} )\bigg),
    \end{eqnarray*}
    womit wir mit $[\dE,i_K] =\dE i_K - (-1)^{k-1} i_K \dE $ die
    Behauptung erhalten. 
\end{proof}

Bezeichnen wir die Kohomologieklassen zu dem Komplex
$\Omega^\bullet(E,\ker \rho)$ mit $H^\bullet(E,\ker \rho)$, so
bedeutet der obige Satz, dass
$$H^\bullet(E,\ker \rho) \cong \frac{\{i_K|\; K \in \Omega(E,\ker
  \rho),\; i_K \mbox{ geschlossen}\}}{\{i_K|\;K \in \Omega(E,\ker
  \rho), \;i_K \mbox{ exakt}\}} .$$

\subsection{Transitive Lie-Algebroide}

Wir wollen in diesem Abschnitt kurz einige Eigenschaften von
$Der_\bullet\Omega(E) $ für ein transitives Lie-Algebroid $E$,
d.h. für Lie-Algebroide $E$ mit surjektivem  Anker,
angeben. Insbesondere gelten also die folgenden Aussagen für $E = TM$
mit der Lie-Klammer auf Vektorfeldern und $\rho = \op{id}$.

\begin{Satz}
    Sei $E$ ein Lie-Algebroid mit surjektivem Anker $\rho: E \ra TM$
    und sei $\der \in Der_k\Omega(E)$. Dann existieren $K \in
    \Omega^k(E,E)$ und $L \in \Omega^{k+1}(E,E)$, so dass
    $$\der = \Lie_K + i_L.$$
    $D$ ist algebraisch genau dann, wenn $K \in
    \Omega^k(E,\ker \rho)$.
\end{Satz}
\begin{proof}
    Seien $s_1,\ldots,s_k \in \sect(E)$. Dann ist durch $f \mapsto
    \der(f)(s_1,\ldots,s_k)$ eine Derivation von $C^\infty(M)$
    definiert.  Wir finden deshalb ein Vektorfeld $X(s_1,\ldots,s_k)
    \in \sect(TM)$ mit $$\der(f)(s_1,\ldots,s_k) = X(s_1\ldots
    ,s_k)(f).$$
    Offensichtlich ist $X$ alternierend und
    $C^\infty(M)$-linear in den $s_i$. Ist nun $\rho: E \ra TM$
    surjektiv, so finden wir ein $K \in \Omega^k(E,E)$
    mit\footnote{Lokal ist das wegen der Linearität von $\rho$ klar.
      Durch eine Partition der Eins erhalten wir die Aussage auch
      global, wobei wir wieder die Linearität von $\rho$ verwenden. }
    $$X(s_1,\ldots,s_k)(f) = \rho(K(s_1,\ldots,s_k))(f) = \dE f \circ
    K (s_1,\ldots ,s_k) = \Lie_K f (s_1,\ldots,s_k).$$
    Damit folgt
    $$(\der - \Lie_K) |_{C^\infty(M)} = 0,$$
    d.h. wir finden ein $L \in
    \Omega^{k+1}(E,E)$ mit
    $$\der = \Lie_K + i_L.$$
    Durch die Konstruktion ist klar, dass für
    $D$ algebraisch notwendigerweise $K \in \Omega(E,\ker \rho)$ sein
    muss. Umgekehrt ist klar, dass $\Lie (\Omega(E,\ker \rho))
    \subseteq \{\mbox{algebraische Derivationen} \} $.
\end{proof}

Ist nun $\der = \Lie_K + i_L \in Der_k\Omega(E)$ geschlossen, so ist
auch $i_L$ geschlossen und damit $L \in \Omega^{k+1}(E,\ker \rho ).$
Ist außerdem $\der = \Lie_K' + i_L'$ eine andere Darstellung, so folgt
$$
i_L - i_L' = [\dE, i_K'-i_K],$$
so dass $i_L$ und $i_L'$ das
gleiche Element in $H_\dE^{k+1}(\Omega(E))$ repräsentieren. Insgesamt
erhalten wir damit \cite[Corollary 4]{crainic.moerdijk:2004a}:
\begin{Folgerung}
    Für Lie-Algebroide mit surjektivem Anker gilt
    $$H_\dE^\bullet(\Omega(E)) \cong \frac{\{i_K|\; K \in
      \Omega(E,\ker \rho),\; i_K \mbox{ geschlossen}\}}{\{i_K|\;K \in
      \Omega(E,\ker \rho E),\;i_K \mbox{ exakt}\}} \cong
    H^\bullet(E,\ker \rho).$$
\end{Folgerung}

Ist insbesondere $E=TM$ mit der Lieklammer für Vektorfelder und
$\rho=\op{id}$ ergibt sich die
\begin{Folgerung}\label{omegaDerivationen}
    Sei $\der \in Der_k\Omega(M)$. Dann gibt es eindeutig bestimmte
    $K\in \Omega^k(M,TM)= \sect(\Wedge ^k T^\ast M \otimes TM)$ und $L
    \in  \Omega^{k+1}(M,TM)$, so dass $\der = \Lie_K + i_L$.  Es ist
    $L=0$ genau dann, wenn $D$ geschlossen ist, und es ist $K=0$ genau
    dann, wenn $D$ algebraisch ist. Weiter gilt für alle
    Kohomologieklassen
    $$H_\dE^k(\Omega(M))=0.$$
\end{Folgerung}

\subsection{Der Zusammenhang mit den Multiderivationen}
  
Wir werden jetzt die Verbindung zu dem Komplex der Multiderivationen
auf $E$ herstellen, vergleiche auch \cite[Abschnitt
2.5]{crainic.moerdijk:2004a}.
\begin{Satz}
    Die Abbildungen
    $$\invL_k : Der_k\Omega(E) \lra Der^k(E),$$
    die für $k \geq 0$,
    $\der \in Der_k\Omega(E)$ und $s_1,\ldots,s_{k+1} \in \sect(E)$
    durch
    $$\alpha(\,\invL_k(\der)(s_1,\ldots,s_{k+1})\,) = (-1)^{k+1}
    \sum_{i=1}^{k+1} (-1)^i \der(\alpha(s_i))(s_1,\elide{i},s_{k+1}) -
    \der \alpha(s_1,\ldots,s_{k+1})$$
    sowie durch $\invL_{-1}(i_s) = s
    \in Der^{-1}(E)$ definiert sind, liefern einen Isomorphismus von
    gradierten Liealgebren.
\end{Satz}

\begin{proof}
    Wir geben zunächst die zu $\invL$ inverse Abbbildung $\op{L}:
    Der^k(E) \ra Der_k\Omega(E)$ an. Zunächst gilt natürlich $\op{L}_s
    =i_s$. Für $D \in Der^k(E)$ mit $k\geq 0$ reicht es, $\op{L}_D$
    auf Funktionen und Einsformen festzulegen. Für Funktionen $f\in
    C^\infty(M)$ gilt
    $$
    \op{L}_D f(s_1,\ldots,s_k) = \sigma_D(s_1,\ldots,s_k)$$
    und für
    Einsformen $\alpha \in \Omega^1(E)$ braucht man nur die obige
    Formel umzustellen, und man erhält
    $$\op{L}_D \alpha(s_1,\ldots,s_{k+1})=(-1)^{k+1}
    \sum_{i=1}^{k+1}(-1)^i
    \sigma_D(s_1,\elide{i},s_{k+1})(\alpha(s_i))
    -\alpha(D(s_1,\ldots,s_{k+1})).$$
    Seien jetzt $D_1 \in Der^p(E)$,
    $D_2 \in Der^q(E)$ sowie $\omega \in \Omega^k(E)$.  Wir definieren
    $$\sigma_{D_1} \circ \omega\;(s_1,\ldots,s_{p+k}) = \sum (-1)^\pi
    \sigma_{D_1}(s_{\pi(k+1)},\ldots,s_{\pi(k+p)})
    (\omega(s_{\pi(1)},\ldots,s_{\pi(k)}))$$
    wobei die Summe über alle
    $(k,p)$-Shuffles zu bilden ist, sowie
    $$\omega \circ D_1\;(s_1,\ldots,s_{p+k})= \sum (-1)^\pi
    \omega(D_1(s_{\pi(1)},\ldots,s_{\pi(p+1)}),s_{\pi(p+2)},
    \ldots,s_{\pi(p+1)})$$
    mit der Summe über alle
    $(p+1,k-1)$-Shuffles. Damit lässt sich nun nachrechnen, das die
    Abbildung $\op{L}$ durch
    $$
    \op{L}_{D_1}\omega = (-1)^{pk} \sigma_{D_1} \circ \omega -
    \omega \circ D_1 $$
    gegeben ist.
    
    Damit rechnen wir einerseits nach, dass
    \begin{eqnarray*}
        \op{L}_{[D_1,D_2]}\omega &=& (-1)^{pq+pk+qk}(\sigma_{D_1}\circ
        D_2)\circ\omega-(-1)^{pk+qk}(\sigma_{D_2} \circ D_1)\circ
        \omega\\ 
        &&\quad +(-1)^{pk+qk}[\sigma_{D_1},\sigma_{D_2}]\circ \omega
         - (-1)^{pq} \omega \circ (D_1 \circ D_2) + \omega \circ (D_2
        \circ D_1),
    \end{eqnarray*}
    wobei $\sigma_{[D_1,D_2]}= (-1)^{pq} \sigma_{D_1} \circ D_2
    -\sigma_{D_2}\circ D_1 +[\sigma_{D_1},\sigma_{D_2}]$ verwendet
    wurde, sowie andererseits
    \begin{eqnarray*}
        [\op{L}_{D_1},\op{L}_{D_2}]&=&(-1)^{pk+qk}\big((-1)^{qp}
        \sigma_{D_1} \circ(\sigma_{D_2} \circ \omega) -
        \sigma_{D_2}\circ(\sigma_{D_1}\circ
        \omega)\big)\\
        &&\quad+(-1)^{pq+pk}\big((\sigma_{D_1}\circ \omega)\circ
        D_2-\sigma_{D_1}\circ (\omega \circ D_2)\big) \\
        &&\quad-(-1)^{qk}\big((\sigma_{D_2} \circ \omega )\circ D_1 -
        \sigma_{D_2}\circ (\omega \circ
        D_1)\big)\\
        &&\quad-\big((-1)^{pq}(\omega\circ D_1)\circ D_2 - (\omega \circ
        D_2)\circ D_1\big).
    \end{eqnarray*}
    Durch Nachrechnen lassen sich nun die folgenden Gleichungen zeigen
    (vgl. auch \cite[Theo. 2]{gerstenhaber:1963a} für die dritte
    Aussage).
    \begin{enumerate}
    \item$[\sigma_{D_1},\sigma_{D_2}]\circ \omega=(-1)^{pq}
        \sigma_{D_1} \circ(\sigma_{D_2} \circ \omega) -
        \sigma_{D_2}\circ(\sigma_{D_1}\circ \omega)$
    \item $(\sigma_{D_1} \circ \omega )\circ D_2 - \sigma_{D_1}\circ
        (\omega \circ D_2) = (-1)^{qk} (\sigma_{D_1}\circ D_2) \circ
        \omega$
    \item $ (-1)^{pq}(\omega\circ D_1)\circ D_2 - (\omega \circ
        D_2)\circ D_1 = (-1)^{pq} \omega \circ (D_1 \circ D_2) -
        \omega \circ (D_2 \circ D_1)$
    \end{enumerate}
    Damit folgt jetzt
    $$[\op{L}_{D_1},\op{L}_{D_2}]=\op{L}_{[D_1,D_2]},$$
    d.h. $\op{L}$
    bzw.  $\invL$ sind Isomorphismen von gradierten Lie-Algebren.
\end{proof}

\begin{Folgerung}
    Sei $d \in Der_1 \Omega(E)$ mit $[d,d]=0$ und $m = \invL_1(d)$.
    Dann induziert $\invL$ einen Isomorphismus
    $$H_\dif^\bullet(\Omega(E)) \cong H_{m,CM}^\bullet(E)$$
    der
    Kohomologien.
\end{Folgerung}
\begin{proof}
    Zunächst gilt $[\invL(\dif), \invL(\dif)]= \invL([\dif,\dif]) =0$, so
    dass $\delta_{\invL(\dif)}$ tatsächlich ein Differential
    definiert. Weiter ist $\delta_m \circ \invL(D) = [\invL(
    \dif),\invL(D)]=\invL[\dif,D]=\invL \circ \delta_\dif(D)$, womit die
    Behauptung folgt.
\end{proof}

\paragraph{Triviale Deformationen}
Eine Äquivalenz
$$m'_t = \exp(\ad_{D_t}) m_t$$
von Lie-Algebroid Strukturen mit $D_t
= t\, D_1 +t^2 D_2 +\ldots \in Der^0(E)[[t]]$ überträgt sich wieder zu
einer Äquivalenz
$$\dif'_t = \exp(\ad_{\op{L}_{D_t}}) \dif_t$$
der zugehörigen Differentiale.

\section{Triviale Deformationen im glatten Fall}

Um die Definitionen, die wir für die Äquivalenz von Deformationen im
formalen Rahmen gegeben haben, zu motivieren, werden wir in diesem 
Abschnitt glatte Deformationen von Lie-Algebroiden betrachten, da in
diesem Falle der Äquivalenzbegriff offensichtlich ist. Zunächst aber
einige vorbereitende Bemerkungen.

Sei $\Phi:E \ra E$ ein Vektorbündelisomorphismus über $\phi:M \ra M$.
Wie üblich definieren wir die Anwendung von $\Phi$ auf einen Schnitt
$s\in \sect(E)$ durch
$$\Phi^\ast s = \Phi^{-1}\circ s \circ \phi \in \sect(E).$$
Weiter
sei die duale Abbildung $\check{\Phi}: E^\ast \ra E^\ast$ mit
Fußpunktabbildung $\phi^{-1}$ für $\alpha_m \in E_m^\ast$ durch
$$\check{\Phi}(\alpha_m)=\alpha_m\circ \Phi|_{E_{\phi^{-1}(m)}} \in
E^\ast_{\phi^{-1}(m)}$$
definiert, und entsprechend die Anwendung auf
Schnitte $\alpha \in \sect(E^\ast)$ durch
$$\check{\Phi}^\ast\alpha = \(\check{\Phi}\)^{-1}\circ \alpha \circ
\phi^{-1} \in \sect(E).$$
Es gilt also
$$\langle \check{\Phi}(\alpha_m),s_{\phi^{-1}(m)} \rangle = \langle
\alpha_m,\Phi(s_{\phi^{-1}(m)}) \rangle = \langle \alpha_m
,\(\Phi^{-1}\)^\ast s\,|_m\rangle$$
sowie
$$\langle \check{\Phi}^\ast \alpha|_m,s_m\rangle = \langle
\alpha_{\phi^{-1}(m)}, \Phi^{-1}(s_m)\rangle.$$

Sei $D_t \in Der^0(E)$ eine zeitabhängige Derivation mit Symbol
$\sigma_t\in \mathfrak{X}(M)$, wobei $D_t$ jetzt eine glatte
Deformation von $D_0$ sein soll. Der Fluss\footnote{Wir verwenden hier,
  wie in der Literatur üblich, die Bezeichnung Fluss auch für die
  Zeitentwicklungen zu zeitabhängigen Vektorfeldern.}  $\Phi_{t,s}$ zu
$D$ ist definiert als ein Vektorbündelisomorphismus, bestimmt durch
die Gleichungen \cite[Appendix A]{crainic.fernandes:2004a}
\begin{enumerate}
\item $\Phi_{t,s}\circ \Phi_{s,u}=\Phi_{t,u},\qquad
    \Phi_{t,t}=\op{id}$
\item $\frac{\dif}{\dif t}|_{t=s} \Phi_{t,s}^\ast e =D_s(e)$ mit $e \in
    \sect(E)$
\end{enumerate}
Für die Fußpunktabbildung $\phi_{t,s}$ zu $\Phi_{t,s}$ folgt dann
$$\frac{\dif}{\dif t}\Big|_{t=s} \phi_{t,s}(m) = \sigma_s(m),$$
d.h. $\phi_{t,s}$ ist der Fluss zu $\sigma_s$.

\begin{Lemma}
    Sei $D_t\in Der^0(E)$ eine zeitabhängige Derivation, $X_t =
    \invJ^{-1}(D_t)\in \homM{1}{0}(E^\ast)$ das zugehörige
    zeitabhängige lineare
    Vektorfeld auf $E^\ast$ sowie $\op{L}_{D_t} \in Der_0\Omega(E)$
    die durch $D_t$ gegebene zeitabhängige Derivation von $\Omega(E)$.
    Sei weiter $\Phi_{t,s}:E\ra E$ ein Vektorbündelisomorphismus über
    dem Diffeomorphismus $\phi_{s,t}$ mit $\Phi_{t,s}\circ
    \Phi_{s,u}=\Phi_{t,u}$ sowie $\Phi_{t,t}=\op{id}$. Dann sind
    folgende Behauptungen äquivalent
    \begin{enumerate}
    \item $\Phi_{t,s}$ ist der Fluss zu $D_s$, d.h für $e\in \sect(E)$
        gilt
        $$\frac{\dif}{\dif t}\Big|_{t=s} \Phi_{t,s}^\ast e=D_s(e).$$
    \item $\check{\Phi}_{t,s}^{-1}:E^\ast\ra E^\ast$ ist der
        Fluss zu dem Vektorfeld $X_t$, d.h. für $\alpha_m \in E^\ast$
        gilt
        $$\frac{\dif}{\dif
          t}\Big|_{t=s}\Phi_{t,s}^{-1}(\alpha_m) =
        X_s(\alpha_m).$$
    \item Für die Abbildung $\(\check{\Phi}_{t,s}^{-1}\)^\ast:
        \Omega(E) \lra \Omega(E)$ gilt für $\alpha \in \sect(E^\ast)$ 
        $$\frac{\dif}{\dif t}\Big|_{t=s}
        \(\check{\Phi}_{t,s}^{-1}\)^\ast\alpha = \op{L}_{D_s}\alpha.$$
    \end{enumerate}
\end{Lemma}


\begin{proof}
    Wir beweisen das Lemma durch eine Rechnung in lokalen Koordinaten.
    Nehmen wir also an, dass $E = U\times\R^k$ mit $U$ offen in
    $\R^n$. Sei $m = (x^1,\ldots,x^n)$ $= (x^i)$ sowie
    $v=(x^1,\ldots,x^n,v^1,\ldots,v^k)=(x^i,v^\mu)\in E_x$ und dazu
    dual $\alpha=(x^i,\alpha_\mu) \in E_x^\ast$. Ein
    Vektorbündelisomorphismus $\Phi$ ist dann gegeben durch
    $$\Phi(x,v)=(\phi(x),\Psi(x)_\mu^\nu v^\mu)$$
    mit einer linearen
    Abbildung $\Psi(x)$. Weiter schreiben wir $\Phi^{-1}$ als 
    $$\Phi^{-1}(x,v)=(\phi^{-1}(x),\tilde{\Psi}(x)_\mu^\nu v^\mu)$$
    und es folgt $\tilde{\Psi}(\phi(x))=(\Psi(x))^{-1}$.
    Sei nun $r = (r^\mu) \in  \sect(E)$. Dann folgt
    \begin{eqnarray*}
        \dnd_{t=s} \Phi^{-1}_{t,s}\big(r(\phi_{t,s}(x))\big)&=&
        \dnd_{t=s}\tilde{\Psi}_{t,s}\big(\phi_{t,s}(x)\big)_\mu^\nu
        r^\mu(\phi_{s,t}(x)) \\
        &=&\(\dnd_{t=s}\tilde{\Psi}_{t,s}\big(\phi_{t,s}(x)
        \big)_\mu^\nu \) r^\mu(x) + \frac{\partial r^\mu}{\partial
          x^i}(x)\sigma_s^i(x)\\ 
        &=& - \(\dnd_{t=s} \Psi_{t,s}(x)_\mu^\nu\) r^\mu(x) +
        \frac{\partial r^\mu}{\partial x^i}(x) \sigma_s^i(x).
    \end{eqnarray*}
    Weiter erhalten wir für $\alpha \in E_x^\ast$
    \begin{eqnarray*}
        \dnd_{t=s}\check{\Phi}^{-1}_{t,s}(\alpha) &=& 
        \(\dnd_{t=s}\phi_{t,s}(x),\alpha_\nu \dnd_{t=s}
        \tilde{\Psi}_{t,s}(\phi_{t,s}(x))_\mu^\nu\) \\
        &=& \(\sigma_s(x), \alpha_\nu
        \dnd_{t=s}\tilde{\Psi}_{t,s}\big(\phi_{t,s}(x)\big)_\mu^\nu
        \)\\
        &=:&  X_s(\alpha).
    \end{eqnarray*}
    Für die Funktion $\hat{r} \in C^\infty(M)$ gilt $r(\alpha) =
    \alpha_\nu r^\nu(x)$ und damit $\frac{\partial \hat{r}}{\partial
      x^i} = \alpha_\nu \frac{\partial r^\nu}{\partial x^i}$ sowie
    $\frac{\partial \hat{r}}{\partial \alpha_\nu} =  r^\nu$. 
    Wir erhalten somit
    \begin{eqnarray*}
        i_{X_s(\alpha)}\dif \hat{r} &=& \alpha_\nu \frac{\partial
          r^\nu}{\partial x^i} \sigma^i_s(x) + \alpha_\nu
        \(\dnd_{t=s}\tilde{\Psi}_{t,s}\big(\phi_{t,s}(x)\big)_\mu^\nu
        \) r^\mu(x)\\
        &=& \Big\langle\alpha,\dnd_{t=s} \Phi_{t,s}^\ast
        r\Big\rangle\Big|_x 
    \end{eqnarray*}
    Damit folgt wegen $\invJ(X_s)(r) = \invJ_0(i_{X_s} \dif \hat{r})$
    die Äquivalenz der ersten beiden Aussagen.  Für einen Schnitt
    $\alpha(x) = \alpha_\mu(x) \in \sect(E^\ast)$ gilt weiter
    \begin{eqnarray*}
        \dnd_{t=s} \check{\Phi}_{t,s}\(\alpha(\phi_{t,s}(X))\) &=& 
         \dnd_{t=s} \alpha_\nu(\phi_{t,s}(x)) \Psi_{t,s}(x)_\mu^\nu \\
         &=& \frac{\partial \alpha_\mu}{\partial x^i}(x)\sigma_s^i(x)
         + \alpha_\nu(x) \dnd_{t=s} \Psi_{t,s}(x)_\mu^\nu.
     \end{eqnarray*}
     Daraus ergibt sich nun
     \begin{eqnarray*}
         \Big\langle\dnd_{t=s} \(\check{\Phi}_{t,s}^{-1}\)^\ast
         \alpha\,,\; 
           r \Big\rangle \Big|_x &=&
         \sigma_s(x)(\alpha(r)) +
         \alpha_\nu(x)\dnd_{t=s}\Psi_{t,s}(x)_\mu^\nu - \sigma_s^i(x)
         \alpha_\mu(x) \frac{\partial r^\mu}{\partial x^i}(x)\\
         &=& \sigma_s(x)(\alpha(r)) - \alpha\(\dnd_{t=s}
         \Psi_{t,s}^\ast r\).
     \end{eqnarray*}
     Erinnern wir uns nun an die Definition $\scal{\op{L}_D,r} =
     \sigma_D(\alpha(r)) - \alpha(D(r))$ für $D \in Der^0(E)$, so
     folgt auch die Äquivalenz der ersten und dritten Aussage.
\end{proof}

Für glatte Deformationen haben wir den folgenden natürlichen
Äquivalenzbegriff.
\begin{Definition}
    Zwei glatte Deformationen $m_t$ und $m'_t$ einer
    Lie-Algebroid-Struktur $m = m_0 \in Der^1(E)$ heißen äquivalent, wenn
    die dadurch gegebenen Lie-Algebroide $(E,m_t,\rho_t)$ und
    $(E,m'_t,\rho'_t)$ für alle $t$ isomorph sind, d.h wenn es einen
    zeitabhängigen Vektorbündelisomorphismus $\Phi_t$ gibt, so dass
    die Gleichungen
    $$\Phi_t^\ast(m'_t(s_1,s_2))=m_t(\Phi_t^\ast s_1,\Phi_t^\ast s_2)$$
    sowie
    $$
    \rho_t \circ \Phi = T\phi_t \circ \rho'_t$$
    für alle $t$ gelten, wobei $\phi_t:M\ra M$ die Fußpunktabbildung
    zu $\Phi_t$ ist.
    Eine Deformation $m_t$ von $m_0 \in Der^1(E)$ heißt trivial, wenn
    $m_t$ äquivalent zu $m_0$ ist.
\end{Definition} 

Seien jetzt also $m_t$ und $m'_t$ zwei äquivalente glatte
Deformationen der Lie-Algebroid-Struktur $m_0$ mit der
Äquivalenztransformation $\Phi_t$.  Mit einer kleinen Rechnung
zeigt man, dass dies äquivalent dazu ist, dass für $\pi_t =
\invJ^{-1}(m_t)$ und $\pi'_t = \invJ^{-1}(m'_t)$ die Gleichung
$$\(\check{\Phi}_t^{-1} \)^\ast \pi'_t = \pi_t$$
gilt. Ebenso ist mit $\dif_t = \op{L}_{m_t}$ und $\dif'_t =
\op{L}_{m'_t}$ dann auch die Aussage 
$$(\check{\Phi}_t^{-1})^\ast \circ \dif'_t = \dif_t \circ
(\check{\Phi}^{-1}_t)^\ast$$
äquivalent zu den beiden oben genannten.
Wir definieren die zeitabhängige Derivation $D_t \in Der^0(E)$ für
$e\in \sect(E)$ durch
$$D_s(e) = \dnd_{t=s} \Phi^\ast_t(e)$$
und setzen $X_t =
\invJ^{-1}(D_t)$. Seien weiter $m_t = m_0 + t m_1 +\ldots$ und $m'_t =
m_0 +t m'_1 +\ldots$ die formalen Taylorentwicklungen.  Ableiten nach
$t$ bei $t=0$ liefert dann jeweils
\begin{enumerate}
\item
    $\Phi_t^\ast(m'_t(s_1,s_2))= m_t(\Phi_t^\ast s_1,\Phi_t^\ast
    s_2)$\\[5mm]
    $
    \begin{array}{lcl}
        \frac{\dif}{\dif t}|_{t=0}\qquad\leadsto\qquad
        (m'_1 - m_1)(s_1,s_2)&=& - D_0(m_0(s_1,s_2))
        + m_0(D_0(s_1),s_2)\\[2mm]
        &&\quad\quad+ m_0(s_1,D_0(s_2))\\[3mm]
        &=& [m_0,D_0](s_1,s_2) = \delta_{m_0}(D_0)(s_1,s_2)
    \end{array}
    $\\[3mm]
    
\item
    $\(\check{\Phi}_t^{-1} \)^\ast \pi'_t = \pi_t$\\[5mm]
    $\frac{\dif}{\dif t}|_{t=0}\qquad\leadsto\qquad
    \pi'_1 - \pi_1 = [\pi_0,X_0] = \delta_{\pi_0}(X_0)$\\[3mm]
    
\item $(\check{\Phi}_t^{-1})^\ast \circ \dif'_t = \dif_t \circ
    (\check{\Phi}^{-1}_t)^\ast$ \\[5mm]
    $\frac{\dif}{\dif t}|_{t=0}\qquad\leadsto\qquad \dif'_t -\dif_t =
    \dif_0 \op{L}_{D_0} - \op{L}_{D_0}\dif_0 =
    [\dif_0,\op{L}_{D_0}]=\delta_{\dif_0}(\op{L}_{D_0})$
\end{enumerate} 
In erster Ordnung haben wir damit auch die formale Äquivalenz in allen
drei Betrachtungsweisen wiedergewonnen. Durch das Bilden von höheren
Ableitungen sieht man, dass gilt:
\begin{Satz}
Die formale Äquivalenz der formalen
Taylorreihen ist eine notwendige Bedingung für die glatte Äquivalenz.
\end{Satz}
Allerdings ist auch dann, wenn die formalen Taylorreihen zweier
glatter Deformationen $m_t$ und $m'_t$ formal äquivalent sind,
keineswegs gesagt, dass $m_t$ und $m'_t$ auch im glatten Sinne
äquivalent sind.

\newpage

\chapter{Deformation von Dirac-Strukturen}

Im ersten Kapitel haben wir das Vektorbündel $E = TM\oplus T^\ast M$
mit der darauf kanonisch gegebenen symmetrischen Bilinearform
betrachtet und auf den Schnitten $\sect(E)$ eine $\R$-bilineare
Verknüpfung, die Courant-Klammer, eingeführt.  Weiter betrachteten wir
maximal isotrope Unterbündel $L$ von $E$, so dass $\sect(L)$ unter der
Courant-Klammer abgeschlossen war. Solche Unterbündel haben wir
Dirac-Strukturen genannt. In diesem Kapitel soll nun eine
Deformationstheorie für Dirac-Strukturen entwickelt werden.  Zuvor
wollen wir jedoch unsere Betrachtungen noch weiter verallgemeinern und
von dem speziellen Vektorbündel $TM \oplus T^\ast M$ zu einem
beliebigen Vektorbündel $E \ra M$ mit einer nicht ausgearteten
Bilinearform und einer $\R$-bilinearen Verknüpfung auf $\sect(E)$, für
die wir die in Lemma \ref{eigenCWK} angegebenen Eigenschaften fordern,
übergehen.
 
\section{Courant-Algebroide}

\begin{Definition}\label{courantalg}
    Ein Courant-Algebroid ist ein Vektorbündel $E$ zusammen mit einer
    nichtausgearteten symmetrischen Bilinearform $\scal{\dcd}$, einer
    bilinearen Verknüpfung $[\dcd]_\C : \sect(E) \times \sect(E) \ra
    \sect(E)$ und einem Vektorbündelhomomorphismus $\rho: E \ra TM$,
    dem Anker, so dass für alle $e_1,e_2,e_3 \in \sect(E)$ und $f\in
    C^\infty(M)$ folgende Bedingungen gelten:
    \begin{enumerate}
    \item Jacobi-Idendität in der Form
        $$[e_1,[e_2,e_3]_\C]_\C=[e_1,e_2]_\C,e_3]_\C +
        [e_2,[e_1,e_3]_\C]_\C.$$  
    \item $[e_1,e_2]_\C + [e_2,e_1]_\C= \D \scal{e_1,e_2}$, wobei $\D:
        C^\infty(M) \ra \sect(E)$ durch
        $$\scal{\D f,e} = \rho(e) f$$
        definiert ist.
    \item $\rho(e_1)\scal{e_2,e_3}=
        \scal{[e_1,e_2]_\C,e_3}+\scal{e_2,[e_1,e_3]_\C}$
\end{enumerate}
\end{Definition}

\begin{Lemma}\label{uchino}
    Seien $e_i \in \sect(E)$ und $f \in C^\infty(M)$. Dann folgt mit den
    Eigenschaften 2 und 3 aus Definition \ref{courantalg} bereits
    die Leibniz-Regel
    $$[e_1,f e_2]_\C = f [e_1,e_2]_\C + (\rho(e_1) f) e_2.$$
    Weiter
    gilt wie im Falle von Lie-Algebroiden die Gleichung
    $$\rho([e_1,e_2]_\C)= [\rho(e_1),\rho(e_2)].$$
\end{Lemma}
Für einen Beweis der ersten Aussage siehe
\cite{uchino:2002a,kosmann-schwarzbach:2003a}. Die zweite Aussage
zeigt man genauso wie bei Lie-Algebroiden, siehe dazu z.B.
\cite{kosmann-schwarzbach.magri:1990a}.

\begin{Bemerkung}
    Man rechnet leicht nach, dass im ersten Argument die Gleichung
    $$[f e_1, e_2]_\C = f [e_1,e_2]_\C - (\rho(e_2) f) e_1 + \scal{e_1,e_2}
    \D f$$
    gilt. Weiter kann man die dritte Bedingung von Definition
    \ref{courantalg} auch schreiben als
    $$\scal{[e_1,e_2]_\C+[e_1,e_2]_\C,e_3}=\rho(e_3)\scal{e_1,e_2}.$$
\end{Bemerkung}

\begin{Lemma}[{\cite[Lemma 2.6.2]{roytenberg:1999a}}]
    In einem Courant-Algebroid gelten die folgenden Gleichungen:
\begin{eqnarray*}
        [e, \D f]_\C &=& \D \scal{e,\D f}\\
        {[\D f, e]_\C} &=& 0\\
        \rho \circ \D &=& 0,\quad \text{d.h.}\quad \scal{\D f,\D g}=0
        \quad \forall f,g \in C^\infty(M)
\end{eqnarray*}
\end{Lemma}
\begin{proof}
    Seien  $e_1,e_2 \in  \sect(E)$, dann gilt
    \begin{eqnarray*}
        \scal{[e_1,\D f]_\C,e_2} &=& -\scal{\D f,[e_1,e_2]_\C} +
        \rho(e_1)\scal{\D f,e_2} \\
        &=& -\rho([e_1,e_2]_\C)f + \rho(e_1)\rho(e_2)f \\
        &=& \rho(e_2)\rho(e_1)f \\
        &=& \rho(e_2) \scal{\D f ,e_1} \\
        &=& \scal{\D\scal{\D f, e_1},e_2}
    \end{eqnarray*}
    womit wir die erste Gleichung gezeigt haben. Damit folgt jetzt auch
    $$[\D f,e_1]_\C = - [e_1,\D f ]_\C + \D\scal{\D f,e_1} = 0.$$
    Für  die letzte Aussage ist
    $$\rho(\D f) g = 0\quad \forall f,g \in C^\infty(M)$$
    zu zeigen. Wir
    können aber zumindest lokal jede Funktion $g \in C^\infty(M)$ als
    $\scal{e_1,e_2}$ mit $e_1,e_2 \in \sect(E)$ schreiben. Damit
    folgt nun
    $$\rho(\D f)\scal{e_1,e_2} = \scal{[\D f, e_1]_\C,e_2} +
    \scal{e_1,[\D f,e_2]_\C} = 0.$$ 
\end{proof}

\begin{Bemerkung}
    Natürlich ist $\stCA$ nach Lemma \ref{eigenCWK} mit den
    entsprechenden Klammern ein Courant-Algebroid. Wir werden deshalb
    auch im Allgemeinen die Klammer $[\dcd]_\C$ Courant-Klammer nennen.
\end{Bemerkung}

\begin{Definition}
    Sei $E$ ein Courant-Algebroid mit gerader Faserdimension und
    maximal indefiniter Bilinearform\footnote{Mit einer maximal
      indefiniten Bilinearform $\scal{\dcd}$ auf einem
      $2n$-dimensionalen Vektorraum meinen wir eine Bilinearform der
      Signatur Null. $\scal{\dcd}$ kann dann durch geeignete Wahl der
      Basen als Matrizen der Form $\scriptstyle{\binom{1\;\;\;
          0\,}{0\:\, -1}}$ oder $\scriptstyle{\binom{0\;\;1}{1\;\;0}}$
      mit $n$-dimensionalen Blöcken dargestellt werden.}.  Eine
    verallgemeinerte Dirac-Struktur $L \subseteq E$ ist ein
    Untervektorbündel von $E$, das bezüglich der gegebenen
    Bilinearform maximal isotrop ist. Eine verallgemeinerte
    Dirac-Struktur heißt integrabel oder kurz Dirac-Struktur, wenn
    $\sect(L)$ unter der Courant-Klammer abgeschlossen ist,
    $$[\sect(L),\sect(L)]_\C\subseteq \sect(L).$$
\end{Definition}

Für ein isotropes Unterbündel $L \subseteq E$ folgt mit der dritten
Forderung in Definition \ref{courantalg} für $s_1,s_2 \in \sect(L)$,
dass
$$[s_1,s_2]_\C + [s_2,s_1]_\C = \D\scal{s_1,s_2} = 0.$$
Ist $L$ sogar
eine Dirac-Struktur, können wir wegen der Integrabilität die
Courant-Klammer auf $L$ einschränken. Diese eingeschränkte Klammer ist
antisymmetrisch und es gelten nach wie vor die Jacobi-Identität und
die Leibniz-Regel. Damit gilt:
\begin{Lemma} 
Eine Dirac-Struktur  $L$ zusammen mit den
Einschränkungen von Courant-Klammer und Anker auf $L$ ist ein
Lie-Algebroid.
\end{Lemma}

Im folgenden werden wir, wenn nicht ausdrücklich etwas anderes erwähnt
wird, nur Courant-Algebroide mit gerader
Faserdimension und maximal indefiniter Bilinearform
betrachten.

\section{Glatte Deformation von Dirac-Strukturen}
Wir wollen jetzt Dirac-Strukturen $L_t$ betrachten, die glatt von
einem Parameter $t \in I$ abhängen, wobei $I \subseteq \R$ ein offenes
Intervall ist.  Um dies zu präzisieren, sei zunächst $\pi:E \ra M$ ein
beliebiges Vektorbündel. Wir betrachten das mit der Projektion $pr: M
\times I \ra M$ zurückgeholte Bündel $pr^\sharp E$, wobei
$$pr^\sharp E = \{((m,t),v)\in (M\times \R) \times E\;|\; m = \pi(v)\},$$
und definieren eine glatte Familie von Unterbündeln $L_t \in E$ als
ein glattes Unterbündel $\mathfrak{L} \subset pr^\sharp E$, wobei $L_t =
\mathfrak{L}|_{M\times\{t\}}$ gilt. Äquivalent können wir eine Familie
$L_t$ auch glatt nennen, wenn für eine beliebige Metrik der
Orthogonalprojektor $P_t$ auf $L_t$ glatt von $t$ abhängt, bzw. wenn
der Projektor $P:pr^\sharp E \ra pr^\sharp E$, definiert durch
$P\bigl((m,t),v\bigr) = \bigl((t,m),P_t(v)\bigr)$ glatt ist. Im
weiteren werden wir immer annehmen, dass das Intervall $I$ die Null
enthält.

\begin{Lemma}\label{ortho}
    Sei $V$ ein Vektorraum, $\dim V = 2n$, mit einer
    nicht-ausgearteten, maximal indefiniten Bilinearform $(\dcd)$
    sowie einem Skalarprodukt $g$. Seien weiter $L_1, L_2 \subseteq V$
    maximal isotrope Unterräume, so dass
    \begin{enumerateR}
    \item $V = L_1 \oplus L_2$,
    \item $L_1^{\bot_g} = L_2$.
    \end{enumerateR}
    Sei weiter $P:V \ra V$ der $g$-Orthogonalprojektor auf $L_1$, also
    $\op{im} P = L_1$, $\op{im} (\op{id}-P) = L_2$.  Bezeichnet nun
    $P^T$ den bezüglich $(\dcd)$ adjungierten Projektor zu $P$, also
    $(P v,w) = (v, P^T w)$ $\forall v,w \in V$, dann gilt
    $$P^T = \op{id} - P,$$
    d.h. $P^T$ ist der $g$-Orthogonalprojektor
    auf $L_2$.
\end{Lemma}
\begin{proof}
Sei $v \in L_1$, $w \in V$. Dann gilt
$$(P^T v, w) = (v, P w) = 0$$
da $v,Pw \in L_1$ und es folgt $P^T v = 0$ für alle $v \in L_1$.
Sei nun $v \in L_2$, $w \in V$ mit $w = (w_1,w_2) \in L_1 \oplus
L_2$. Jetzt rechnet man
$$(P^T v, w) = (v, P(w_1 + w_2)) = (v,w_1) = (v,w_1) + (v,w_2) =
(v,w),$$
womit  $P^T v = v $ für alle $v \in L_2$ folgt.  Damit ist
gezeigt, dass $P^T$ der $g$-Orthogonalprojektor auf $L_2$ ist.
\end{proof}

Aus der symplektischen Geometrie ist bekannt, dass zu einem
Lagrangeschen Unterbündel $L$ immer ein Lagrangesches
Komplementärbündel $L'$ existiert. Um das zu zeigen, konstruiert man
eine mit der symplektischen Form kompatible, fast komplexe Struktur $J$
und definiert $L' = J(L)$. 
Auf gleiche Weise erhalten wir in unserem Fall ein
entsprechendes Ergebnis.

\begin{Lemma}\label{fcs}
Sei $E$ ein Vektorbündel mit gerader Faserdimension und einer
symmetrischen, nicht-ausgearteten, maximal indefiniten Bilinearform
$(\dcd)$. Dann gibt es eine Metrik\footnote{Mit Metrik ist eine
  positiv definite Bilinearform gemeint. Verlangen wir nur die
  Nichtausgeartetheit schreiben wir Pseudometrik.} $g$ und einen
Vektorbündelisomorphismus $J:E\ra E$ mit $J^2 = \op{id}$, so dass die
Gleichung
$$g(e_1,e_2)=(e_1,J e_2)\quad \forall e_1,e_2 \in \sect(E)$$
gilt und $J$ eine Isometrie von $(\dcd)$ und damit auch von $g$ ist.
\end{Lemma}
\begin{proof}
    Wir wählen eine Metrik $k$ auf $E$ und definieren $A:E\ra E$ durch
    $k(A e_1,e_2)=(e_1,e_2)$. Dann gilt $A^\ast = A$, wobei ${}^\ast$
    die Adjunktion bezüglich der Metrik $k$ bezeichnet. Man beachte,
    dass $A$ nicht positiv definit ist, da ja auch $(\dcd)$ nicht
    positiv definit ist. Wir definieren jetzt 
    (siehe "`Polarzerlegung"', \cite[Abschnitt
    25.20]{michor:2004a})
    $$B = |A| = \sqrt{A A^\ast} = \exp\Bigl(\frac{1}{2} \log(A
    A^\ast)\Bigr),$$
    und erhalten damit einen Vektorbündelisomorphismus $B$, wobei die
    Glattheit von $B$ aus der Invertierbarkeit von $A$ folgt.
    Dann ist $B$ $k$-symmetrisch,
    $B^\ast = B$, und positiv definit, somit also $B \neq A$. Aus der
    Definition von $B$ folgt, dass $A$ mit $B$ und folglich auch mit
    $B^{-1}$ kommutiert.  Wir setzen jetzt $J = B^{-1} A$. Dann
    kommutiert auch $J$ mit $A$ und $B$ und wir rechnen
    $$J J^\ast = B^{-1} A A^\ast (B^{-1})^\ast = B^{-1} B^2
    (B^{-1})^\ast = \op{id},$$
    also $J^\ast = J^{-1}$.  Damit gilt
    weiter
    $$J^{-1} = J^\ast = (B^{-1} A)^\ast = A B^{-1} = B^{-1} A = J$$
    und so $J^2 = \op{id}$.  Als nächstes rechnen wir
    $$(J e_1,J e_2) = k(A J e_1,J e_2) = k(J A e_1, J e_2) = k(A e_1,
    J^\ast J e_2) = k(A e_1,e_2) = (e_1,e_2),$$
    womit gezeigt wäre,
    dass $J$ eine Isometrie bezüglich unserer indefiniten Bilinearform
    ist.  Schließlich setzen wir $g(e_1,e_2) = (e_1,J e_2)$. Dann ist
    $g$ symmetrisch und für alle $m \in M$, $v \in E|_m$ mit $v \neq
    0$ gilt
    $$g(v,v) = (v, J v) = k(A v, J v) = k(B J v,J v) \geq 0,$$
    da $B$
    positiv definit ist.
\end{proof}  

\begin{Folgerung}\label{lagKomp}
    Sei $E$ wie oben und sei $L \subset E$ ein maximal isotropes
    Unterbündel von $E$. Wir wählen nach Lemma \ref{fcs} einen
    Vektorbündelisomorphismus $J$ und eine Metrik $g$ mit $g(e_1,e_2)
    = (e_1,J e_2)$. Dann ist auch $J(L)$ maximal isotrop und es gilt
    $$E = L \oplus J(L)$$
    sowie
    $$L^{\bot_g} = J(L).$$
\end{Folgerung}
\begin{proof}
    Sei $e_1 \in \sect(L)$ und $e_2 = J(e'_2)$ für ein $e'_2 \in
    \sect(L)$. Damit folgt
    $$g(e_1,e_2) = g(e_1, J e'_2) = (e_1, J^2 e'_2) = (e_1,e'_2) = 0$$
    und somit aus Dimensionsgründen $J(L) = L^{\bot_g}$. Dass $J(L)$
    isotrop ist, ist klar, da $J$ eine Isometrie ist, und maximal isotrop
    folgt wieder aus Dimensionsgründen.
\end{proof}

\begin{Bemerkung}\label{bemOrtho}
    Ist $P$ der $g$-Orthogonalprojektor auf $L$, dann gilt nach Lemma
    \ref{ortho}, dass $P^T$ der $g$-Orthogonalprojektor auf $J(L)$
    ist.
\end{Bemerkung}

\begin{Satz}\label{FedosovTrick}
    Sei $E$ ein Vektorbündel, $I \subseteq \R$ ein offenes Intervall
    und sei $L_t \subset E$ mit $t \in I$ ein glatt von $t$-abhängiges
    Unterbündel. Dann gibt es einen glatt von $t$ abhängigen
    Vektorbündeliso\-morphismus $U_t:E \ra E$, so dass
    $$L_t = U_t(L_0).$$
    Ist $E$ ein Courant-Algebroid mit Bilinearform
    $(\dcd)$ und $L_t$ eine verallgemeinerte Dirac-Struktur, dann kann
    zusätzlich erreicht werden, dass $U_t$ eine Isometrie von $(\dcd)$
    ist.
\end{Satz}

\begin{proof}
    Der Beweis verläuft im wesentlichen nach \cite[Lemma
    1.1.5]{fedosov:1996a}.  Ist $L_t$ eine verallgemeinerte
    Dirac-Struktur, dann finden wir nach Folgerung \ref{lagKomp} einen
    Vektorbündel\-iso\-morphismus $J$ und eine Metrik $g$, so dass $E$
    für alle $t \in I$ die $g$-orthogonale direkte Summe von $L_t$ und
    $J(L_t)$ ist,
    $$E = L_t \oplus J(L_t) \quad \mbox{und} \qquad L_t^{\bot_g} =
    J(L_t)$$
    Falls $L_t$ ein beliebiges Unterbündel in einem
    Vektorbündel $E$ ist, so wählen wir eine nicht weiter bestimmte
    Metrik auf $E$.  Sei $P_t: E \ra E$ der $g$-Orthogonalprojektor
    auf $L_t$. Es gilt dann $P_t^\ast = P_t$, wobei ${}^\ast$ die
    Adjunktion bezüglich der Metrik $g$ bezeichnet. Ist $L_t$ eine
    verallgemeinerte Dirac-Struktur, dann haben wir nach Bemerkung
    \ref{bemOrtho} weiter die Gleichung $P_t^T = \op{id}-P_t$.
    
    Wir betrachten jetzt die Differentialgleichung
    $$\dot{U}_t = [\dot{P}_t,P_t]U_t$$
    mit der Anfangsbedingung $U_0 =
    \op{id}$.  Lokal ist diese Gleichung von der Form
    $$
    \frac{\dif}{\dif t} U(t,x) = A(t,x) U(t,x)$$
    mit $x \in V
    \subseteq \R^n$ und $A: I \times V \ra M_k(\R)$, und hat eine
    eindeutig bestimmte, für alle $t \in I$ definierte glatte Lösung
    $U_t$, die außerdem auch glatt von $x$ abhängt, siehe z.B.
    \cite[Abschnitt IV.1]{lang:1999a}. Aufgrund der Eindeutigkeit
    fügen sich die lokalen Lösungen zu einer globalen Lösung $U_t:E \ra
    E$ der obigen Gleichung zusammen. Um zu sehen, dass $U_t$
    invertierbar ist, betrachten wir die Gleichung
    $$\dot{V}_t = V_t [P_t, \dot{P}_t]$$
    mit der Anfangsbedingung $V_0
    = \op{id}$, die auch eine für alle $t \in I$ definierte, glatte
    Lösung hat. Dann folgt 
    \begin{eqnarray*}
        \frac{\dif}{\dif t} (U_t V_t) &=& \dot{U}_t V_t +U_t \dot{V}_t \\
        &=& [\dot{P}_t,P_t] U_t V_t + U_t V_t [P_t,\dot{P}_t]\\
        &=& \bigl[[\dot{P}_t,P_t],U_t V_t \bigr]
    \end{eqnarray*}
    und $U_t V_t$ ist damit die eindeutig bestimmte Lösung der
    Gleichung
    $$\frac{\dif}{\dif t} M_t = \bigl[[\dot{P}_t,P_t],M_t \bigr]$$
    mit
    $M_0 = \op{id}$. Offensichtlich ist aber auch $M_t = \op{id}$ eine
    Lösung, so dass $U_t V_t = \op{id}$ folgt. Auf ähnliche Weise
    erhalten wir auch $V_t U_t = \op{id}$, insgesamt gilt damit
    $U_t^{-1} = V_t$.  Da $P_t$ ein $g$-Orthogonalprojektor ist, gilt
    $P_t^\ast = P_t$ und es folgt
    $$(\dot{U}_t)^\ast = U_t^\ast [P_t^\ast,(\dot{P}_t)^\ast]$$
    bzw.
    $$
    (U_t^\ast)\dot{} = U_t^\ast [P_t, \dot{P}_t],$$
    so dass
    $U_t^\ast = V_t = U_t^{-1}$.  War $L_t$ eine verallgemeinerte
    Dirac-Struktur, erhalten wir mit $P_t^T = \op{id} - P_t$ weiter
    $$
    (U_t^T)\dot{} = U_t^T [\op{id}-P_t,- \dot{P}_t] = U_t^T
    [P_t,\dot{P}_t] $$
    so dass auch $U_t^T = U_t^\ast = U_t^{-1}$ gilt.
    Wir müssen jetzt noch zeigen, dass tatsächlich $P_t = U_t P_0
    U_t^{-1}$ gilt.  Da $P_t$ ein Projektor ist, $P_t^2 = P_t$, folgen
    durch Ableiten die Gleichungen $\dot{P}_t = \dot{P}_t P_t + P_t
    \dot{P}_t$ sowie $P_t \dot{P}_t P_t = 0$. Damit rechnet man nach,
    dass $P_t$ die Differentialgleichung
    $$[\dot{U}_t U_t^{-1},P_t]=\dot{P}_t$$
    erfüllt. Man zeigt aber
    auch, dass $U_t P_0 U_t^{-1}$ die obige Gleichung zur selben
    Anfangsbedingung löst, so dass wegen der Eindeutigkeit der Lösung
    $P_t = U_t P_0 U_t^{-1}$ folgt.
\end{proof}

Ist $E \ra M$ ein Vektorbündel und $\Phi:E\ra E$ ein
Vektorbündelisomorphismus über einem Diffeomorphismus $\phi:M \ra M$,
dann definieren wir wie üblich für einen Schnitt $s:M \ra E$ den
Push-forward durch $\Phi_\ast s = \Phi \circ s \circ \phi^{-1}$ und
den Pull-back durch $\Phi^\ast s = \Phi^{-1} \circ s \circ \phi$.

\begin{Lemma}
    Sei $E$ ein Courant-Algebroid mit Bilinearform $\scal{\dcd}$,
    Courant-Klammer $[\dcd]$ und Anker $\rho$. Weiter sei $\Phi_t:E
    \ra E$ ein glatt von $t$ abhängiger Vektorbündelisomorphismus. Wir
    definieren die $t$-abhängigen Klammern
    $$\scal{e_1,e_2}_t = \scal{\Phi_t^\ast e_1, \Phi_t^\ast e_2}$$
    sowie
    $$
    [e_1,e_2]_t = (\Phi_t^{-1})^\ast [\phi_t^\ast e_1,\phi_t^\ast e_2]$$
    und einen
    $t$-abhängigen Anker
    $$\rho_t = T\phi_t^{-1} \circ \rho \circ \Phi_t.$$
    Dann ist für jedes $t$ das
    Vektorbündel $E$ zusammen mit den Klammern $\scal{\dcd}_t$ und
    $[\dcd]_t$ sowie dem Anker $\rho_t$ ein Courant-Algebroid.
\end{Lemma}
Ist $\Phi_0 = \op{id}$, so nennen wir
$(E,\scal{\dcd}_t,[\dcd]_t,\rho_t)$ das durch $\Phi_t$ deformierte
Courant-Algebroid. Mit Satz \ref{FedosovTrick} erhalten wir sofort das
folgenden Lemma.
\begin{Lemma}\label{smoothDiracDef}
    Sei $E$ ein Courant-Algebroid und sei $L_t$ ein glatt von $t$
    abhängiges Untervektorbündel in $E$. Sei $U_t$ die Isometrie aus
    Satz \ref{FedosovTrick}. Dann ist $L_t$ genau dann für alle $t$
    eine Dirac-Struktur, wenn $L_0$ eine Dirac-Struktur für jedes
    durch $U_t$ deformierte Courant-Algebroid
    $\(E,\scal{\dcd},[\dcd]_t,\rho_t\)$ ist.
\end{Lemma}

\subsection{Triviale Deformationen}

\begin{Definition}
    Sei $E \ra M$ ein Courant-Algebroid. Ein
    Courant-Algebroidautomorphis\-mus ist ein Vektorbündelautomorphismus
    $\Phi:E\ra E$ über einem Diffeomorphismus $\phi:M \ra M$, so dass
    gilt:
\begin{enumerateR}
\item $\Phi$ ist eine Isometrie der Bilinearform $\scal{\dcd}$,
    $$\scal{\Phi^\ast e_1,\Phi^\ast e_2}= \scal{e_1,e_2}.$$
\item $\Phi$ ist natürlich bezüglich der Courant-Klammer,
    $$
    [\Phi^\ast e_1,\Phi^\ast e_2]_\C = \Phi^\ast [e_1,e_2]_\C. $$
\end{enumerateR}
\end{Definition}

\begin{Lemma}\label{couAutA}
    Sei $\Phi:E \ra E$ ein Courant-Algebroidautomorphismus. Dann gilt
    für den Anker $\rho$ die Gleichung
    $$
    \rho \circ \Phi = T\phi \circ \rho.$$
\end{Lemma}
\begin{proof}
    Wir rechnen einerseits nach, dass
    $$
    [\Phi^\ast e_1, \Phi^\ast (f e_2)]_\C = [\Phi^\ast e_1, \phi^\ast
    f\, \Phi^\ast e_2]_\C = \Phi^\ast( f [e_1,e_2]_\C)+\rho(\Phi^\ast
    e_1)(\phi^\ast f) \Phi^\ast e_2
    $$
    gilt. Andererseits haben wir auch die Gleichng
    $$
    [\Phi^\ast e_1, \Phi^\ast (f e_2)]_\C = \Phi^\ast [e_1,f
    e_2]_\C = \Phi^\ast(f [e_1,e_2]_\C) + \phi^\ast(\rho(e_1)
    f)\Phi^\ast e_2,$$
     womit wir jetzt
    \begin{eqnarray*}
        \phi^\ast(\rho(e_1))(\phi^\ast f) &=& \phi^\ast(\rho(e_1)f) \\
        &=& \rho(\Phi^\ast e_1)(\phi^\ast f)
    \end{eqnarray*}
    erhalten, d.h.
    $$T\phi^{-1} \circ \rho(e_1) \circ \phi = \rho(\Phi^{-1}\circ
    e_1\circ \phi),$$
    was zu zeigen war.
\end{proof}

\begin{Satz}\label{stCAautos}
    Für $E = \stCA$ mit der kanonischen Courant-Algebroidstruktur ist
    jeder Automorphismus $\Phi$ von der Form
    $$
    \Phi = \tau_B \circ (T\phi,T_\ast \phi),
    $$
    wobei $\tau_B$ eine Eichtransformation zu einer geschlossenen
    Zweiform $B \in \Omega^2(M)$  und $\phi:M \ra M$ ein
    Diffeomorphismus von $M$ ist. $B$ und $\phi$ sind dabei eindeutig
    bestimmt. Die Automorphismengruppe von $\stCA$ ist damit isomorph
    zu dem semidirekten Produkt $\mathcal{Z}^2(M) \rtimes
    \mathcal{D}\text{\itshape{iff}\/}(M)$ mit $\mathcal{Z}^2(M) =
    \op{ker}(\dif_{|\Omega^2(M)})$, wobei die Verknüpfung durch
    $$(B,\phi)(C,\psi) = (B + (\phi^{-1})^\ast C, \phi \circ \psi)$$
    gegeben ist.
\end{Satz}
\begin{proof}
    Sei zunächst $\Phi = (\Phi_1,\Phi_2):\stCA \ra \stCA$ ein
    Automorphismus über der Identität. Mit Lemma \ref{couAutA} folgt
    dann $\Phi_1(X,\alpha) = \rho(\Phi(X,\alpha))=\rho(X,\alpha)=
    X$.  Da $\Phi$ eine Isometrie der kanonischen symmetrischen
    Bilinearform auf $\stCA$ ist, muss folgende Gleichung gelten:
    $$\scal{X,\Phi_2(Y,\beta)} + \scal{Y,\Phi_2(X,\alpha)} =
    \scal{X,\beta}+\scal{Y,\alpha}.$$
    Schreiben wir $\Phi$ als Blockmatrix bezüglich der direkten Summe
    $TM\oplus T^\ast M$,
    $$\Phi = \(\begin{array}{cc}
        \Phi_{11}&\Phi_{12}\\
        \Phi_{21}&\Phi_{22} \end{array}\),
    $$
    dann folgt, indem wir $X = 0$ setzen, die Bedingung
    $\Phi_{22}(\alpha) = \alpha$, und weiter, wenn wir $\alpha = \beta
    = 0$ setzen, die Gleichung
    $$\scal{X,\Phi_2(Y,0)} + \scal{Y,\Phi_2(X,0)} =0,$$
    d.h.
    $$
    \Phi_{21} + \Phi_{21}^\ast = 0
    $$
    Also gilt $\Phi_{21} = B \in \Omega^2(M)$ und damit $\Phi =
    \tau_B$. Mit Lemma \ref{eichtrafo} folgt weiter $\dif B = 0$, da
    $\tau_B$ ein Automorphismus der Courant-Klammer sein muss.
    Insgesamt haben wir jetzt
    $$
    \Phi = \(\begin{array}{cc}
        \op{id}&0\\
        B&\op{id} \end{array}\).
    $$
    Ist nun $\Phi$ ein Automorphismus über einem Diffeomorphismus
    $\phi:M\ra M$, dann ist die Verknüpfung $(T\phi^{-1},T^\ast \phi)
    \circ \Phi$ wie eben gezeigt eine Eichtransformation $\tau_B$ zu
    einer geschlossenen Zweiform $B$. Die Eindeutigkeit von $B$ und
    $\phi$ ist klar, und die Behauptung über die Verknüpfung in dem
    semidirekten Produkt $\mathcal{Z}^2(M) \rtimes
    \mathcal{D}\text{\itshape{iff}\/}(M)$ folgt leicht aus Lemma
    \ref{eichequi}.
    \end{proof} 
    
\paragraph{Triviale Deformationen}

Um zu klären, was wir als triviale glatte Deformationen bezeichnen
wollen, lassen wir uns von den beiden Spezialfällen für
Dirac-Strukturen, nämlich symplektische Mannigfaltigkeiten und
Poisson-Mannigfaltigkeiten, leiten. Eine Deformation $\omega_t$ einer
symplektischen Form $\omega_0$ auf $M$ heißt trivial, wenn es eine
glatte Familie von Diffeomorphismen $\phi_t:M\ra M$ gibt, so dass
$\omega_t = \phi^\ast \omega_0$. Entsprechendes gilt für triviale
Deformationen $\pi_t$ von Poisson-Strukturen. Wir wissen aber, dass
die Gleichungen
$$\op{graph}(\phi_t^\ast \omega_0)=\F\phi_t(\op{graph}( \omega_0))$$ 
bzw.
$$\op{graph}(\phi_t^\ast \pi_0)=\F\phi_t(\op{graph}( \pi_0))$$
gelten. Deshalb
manchen wir für Dirac-Strukturen in $\stCA$ folgende
\begin{Definition}
    Zwei glatte Deformationen $L_t$ und $L'_t$ von $L_0 \in \stCA$
    heißen äquivalent, wenn es glatt von $t$ abhängige
    Diffeomorphismen $\phi_t:M\ra M$ gibt, so dass
    $$L'_t = \F\phi_t(L_t).$$
    Eine glatte Deformation $L_t$ von
    $L_0\in\stCA$ heißt trivial, wenn $L_t$ äquivalent zu $L_0$ ist,
    $$L_t = \F\phi_t(L_0).$$
\end{Definition}
Man beachte, dass dies nach Lemma $\ref{stCAautos}$ nicht damit
übereinstimmt, zu verlangen, dass es einen
Courant-Algebroidautomorphismus $\Phi_t$ gibt, so dass $L_t =
\Phi_t(L_0)$. In diesem Fall wären ja zusätzlich noch die
Eichtransformationen zu geschlossenen Zweiformen dabei, was
beispielsweise dazu führen würde, dass je zwei Deformationen der Form
$L_t = \op{graph}(\omega_t)$ und $L'_t = \op{graph}(\omega'_t)$
äquivalent wären, wobei die Äquivalenztransformation durch
$\tau_{\omega_t-\omega'_t}$ gegeben ist.

\begin{Bemerkung}
    Im Fall eines allgemeinen Courant-Algebroids $E\ra M$ haben wir
    keinen kanonischen Lift mehr von Diffeomorphismen von $M$ zu
    Automorphismen von $E$. Damit sind zwar weiterhin die
    Automorphismen über der Identität ausgezeichnet, eine Darstellung
    der Automorphismengruppe als semidirektes Produkt wie in Lemma
    \ref{stCAautos} existiert jedoch für ein beliebiges
    Courant-Algebroid nicht. Im allgemeinen Fall käme für die Menge der
    Äquivalenztransformationen deshalb nur die ganze
    Automorphismenruppe in Frage. Wie wir aber gerade gesehen haben,
    ist diese im Falle $E = \stCA$ zu groß, sofern wir für
    symplektische und Poisson-Mannigfaltigkeiten bekannte Ergebnisse
    reproduzieren wollen.  Bei der Definition von äquivalenten und
    trivialen Deformationen beschränken wir uns deshalb auf den Fall
    $E = \stCA$.
\end{Bemerkung}

\begin{Bemerkung}
    Haben wir eine glatte Deformation $L_t$ einer Dirac-Struktur $L =
    L_0$ gegeben, so liefert dies nach Lemma \ref{smoothDiracDef}
    insbesondere eine Deformation des Lie-Algebroids $L_0$. Es gibt
    jedoch keinen direkten Zusammenhang zwischen den Deformationen von
    $L$ als Dirac-Struktur und den Lie-Algebroiddeformationen von $L$.
    Beispielsweise ist durch eine geschlossenen Zweiform $\omega$ mit
    $L_t = \op{graph}(t \omega)$ eine glatte Dirac-Deformation von
    $L_0 = TM$ gegeben, die aber, außer für $\omega=0$, keine triviale
    Deformation sein kann, da ja $TM$ unter den Abbildungen $\F\phi$
    für einen Diffeomorphismus $\phi$ von $M$ erhalten bleibt.
    Andererseits wissen wir aber aus Folgerung \ref{TMdef}, dass es
    nur triviale Deformationen von $TM$ als Lie-Algebroid gibt.
\end{Bemerkung}

\section{Formale Deformation von Dirac-Strukturen}

Bevor wir mit der Formulierung einer formalen Deformationstheorie
beginnen können, müssen wir zunächst eine dafür geeignete Beschreibung
unseres Problems finden. Dies ist notwendig, da wir formale Reihen von
den zu deformierenden Objekten bilden können müssen, was aber für
Dirac-Strukturen als Unterbündel in einem Courant-Algebroid auf
direktem Weg nicht möglich ist. Viele Aussagen in den folgenden
Abschnitten sind jedoch unabhängig davon, ob wir formale oder glatte
Deformationen betrachten. Außerdem werden wir teilweise auch weiterhin
glatte Deformationen betrachten, um Definitionen für den formalen Fall zu
motivieren. 

\subsection{Deformierte Dirac-Strukturen als Graphen}\label{graphDef}
Sei $L_t$ eine glatt von $t$ abhängige Dirac-Struktur in einem
Courant-Algebroid $E \ra M$. Wir finden dann eine Metrik $g$, so dass
das $g$-orthogonale Komplement $L^\perp$ von $L$ maximal isotrop ist.
Mit Hilfe der Bilinearform $\scal{\dcd}$ können wir $L^\perp$ mit
$L^\ast$ identifizieren und erhalten dadurch einen
Isomorphismus\footnote{Der Isomorphismus ist natürlich von der Wahl
  des Komplements $L^\perp$ abhängig} $E \cong L\oplus L^\ast$. Ist
$M$ kompakt, so kann $L_t$ für kleine $t$ als Graph einer Abbildung
$\omega_t:L \ra L^\ast$ beschrieben werden, im Allgemeinen ist dies
zumindest lokal möglich.  Die Isotropie von $L_t$
bedeutet gerade $\omega_t \in \sect(\Wedge^2 L^\ast)$, wobei wir
wieder die Zweiform $\omega_t$ auf $L$ durch $\omega(s_1)(s_2) =
\omega(s_1,s_2)$ für $s_1,s_2 \in \sect(L)$ mit einem Element
$\sect\big(\op{Hom}(L,L^\ast)\big)$ identifizieren.
Weiter ist
$\op{graph}(\omega)$ genau dann integrabel, wenn für alle $s_1,s_2,s_3
\in \sect(L)$ gilt
\begin{eqnarray*}
0 &=& \scal{[s_1+\omega(s_1),s_2+\omega(s_2)]_\C,s_3+\omega(s_3)} \\
&=& \scal{[s_1,s_2]_\C,s_3} \\
&& + \scal{[s_1,\omega(s_2)]_\C,s_3} + \scal{[\omega(s_1),s_2]_\C,s_3} +
\scal{[s_1,s_2]_\C,\omega(s_3)} \\
&& + \scal{[s_1,\omega(s_2)]_\C,\omega(s_3)}+\scal{[\omega(s_1),s_2]_\C,
  \omega(s_3)} +
\scal{[\omega(s_1),\omega(s_2)]_\C,s_3} \\
&& + \scal{[\omega(s_1),\omega(s_2)]_\C,\omega(s_3)}.
\end{eqnarray*}
Wir werden die einzelnen Ordnungen in $\omega$ getrennt untersuchen.
Zunächst ist der erste Term $\scal{[s_1,s_2]_\C,s_3}=0$, da ja $L$
integrabel ist. Die in $ \omega$ linearen Terme sind durch das
folgende Lemma erfasst.
\begin{Lemma}
Bezeichnet $\dif_L$ das Lie-Algebroiddifferential von $L$, dann gilt
$$\dif_L\omega(s_1,s_2,s_3) = \scal{[s_1,\omega(s_2)]_\C,s_3} +
\scal{[\omega(s_1),s_2]_\C,s_3} + \scal{[s_1,s_2]_\C,\omega(s_3)}.$$
\end{Lemma}
\begin{proof}
    Mit Hilfe der in einem Courant-Algebroid gültigen Identitäten
    rechnen wir:
    \begin{eqnarray*}
        \dif_L\omega(s_1,s_2,s_3) &=& \rho(s_1)\omega(s_2,s_3)+\rho(s_2)
        \omega(s_3,s_1) + 
        \rho(s_3) \omega(s_1,s_2) \\
        &&\quad - \omega([s_1,s_2]_\C,s_3) + \omega([s_1,s_3]_\C,s_2) -
        \omega([s_2,s_3]_\C,s_1) 
        \\
        &=& \rho(s_1)\scal{\omega(s_2),s_3} -
        \rho(s_2)\scal{\omega(s_1),s_3} + 
        \rho(s_3) \scal{\omega(s_1),s_2} \\
        &&\quad - \omega([s_1,s_2]_\C,g) + \omega([s_1,g]_\C,s_2) -
        \omega([s_2,s_3]_\C,e)\\ 
        &=& \scal{[s_1,\omega(s_2)]_\C,s_3} + \scal{\omega(s_2),[s_1,s_3]_\C} +
        \scal{[\omega(s_1),s_2]_\C,s_3}\\
        &&\qquad - \scal{\omega(s_1),[s_2,s_3]_\C}\\
        &&\quad - \omega([s_1,s_2]_\C,s_3) + \omega([s_1,s_3]_\C,s_2) -
        \omega([s_2,s_3]_\C,s_1)\\
        &=& \scal{[s_1,\omega(s_2)]_\C,s_3} + \scal{[\omega(s_1),s_2]_\C,s_3} +
        \scal{[s_1,s_2]_\C,\omega(s_3)}.
    \end{eqnarray*}
\end{proof}
Um die in $\omega$ quadratischen Terme systematisch untersuchen zu
können, müssen wir etwas mehr arbeiten. In dem folgenden Abschnitt
werden wir eine Klammer auf $L^\ast$ definieren, mit deren Hilfe sich
die Gleichung, die eine Deformation $\omega_t$ erfüllen muss, auf
einfache Weise schreiben lässt.

\subsection{Eine verallgemeinerte Schouten-Nijenhuis-Klammer}

Falls auch $L^\ast$ eine Dirac-Struktur und damit ein Lie-Algebroid
ist, verschwindet der in $\omega$ kubische Term. In diesem Fall ist
$\op{graph}(\omega)$ genau dann eine Dirac-Struktur, wenn die
Gleichung
$$d_L \omega + \frac{1}{2}[\omega,\omega]_\ast = 0$$
erfüllt ist
\cite{liu.weinstein.xu:1997a}, wobei $[\dcd]_\ast$ die
Schouten-Nijenhuis-Klammer auf $L^\ast$ bezeichnet, siehe Definition
\ref{defSchouten}.

Es stellt sich die Frage, ob man zu jeder Dirac-Struktur $L$ eine
komplementäre Dirac-Struktur $L'$ finden kann. Im Allgemeinen ist dies
nicht möglich, wie folgendes Beispiel, welches uns von Henrique
Bursztyn mitgeteilt wurde, zeigt. Sei $E = \stCA$ und $\phi$ eine
geschlossene $3$-Form auf $M$. Wir betrachten die "`$\phi$-twisted"'
Courant-Klammer
$$[(X,\alpha),(Y,\beta)]_\phi = [(X,\alpha),(Y,\beta)] +
\phi(X,Y,\cdot),$$
wobei die Klammer auf der rechten Seite die
Standart-Courant-Klammer ist. $E$ zusammen mit $[\dcd]_\phi$, der
Standard-Bilinearform und der Projektion auf $TM$ als Anker ist dann
ebenfalls ein Courant-Algebroid, siehe \cite{severa.weinstein:2001a}.
In $E$ ist $L = T^\ast M$ eine Dirac-Struktur, $TM$ ist jedoch unter
$[\dcd]_\phi$ nicht mehr abgeschlossen. Jedes isotrope Unterbündel
$L'$ komplementär zu $T^\ast M$ ist Graph einer Zweiform $\omega \in
\Omega^2(M)$. Damit $L'$ eine Dirac-Struktur ist, muss die Bedingung
$\dif \omega = \phi$ erfüllt sein, siehe ebenfalls
\cite{severa.weinstein:2001a}. Ist $\phi$ nicht exakt, finden wir
somit keine zu $T^\ast M$ transversale Dirac-Struktur.

Im Allgemeinen können wir also nicht davon ausgehen, dass $L^\ast$ ein
Lie-Algebroid ist und die Courant-Klammer auf 
$L\oplus L^\ast$ kann nicht einfach auf  $L^\ast$ eingeschränkt werden. Wir
können jedoch durch
$$[\alpha,\beta]_\ast = \op{pr}_{L^\ast}([\alpha,\beta]_\C)$$
eine  Verknüpfung
$[\dcd]_\ast:\sect(L^\ast) \times \sect(L^\ast) \ra \sect(L^\ast)$
definieren, wobei $\op{pr}_{L^\ast}: L\oplus L^\ast \ra L^\ast$ die
Projektion auf $L^\ast$ bezeichnet. 
\begin{Lemma}\label{dualKlammer}
    Die Klammer $[\dcd]_\ast$ ist antisymmetrisch und erfüllt eine
    Leibniz-Regel:
    $$[\alpha, f \beta]_\ast = f [\alpha, \beta]_\ast +
    \rho(\alpha)f\; \beta \qquad \forall \alpha, \beta \in
    \sect(L^\ast),\; f\in C^\infty(M).$$
\end{Lemma}
\begin{proof}
    Mit der Isotropie von $L^\ast$ folgt
    $$0 = \rho(e)\scal{\alpha,\beta} =
    \scal{e,[\alpha,\beta]_\C+[\beta,\alpha]_\C}$$
    und damit die
    Antisymmetrie von $[\dcd]_\ast$. Die Leibniz-Regel folgt direkt
    aus der Leibniz-Regel für die Courant-Klammer.
\end{proof}
\begin{Bemerkung}
    Im Allgemeinen wird die Jacobi-Identität für die Klammer
    $[\dcd]_\ast$ nicht gelten. Wir werden später sehen (Lemma
    \ref{quasiLemma}), dass die Jacobi-Identität für $[\dcd]_\ast$
    genau dann gilt, wenn $L^\ast$ eine Dirac-Struktur ist.
\end{Bemerkung}

Wie bei den Lie-Algebroiden (vgl. \ref{defSchouten}) erweitern wir
jetzt diese Klammer auf alle Formen $\omega \in \sect(\Wedge^\bullet
L^\ast)$, indem wir
$$[f,g]_\ast = 0\quad \forall f,g \in C^\infty(M),$$
$$[\alpha,f]_\ast = \rho(\alpha)f = -[f,\alpha]_\ast\quad \forall
\alpha \in \sect(L^\ast),\; f\in C^\infty(M)$$
und für $\omega \in
\sect(\Wedge^k L^\ast)$, $\mu \in \sect(\Wedge^l L^\ast)$ sowie $\eta
\in \sect(\Wedge^\bullet L^\ast)$ die Leibniz-Regel
$$[\omega,\mu \wedge \eta]_\ast = [\omega,\mu]_\ast\wedge \eta +
(-1)^{(k-1)l} \mu \wedge [\omega,\eta]_\ast$$
fordern. Lokal ergeben
sich für Formen $\omega = \alpha_1 \wedge \ldots \wedge \alpha_k$,
$\mu = \beta_1 \wedge \ldots \wedge \beta_l$ und $f \in C^\infty$ die
Gleichungen
$$[f,\omega]_\ast= \sum_{i=1}^{k} (-1)^{i}\rho(\alpha_i)f \alpha_1
\wedge \elide{i} \wedge \alpha_k$$
und
$$[\omega,\mu]_\ast = \sum_{i=1}^k \sum_{j=1}^l (-1)^{i+j}
[\alpha_i,\beta_j]_\ast\wedge
\alpha_1\elide{i}\wedge\alpha_k\wedge\beta_1\wedge\elide{j}
\wedge\beta_l.$$
Wie im Falle der Schouten-Nijenhuis-Klammer bei
Lie-Algebroiden überlegt man sich, dass
$$[\dcd]_\ast:\sect(\Wedge^k L^\ast)\times \sect(\Wedge^l L^\ast) \ra
\sect(\Wedge^{k+l-1} L^\ast)$$
dadurch wohldefiniert ist. Die
Bedeutung dieser Klammer für uns ergibt sich aus dem dritten Teil des
folgenden Lemmas.

\begin{Lemma}
    Sei $\alpha \in \sect(L^\ast)$, $\omega,\mu \in \sect(\Wedge^2
    L^\ast)$ und $f \in C^\infty(M)$ sowie $s_1,s_2,s_3 \in \sect(L)$.
    Dann gilt
    \begin{enumerateR}
    \item $[\omega,f]_\ast(s_1) = [\omega(s_1),f]_\ast$
    \item $[\omega,\alpha]_\ast(s_1,s_2) =
        \scal{[\omega(s_1),\alpha]_\ast,s_2} 
        - \scal{[\omega(s_2),\alpha]_\ast,s_1} + \rho(\alpha)
        \omega(s_1,s_2)$
    \item $\begin{array}[t]{@{}lcl} [\omega,\mu]_\ast(s_1,s_2,s_3)&=&
            \scal{[\omega(s_1),s_2]_\C,\mu(s_3)} +
            \scal{[s_1,\omega(s_2)]_\C,\mu(s_3)} \\[1.5mm]
            &&+\scal{[\omega(s_1),\mu(s_2)]_\C,s_3}
             + \scal{[\mu(s_1),s_2]_\C,\omega(s_3)}\\[1.5mm] 
            &&+\scal{[s_1,\mu(s_2)]_\C,\omega(s_3)} +
            \scal{[\mu(s_1),\omega(s_2)]_\C,s_3} \\
        \end{array}$
    \end{enumerateR}
\end{Lemma}
\begin{proof}
\begin{enumerateR}
    \item Zunächst überlegt man sich, dass sowohl der Ausdruck links
        des Gleichheitszeichens als auch der auf der rechten Seite
        funktionenlinear in $\omega$ ist. Deshalb genügt es, die
        Behauptung für $\omega = \mu \wedge \eta$ zu zeigen:
        $$[\mu \wedge \eta, f]_\ast(s) = -(\rho(\mu)f)\eta(s) +
        (\rho(\eta)f)\mu(s) = [\mu(s) \eta - \eta(s) \mu,f]_\ast =
        [(\mu\wedge\eta)(s),f]_\ast.$$
    \item Nach Konstruktion gilt für die linke Seite folgende
        Leibniz-Regel:
        $$[f \omega,\alpha]_\ast = f [\omega,\alpha]_\ast +
        (\rho(\alpha)f)\omega.$$
        Mit einer kleinen Rechnung überzeugt
        man sich davon, dass die selbe Leibniz-Regel auch für die
        rechte Seite gilt, so dass es wieder genügt, die Behauptung für
        $\omega = \mu \wedge \eta$ zu zeigen. Dann ergibt sich für die
        linke Seite
        \begin{eqnarray*}
            \lefteqn{[\mu\wedge\eta,\alpha]_\ast(s_1,s_2) } \hspace{10mm}\\
            &=&([\mu,\alpha]_\ast\wedge\eta - 
            [\eta,\alpha]_\ast\wedge\mu)(s_1,s_2) \\
            &=& \scal{[\mu,\alpha]_\C,s_1}\eta(s_2) -
            \scal{[\mu,\alpha]_\C,s_2}\eta(s_1)
            -\scal{[\eta,\alpha]_\C,s_1}\mu(s_2)
            + \scal{[\eta,\alpha]_\C,s_2}\mu(s_1).  
        \end{eqnarray*}
        Jetzt zur rechten Seite:
        \begin{eqnarray*}
            \lefteqn{\scal{[(\mu\wedge\eta)(s_1),\alpha]_\ast,s_2}}
            \hspace{7mm} \\
            &=&
            \scal{[\mu(s_1)\eta - \eta(s_1)\mu,\alpha]_\C,s_2}\\
            &=&
            \scal{\mu(s_1)[\eta,\alpha]_\C-\bigl(\rho(\alpha)\mu(s_1)\bigr)
              \eta,s_2}            
            -\scal{\eta(s_1)[\mu,\alpha]_\C+\bigl(\rho(\alpha)\eta(s_1)\bigr)
              \mu,s_2}\\
           &=&
           \scal{[\eta,\alpha]_\C,s_2}\mu(s_1)-
           \scal{[\mu,\alpha]_\C,s_2}\eta(s_1) +
           \bigl(\rho(\alpha)\eta(s_1)\bigr)\mu(s_2) - 
           \bigl(\rho(\alpha)\mu(s_1)\bigl) \eta(s_2),
       \end{eqnarray*}
       \begin{eqnarray*}
           \lefteqn{\scal{[(\mu\wedge\eta)(s_2),\alpha]_\ast,s_1}}
           \hspace{7mm} \\
           &=&\scal{[\eta,\alpha]_\C,s_1}\mu(s_2)-
           \scal{\mu,\alpha]_\C,s_1}\eta(s_2)
           - \bigl(\rho(\alpha)\eta(s_2)\bigr)\mu(s_1) -
           \bigl(\rho(\alpha)\mu(s_2)\bigl) \eta(s_1).
       \end{eqnarray*}
       Damit folgt
       \begin{eqnarray*}
           \lefteqn{[\mu\wedge\eta,\alpha]_\ast(s_1,s_2) -
           \scal{[(\mu\wedge\eta)(s_1),\alpha]_\ast,s_2}+
           \scal{[(\mu\wedge\eta)(s_2),\alpha]_\ast,s_1}}\\ &\quad=&
           -\bigl(\rho(\alpha)\eta(s_1)\bigr)\mu(s_2) +
           \bigl(\rho(\alpha)\mu(s_1)\bigl) \eta(s_2)
           +\bigl(\rho(\alpha)\eta(s_2)\bigr)\mu(s_1) -
           \bigl(\rho(\alpha)\mu(s_2)\bigl) \eta(s_1) \\
           &\quad=& \rho(\alpha)\big(\mu(s_1)\eta(s_2) -
           \mu(s_2)\eta(s_1)\big) \\
           &\quad=& \rho(\alpha)(\mu\wedge\eta)(s_1,s_2),
       \end{eqnarray*}
       womit die zweite Aussage gezeigt ist.
   \item
    Für die linke Seite gelten die Leibniz-Regeln
    $$[\omega, f \mu]_\ast = f [\omega,\mu]_\ast + [\omega,f]_\ast \mu$$
    sowie
    $$[f \omega, \mu]_\ast = f [\omega,\mu]_\ast + [f,\mu]_\ast
    \omega.$$
    Mit Hilfe des bisher gezeigten finden wir, dass für die
    rechte Seite die gleichen Leibniz-Regeln gelten, sodass es wieder
    reicht, die Behauptung auf Formen $\omega = \alpha_1 \wedge
    \alpha_2$, $\mu = \beta_1 \wedge \beta_2$ zu überprüfen.  Wir
    rechnen also für die linke Seite
    \begin{eqnarray*}
        \lefteqn{[\alpha_1\wedge\alpha_2,\beta_1\wedge\beta_2]_\ast
          (s_1,s_2,s_3)} \hspace{10mm}\\  
        &=&([\alpha_1,\beta_1]_\ast\wedge
        \alpha_2\wedge\beta_2)(s_1,s_2,s_3) 
        - ([\alpha_1,\beta_2]_\ast\wedge
        \alpha_2\wedge\beta_1)(s_1,s_2,s_3)\\
        &&-
        ([\alpha_2,\beta_1]_\ast\wedge \alpha_1\wedge\beta_2)(s_1,s_2,s_3)
        + ([\alpha_2,\beta_2]_\ast\wedge
        \alpha_1\wedge\beta_1)(s_1,s_2,s_3).
    \end{eqnarray*} 
    Für die ersten drei Terme
    auf der rechten Seite der zu zeigenden Gleichung ergibt sich 
    \begin{eqnarray*}
        \lefteqn{\scal{[(\alpha_1 \wedge \alpha_2)(s_1),s_2]_\ast,(\beta_1
            \wedge \beta_2)(s_3)}} \\
        &\qquad=&\alpha_1(s_1)\beta_1(s_3)
        \bigl(\rho(\alpha_2)(\beta_2(s_2))\bigr) - \alpha_1(s_1)
        \beta_1(s_3)\scal{s_2,[\alpha_2,\beta_2]_\C}\\
        && -\alpha_1(s_1)\beta_2(s_3)
        \bigl(\rho(\alpha_2)(\beta_1(s_2))\bigr) + \alpha_1(s_1)
        \beta_2(s_3)\scal{s_2,[\alpha_2,\beta_1]_\C}\\
        && +\alpha_2(s_2)\beta_1(s_3)
        \bigl(\rho(\beta_2)(\alpha_1(s_1))\bigr) -
        \alpha_2(s_2)\beta_2(s_3) 
        \bigl(\rho(\beta_1)(\alpha_1(s_1))\bigr)\\
        &&-\alpha_2(s_1)\beta_1(s_3)
        \bigl(\rho(\alpha_1)(\beta_2(s_2))\bigr) + \alpha_2(s_1)
        \beta_1(s_3)\scal{s_2,[\alpha_1,\beta_2]_\C}\\
        && +\alpha_2(s_1)\beta_2(s_3)
        \bigl(\rho(\alpha_1)(\beta_1(s_2))\bigr) - \alpha_2(s_1)
        \beta_2(s_3)\scal{s_2,[\alpha_1,\beta_1]_\C}\\
        && -\alpha_1(s_2)\beta_1(s_3)
        \bigl(\rho(\beta_2)(\alpha_2(s_1))\bigr) +
        \alpha_1(s_2)\beta_2(s_3) 
        \bigl(\rho(\beta_1)(\alpha_2(s_1))\bigr)\\
    \end{eqnarray*}
    sowie
    \begin{eqnarray*}
        \lefteqn{\scal{[s_1,(\alpha_1 \wedge \alpha_2)(s_2)]_\ast,(\beta_1
            \wedge \beta_2)(s_3)}}  \\
        &\qquad=&\alpha_1(s_2)\beta_1(s_3)
        \bigl(\rho(\beta_2)(\alpha_2(s_1))\bigr) + \alpha_1(s_2)
        \beta_1(s_3)\scal{s_1,[\alpha_2,\beta_2]_\C}\\
        && -\alpha_1(s_2)\beta_1(s_3)
        \bigl(\rho(\alpha_2)(\beta_2(s_1))\bigr) - \alpha_1(s_2)
        \beta_2(s_3)\scal{s_1,[\alpha_2,\beta_1]_\C}\\
        && + \alpha_1(s_2)\beta_2(s_3) 
        \bigl(\rho(\alpha_2)(\beta_1(s_1))\bigr) - 
        \alpha_1(s_2)\beta_2(s_3)
        \bigl(\rho(\beta_1)(\alpha_2(s_1))\bigr)\\
        &&-\alpha_2(s_2)\beta_1(s_3)
        \bigl(\rho(\beta_2)(\alpha_1(s_1))\bigr) - \alpha_2(s_2)
        \beta_1(s_3)\scal{s_1,[\alpha_1,\beta_2]_\C}\\
        && +\alpha_2(s_2)\beta_1(s_3)
        \bigl(\rho(\alpha_1)(\beta_2(s_1))\bigr) + \alpha_2(s_2)
        \beta_2(s_3)\scal{s_1,[\alpha_1,\beta_1]_\C}\\
        && -\alpha_2(s_2)\beta_2(s_3)
        \bigl(\rho(\alpha_1)(\beta_1(s_1))\bigr) +
        \alpha_2(s_2)\beta_2(s_3) 
        \bigl(\rho(\beta_1)(\alpha_1(s_1))\bigr)\\
    \end{eqnarray*}
    und
    \begin{eqnarray*}
        \lefteqn{\scal{[(\alpha_1 \wedge \alpha_2)(s_1),(\beta_1
            \wedge \beta_2)(s_2)]_\ast,s_3}}  \\
        &\qquad=&\alpha_1(s_1)\beta_2(s_3)
        \bigl(\rho(\alpha_2)(\beta_1(s_2))\bigr) + \alpha_1(s_1)
        \beta_1(s_2)\scal{s_3,[\alpha_2,\beta_2]_\C}\\
        && -\alpha_2(s_3)\beta_1(s_2)
        \bigl(\rho(\beta_2)(\alpha_1(s_1))\bigr) - \alpha_1(s_1)
        \beta_2(s_2)\scal{s_3,[\alpha_2,\beta_1]_\C}\\
        && + \alpha_2(s_3)\beta_2(s_2) 
        \bigl(\rho(\beta_1)(\alpha_1(s_1))\bigr) - 
        \alpha_1(s_1)\beta_1(s_3)
        \bigl(\rho(\alpha_2)(\beta_2(s_2))\bigr)\\
        &&-\alpha_2(s_1)\beta_2(s_3)
        \bigl(\rho(\alpha_1)(\beta_1(s_2))\bigr) - \alpha_2(s_1)
        \beta_1(s_2)\scal{s_3,[\alpha_1,\beta_2]_\C}\\
        && +\alpha_1(s_3)\beta_1(s_2)
        \bigl(\rho(\beta_2)(\alpha_2(s_1))\bigr) - \alpha_2(s_1)
        \beta_1(s_2)\scal{s_3,[\alpha_1,\beta_2]_\C}\\
        && -\alpha_1(s_3)\beta_2(s_2)
        \bigl(\rho(\beta_1)(\alpha_2(s_1))\bigr) +
        \alpha_2(s_1)\beta_1(s_3) 
        \bigl(\rho(\alpha_1)(\beta_2(s_2))\bigr).\\
    \end{eqnarray*}
    Die verbleibenden drei Terme erhält man jetzt aus den obigen
    Gleichungen durch Vertauschen der $\alpha$'s und $\beta$'s.
    Summiert man nun alles auf der rechten Seite auf, so bleiben nur
    Terme der Form $\alpha_1(s_1)
    \beta_1(s_3)\scal{s_2,[\alpha_2,\beta_2]_\C}$ usw. übrig, und man
    erkennt, dass diese verbleibenden Terme gerade die ausmultiplizierte
    Version der oben berechneten linke Seiten darstellen.
   \end{enumerateR}
\end{proof}
Wir haben also insbesondere die Gleichung
$$
\tfrac{1}{2}[\omega,\omega]_\ast(s_1,s_2,s_3)=
\scal{[\omega(s_1),s_2]_\C,\omega(s_3)} +
\scal{[s_1,\omega(s_2)]_\C,\omega(s_3)} +
\scal{[\omega(s_1),\omega(s_1)]_\C,s_3}$$
gezeigt, d.h. die
quadratischen Terme der Deformationsbedingung in Abschnitt
\ref{graphDef} werden durch $\tfrac{1}{2}[\omega,\omega]_\ast$ erfasst.

Ist $L^\ast$ auch eine Dirac-Struktur, so ist $(L,L^\ast)$ ein
Lie-Bialgebroid (siehe \ref{defLieBiAlg})
\cite{liu.weinstein.xu:1997a}. In diesem Fall ist $d_L$ eine
Superderivation vom Grad Eins bezüglich der Schouten-Nijenhuis-Klammer
$[\dcd]_\ast$. Wir werden jetzt zeigen, dass diese Aussage auch im
allgemeinen Fall weiterhin richtig bleibt.

Im folgenden bezeichnen wir mit $\prL$ bzw. $\prLs$ die Projektion auf
$L$ bzw. $L^\ast$ oder benutzen die kürzeren Schreibweisen $e_L :=
\prL(e)$ und $e_{L^\ast}:= \prLs(e)$ für $e \in \sect(L\oplus
L^\ast)$.  Es gilt dann für $s \in \sect(L)$, $e_1,e_2 \in
\sect(L\oplus L^\ast)$ und $f \in C^\infty(M)$
$$\scal{\dL f,s} = \rho(s)f = \scal{\D f,s} = \scal{\prLs(\D f),s} =
\scal{\D f_{L^\ast},s}$$
sowie
$$\scal{e_1|_{L^\ast},e_2}=\scal{\prLs(e_1),e_2} = \scal{e_1,\prL(e_2)} = \scal{e_1,e_2|_L}.$$
\begin{Lemma}\label{courantLemma}
    Sei $s \in \sect(L)$ und $\alpha \in \sect(L^\ast)$. Dann gilt
    \begin{enumerateR}
    \item $[s,\alpha]_{L^\ast} = \Lie_s \alpha$
    \item $[\alpha,s]_{L^\ast} = -i_s \dL \alpha\;$ bzw.
        $\scal{[\alpha,s_1]_\C,s_2} = -\dL \alpha(s_1,s_2)$  für $s_1,s_2
        \in \sect(L)$.
    \end{enumerateR}
    Dabei bezeichnet  $\Lie_s = \dif_L\,i_s + i_s\,\dif_L$ die
    Lieableitung (siehe Def. \ref{defLieAb}) auf dem
    Lie-Alge\-bro\-id $L$.
\end{Lemma}
\begin{proof}
    Die erste Behauptung ergibt sich aus der folgenden Rechnung, wobei
    $s_1,s_2 \in \sect(L)$,
    \begin{eqnarray*}
        \scal{[s_1,\alpha]_\C,s_2} &=& -\scal{\alpha,[s_1,s_2]_\C} +
    \rho(s_1)\alpha(s_2)\\
    &=& \dL \alpha(s_1,s_2) + \rho(s_2)\alpha(s_1)\\
    &=& \scal{i_{s_1}\dL \alpha + \dL i_{s_1} \alpha,s_2},
    \end{eqnarray*}
    die zweite Behauptung folgt jetzt mit
    $$\scal{[\alpha,s_1]_\C,s_2} =
    -\scal{[s_1,\alpha]_\C,s_2}+\rho(s_2)\alpha(s_1) = -\scal{i_{s_1}
      \dL \alpha,s_2}.$$
\end{proof}
 
Mit diesen beiden Lemmata können wir nun folgendes zeigen:
\begin{Lemma}
    Seien $f,g \in C^\infty(M)$ und $\alpha, \beta \in \sect(L^\ast)$.
    Dann gilt
    \begin{enumerateR}
    \item $\dL [f,g]_\ast = 0 = [\dL f,g]_\ast - [f, \dL g]_\ast$
    \item $\dL [\alpha,f]_\ast = [\dL \alpha, f]_\ast + [\alpha, \dL
        f]_\ast$ 
    \item $\dL [\alpha,\beta]_\ast = [\dL \alpha,\beta]_\ast + [\alpha,\dL
        \beta]_\ast$
    \end{enumerateR}
\end{Lemma}

\begin{proof}
    \begin{enumerateR}
    \item   Es gilt
      $$[\dL f,g]_\ast = \rho(\dL f)g = \scal{\D g, \dL f} = \scal{\D g,
        \prLs(\D f)}$$
      sowie
      $$[f,\dL g]_\ast = - \scal{\D f,\prLs(\D g)} = -\scal{\D g,\prL(\D
        f)}$$
      und damit
      $$[\dL f,g]_\ast - [f,\dL g]_\ast = \scal{\D f,\D g} = 0.$$
  \item Sei $s \in \sect(L)$. Dann rechnen wir für die linke Seite
      $$ \scal{\dL [\alpha,f]_\ast,s} = \scal{\dL (\rho(\alpha)f),s} =
      \rho(s)\rho(\alpha) f = - \rho([\alpha,s]_\C)f+ \rho(\alpha)\rho(s)
      f$$
      und für die rechte Seite
      $$\scal{[\dL \alpha, f]_\ast,s} = [\dL \alpha(s),f]_\ast = \rho(\dL
      \alpha(s))f = - \rho([\alpha,s]_{L^\ast})f$$
      sowie
      $$\scal{[\alpha,\dL f]_\ast,s} = - \scal{\dL f,[\alpha,s]_\C} +
      \rho(\alpha)\scal{\dL f,s} = -\rho([\alpha,s]_L)f +
      \rho(\alpha)\rho(s)f.$$
      Damit ist die zweite Aussage gezeigt.
  \item Zunächst erhalten wir mit $s_1,s_2 \in \sect(L)$ für die linke
      Seite:
      \begin{eqnarray*}
          \dL [\alpha,\beta]_\ast(s_1,s_2) &=& \scal{i_{s_1} \dL
            [\alpha,\beta]_\ast,s_2} =
          -\scal{[[\alpha,\beta]_\ast,s_1]_\C,s_2}\\ 
          &=&-\scal{[[\alpha,\beta]_\C,s_1]_\C,s_2} 
          +\scal{[[\alpha,\beta]_L,s_1]_\C,s_2}\\
          &=& -\scal{[\alpha,[\beta,s_1]_\C]_\C,s_2} +
          \scal{[\beta,[\alpha,s_1]_\C]_\C,s_2} \\
          &=& \scal{[\beta,s_1]_\C,[\alpha,s_2]_\C} -
          \rho(\alpha)\scal{[\beta,s_1]_\C,s_2}\\
          &&\qquad-\scal{[\alpha,s_1]_\C,[\beta,s_2]_\C} +
          \rho(\beta)\scal{[\alpha,s_1]_\C,s_2} \\
           &=& \scal{[\beta,s_1]_\C,[\alpha,s_2]_\C} +
          \rho(\alpha)\dL\beta(s_1,s_2)\\
          &&\qquad-\scal{[\alpha,s_1]_\C,[\beta,s_2]_\C} -
          \rho(\beta)\dL\alpha(s_1,s_2)
      \end{eqnarray*}
      Weiter mit der rechten Seite:
      \begin{eqnarray*}
           [\dL \alpha,\beta]_\ast(s_1,s_2) &=& \scal{[\dL
             \alpha(s_1),\beta]_\C,s_2} -\scal{[\dL
             \alpha(s_2),\beta]_\C,s_1} + \rho(\beta)\dL
           \alpha(s_1,s_2)\\
           &=& -\scal{[\beta,\dL \alpha(s_1)]_\C,s_2}  + \scal{[\beta,\dL
             \alpha(s_2)]_\C,s_1} + \rho(\beta)\dL \alpha(s_1,s_2) \\
           &=& \scal{i_{s_1}\dL \alpha,[\beta,s_2]_\C} - \scal{i_{s_2}
             \dL \alpha,[\beta,s_1]_\C} - \rho(\beta) \dL
           \alpha(s_1,s_2)\\ 
           &=& -\scal{[\alpha,s_1]_{L^\ast},[\beta,s_2]_\C} +
           \scal{[\alpha,s_2]_{L^\ast},[\beta,s_1]_\C}-\rho(\beta)\dL
           \alpha(s_1,s_2)
       \end{eqnarray*}
       \begin{eqnarray*}
           [\alpha,\dL \beta]_\ast(s_1,s_2) &=& - [\dL
           \beta,\alpha]_\ast(s_1,s_2) \\
           &=&  \scal{[\beta,s_1]_{L^\ast},[\alpha,s_2]_\C} -
           \scal{[\beta,s_2]_{L^\ast},[\alpha,s_1]_\C}+\rho(\alpha)\dL
           \beta(s_1,s_2)\\
           &=& \scal{[\beta,s_1]_\C,[\alpha,s_2]_L} -
           \scal{[\beta,s_2]_\C,[\alpha,s_1]_L}+\rho(\alpha)\dL
           \beta(s_1,s_2)
       \end{eqnarray*}
       Damit folgt nun die Behauptung.
  \end{enumerateR}
\end{proof}

\begin{Satz}\label{compatibility}
    Das Lie-Algebroiddifferential $d_L$ ist eine Superderivation vom
    Grad Eins bezüglich $[\dcd]_\ast$, d.h. für $\omega \in
    \sect(\Wedge^k L^\ast)$ und $\mu \in \sect(\Wedge^l)$ gilt
    $$\dL [\omega,\mu]_\ast = [\dL \omega,\mu]_\ast +
    (-1)^{k-1}[\omega,\dL \mu]_\ast.$$
\end{Satz}
\begin{proof}
    Definieren wir für $\omega \in \sect(\Wedge^k L^\ast)$ eine
    Abbildung $ad_\omega$ durch $ad_\omega(\mu) =
    [\omega,\mu]_\ast$ für $\mu \in \sect(\Wedge^\bullet L^\ast)$,
    dann ist $ad_\omega$ aufgrund der Leibniz-Regel für $[\dcd]_\ast$
    eine Superderivation bezüglich des Dachproduktes vom Grad $k-1$.
    Weiter ist $\dL$ eine Superderivation vom Grad Eins, und damit ist
    der Superkommutator
    $$[\dL,ad_\omega] = \dL\circ ad_\omega - (-1)^{k-1}
    ad_\omega\circ\dL$$
    eine Derivation vom Grad $k$. Andererseits ist
    aber auch $ad_{\dL \omega}$ eine Superderivation vom Grad $k$, und
    nach dem obigen Lemma stimmen beide auf den Erzeugern
    $C^\infty(M)$ und $\sect(L^\ast)$ von $\sect(\Wedge^\bullet
    L^\ast)$ überein. Damit folgt aber sofort
    $$
    \dL\circ ad_\omega(\mu) - (-1)^{k-1} ad_\omega \circ \dL \mu =
    ad_{\dL \omega}(\mu)$$
    für alle $\omega \in \sect(\Wedge^k
    L^\ast)$ und $\beta \in \sect(\Wedge^\bullet L^\ast)$.
\end{proof}

Wir haben bis jetzt insbesondere gezeigt:
\begin{Satz}[{\cite[Theorem 2.6]{liu.weinstein.xu:1997a}}]
    Seien $L$ und $L'$ transversale Dirac-Strukturen in einem
    Courant-Algebroid $E$. Dann ist $(L,L')$ ein Lie-Bialgebroid,
    wobei $L'$ vermöge der in $E$ gegebenen Bilinearform mit $L^\ast$
    identifiziert wird, und das dadurch definierte Courant-Algebroid
    $L\oplus L'$ stimmt mit $E$ überein.
\end{Satz}

Wir wollen jetzt die Gleichung, die $\omega \in \sect(\Wedge^2
L^\ast)$ erfüllen muss, damit $L = \op{graph}(\omega)$ eine
Dirac-Struktur ist, mit Hilfe der Klammer $[\dcd]_\ast$ formulieren.
Doch sei zuerst noch eine 3-Form $T_\omega \in \sect(\Wedge^3 L^\ast)$
durch
$$T_\omega(s_1,s_2,s_3) =
\scal{[\omega(s_1),\omega(s_2)]_\C,\omega(s_3)}$$
definiert. Die
Antisymmetrie von $T_\omega$ folgt dabei mit der Isotropie von
$L^\ast$, denn es gilt
\begin{eqnarray*}
    T_\omega(s_1,s_2,s_3) &=&
    \scal{[\omega(s_1),\omega(s_2)]_\C,\omega(s_3)}\\ 
    &=& -\scal{[\omega(s_2),\omega(s_1)]_\C,\omega(s_3)} +
      \rho(\omega(s_3))\scal{\omega(s_1),\omega(s_2)}\\
      &=& - T_\omega{(s_2,s_1,s_3)}
\end{eqnarray*}
sowie
\begin{eqnarray*}
    T_\omega(s_1,s_2,s_3) &=&
    \scal{[\omega(s_1),\omega(s_2)]_\C,\omega(s_3)} \\
    &=& - \scal{\omega(s_2),[\omega(s_1),\omega(s_3)]_\C} +
      \rho(\omega(s_1)) \scal{\omega(s_1),\omega(s_3)}\\
      &=& -T_\omega(s_1,s_3,s_2).
\end{eqnarray*}
Aus der Antisymmetrie folgt jetzt auch, dass $T_\omega$ tensoriell
ist, da die $C^\infty(M)$-Linearität im dritten Argument
offensichtlich ist.  Damit haben wir den folgenden Satz gezeigt:
\begin{Satz}\label{diracdeformation}
    Sei $E$ ein Courant-Algebroid mit Dirac-Struktur $L$. Sei weiter
    $L'$ ein isotropes Komplement von $L$. Identifizieren wir $E$ mit
    $L\oplus L^\ast$, dann ist $\op{graph}(\omega)$ für eine Zweiform
    $\omega \in \sect(\Wedge^2 L^\ast)$ genau dann eine
    Dirac-Struktur, wenn die 
    Gleichung
    $$\dif_L \omega + \frac{1}{2}[\omega,\omega]_\ast + T_\omega = 0$$
    erfüllt ist.
\end{Satz}

\subsection{Formale Dirac-Strukturen}
Die Gleichung, die $\omega$ in Lemma \ref{diracdeformation} erfüllen
muss, ist also eine Gleichung in $\sect(\Wedge^3 L^\ast)$.  Damit
haben wir jetzt eine Möglichkeit gefunden, wie wir die formale
Deformationstheorie beschreiben können.

\begin{Definition}
    Seien die Voraussetzungen von Satz \ref{diracdeformation}
    gegeben. Eine formale Deformation der Dirac-Struktur $L =
    L_0$  ist eine formale Potenzreihe
    $$\omega_t = t\; \omega_1 +t^2\,\omega_2 + \ldots \in \sect(\Wedge^2
    L^\ast)[[t]],$$
    so dass die Gleichung
    $$\dif_L \omega + \frac{1}{2}[\omega,\omega]_\ast + T_\omega = 0$$
    in jeder Ordnung von $t$ erfüllt ist.  Eine formale Deformation
    der Ordnung $N$ ist ein formale Reihe $\omega_t =
    t\omega_1+\ldots$, so dass die obige Gleichung bis zur Ordnung $N$
    erfüllt ist.
\end{Definition}

Wählen wir in $E$ ein anderes isotropes Komplement $L'$ zu $L$, dann
erhalten wir natürlich einen anderen Isomorphismus $E\cong L\oplus
L^\ast$ und damit eine andere Courant-Algebroid\-struk\-tur auf $L\oplus
L^\ast$.  Diese beiden Strukturen sind nach Konstruktion isomorph, und
der Isomorphismus $\Phi:  L\oplus L^\ast \ra L \oplus L^\ast$ zwischen
beiden Strukturen ist eingeschränkt auf $L$ die
Identität. Da $\Phi$ eine Isometrie ist, folgt ähnlich wie in Lemma
\ref{stCAautos}, dass $\Phi$ von der Form
$$\Phi(e,\alpha) = \nu_\Lambda(e,\alpha):=(e+\Lambda(\alpha),\alpha)$$
mit $\Lambda \in \sect(\Wedge^2 L)$, aufgefasst als Abbildung
$\Lambda:L^\ast \ra L$, ist. Sei nun $L_t = \op{graph}(\omega_t)$. Man
überlegt sich leicht, dass $\nu_\Lambda(L_t)$ genau dann wieder ein
Graph ist, wenn $(\op{id} + \Lambda \omega_t)$ invertierbar ist (was im
formalen Fall stets gilt) und dass
dann  $\nu_\Lambda(L_t) = \op{graph}(\omega'_t)$ mit
$\omega'_t=\omega_t (\op{id} + \Lambda \omega)^{-1}$ folgt.

Die beiden folgenden Beispiele zeigen, dass unsere Überlegungen in
diesen Spezialfällen die bekannten Ergebnisse reproduzieren.
\subsection{Präsymplektische Mannigfaltigkeiten}\label{graphPreSymp}
Ist $L = \op{graph}(\omega)$ für eine geschlossene $2$-Form $\omega$,
dann ist $T^\ast M$ eine zu $L$ transversale Dirac-Struktur und wir
können $L^\ast$ mit $T^\ast M$ identifizieren.  In diesem Fall ist
$L\oplus L^\ast$ ein Lie-Bialgebroid, wobei die Lie-Klammer auf
$L^\ast$ identisch Null ist. Für eine $t$-abhängige $2$-Form $\eta_t:
L \ra L^\ast$ ist $L_t=\op{graph}(\eta_t)$ eine Dirac-Struktur, wenn
$\dL \eta_t = 0$ gilt. Identifizieren wir $L$ auf offensichtliche
Weise mit $TM$, so wird aus $\dL$ das gewöhnliche deRahm-Differential
auf $M$ und aus $\eta_t$ eine $2$-Form $\eta'_t$ auf $M$. Genauer
gesagt betrachten wir die Eichtransformation $\tau_\omega$, die wegen
der Geschlossenheit von $\omega$ ein Automorphismus von $\stCA$ ist.
Dies liefert einen Lie-Algebroid-Isomorphismus $\tau_\omega|_{TM}:TM
\tilde{\lra} \op{graph}(\omega)$ und es gilt dann $L_t =
\op{graph}(\omega + \eta'_t) = \tau_{\omega+\eta'_t}(TM)$. Wir
erhalten also das bekannte Ergebnis, dass $L_t$ genau dann eine
Dirac-Struktur ist, wenn $\omega+ \eta'_t$ geschlossen ist.

Triviale Deformationen von $L$ sind solche, die sich als
$L_t=\F\phi_t(L)$ für einen Diffeomorphismus $\phi_t$ von $M$ schreiben
lassen.  Dies bedeutet in diesem Fall, dass
$$\omega_t = \phi_t^\ast \omega_0$$
mit $\omega_t = \omega + \eta'_t$. Wir nehmen nun an, dass wir eine
global definierte zeitabhängige Einsform $\beta_t$ finden können, so dass
$$\dif \beta_t = \frac{\dif}{\dif t} \omega_t = \dot{\omega}_t.$$
Falls  $\omega_t$ symplektisch ist, können wir durch 
$$i_{X_t}\omega_t = -\beta_t$$
ein zeitabhängiges Vektorfeld
definieren. War $M$ kompakt, so ist der Fluss (oder genauer die
Zeitentwicklung) $\psi_t$ zu $X_t$ für alle $t$ definiert.  In
Allgemeinen finden wir zumindest für jeden Punkt $p \in M$ eine
Umgebung, auf der der Fluss $\psi_t$ für $t$ in einem Intervall
$[0,\epsilon]$ mit $\epsilon > 0$ definiert ist.  Ableiten von
$\psi_t^\ast \omega_t$ liefert jetzt
$$\frac{\dif}{\dif t} \psi_t^\ast \omega_t = \psi_t^\ast(i_{X_t} \dif
\omega + \dif i_{X_t}\omega + \dot{\omega}_t) = \psi_t^\ast
\dif(i_{X_t}\omega+\beta_t) = 0,$$
und damit $\psi_t^\ast \omega_t = \psi_0^\ast \omega_0 = \omega_0$.

Umgekehrt erhalten wir für eine triviale Deformation $\omega_t =
\phi_t^\ast  \omega_0$ durch  Ableiten nach $t$ die Gleichung
$$\dot{\omega}_t = \phi_t^\ast \dif i_{X_t}\omega_0 = \dif \phi^\ast_t
i_{X_t}\omega_0,$$
also ist $\dot{\omega}_t$ exakt. Durch weiteres Ableiten folgt
schließlich, dass  
$$\frac{\dif^k\omega_t}{\dif t^k}  $$ 
für alle $k>0$ exakt ist.
Wir definieren deshalb:
\begin{Definition}
    Seien $\omega_t$ und $\omega'_t$ zwei formale Deformationen von
    $\omega = \omega_0$. Dann heißen $\omega_t$ und $\omega'_t$
    äquivalent, wenn es einen formalen Diffeomorphismus
    $$\phi_t = \exp(\Lie_{X_t})$$
    mit $X_t = t X_1 + t^2 X_2 +\ldots
    \in \mathcal{X}(M)[[t]]$ gibt, so dass
    $$\omega'_t = \phi_t(\omega_t).$$ Eine Deformation $\omega_t$
    heißt trivial, wenn $\omega_t$ äquivalent zu $\omega_0$ ist,
     $$\omega_t = \phi_t(\omega_0).$$
\end{Definition}
\begin{Bemerkung}
    Man kann zeigen, dass die homogenen Automorphismen der Algebra
    $$(\Omega^\bullet(M)[[t]],\wedge),$$ die in unterster Ordnung mit der
    Identität beginnen, von der Form $\exp(D_t)$ mit einer formalen
    Reihe $D_t = t D_1 + t D_2 +\ldots$ von Derivationen von
    $\Omega^\bullet(M)$ sind \cite{waldmann:2004a}.  Verlangen wir
    weiter, dass $[d,\exp(D_t)] = 0$, so folgt $[d,D_k]=0$ für alle $k
    \geq 1$ und nach Bemerkung \ref{omegaDerivationen} muss dann $D_k
    = \Lie_{X_k}$ für ein Vektorfeld $X_k$ gelten. D.h. die homogenen
    Automorphismen von $(\Omega^\bullet(M)[[t]],\wedge)$, die in
    unterster Ordnung mit der Identität beginnen und die Gleichung
    $\dif \omega = 0$ invariant lassen, sind genau die von der Form
    $\exp(\Lie_{X_t})$.
\end{Bemerkung}
Sei $\omega_t = \omega_0 + t^{k+1} \omega_{k+1} +\ldots$ die
Deformation einer symplektischen Form, die bis zur Ordnung $k$ mit
$\omega_0$ übereinstimmt. Ist $\omega_{k+1}$ exakt, $\omega_k = \dif
\alpha$, so ist $\omega_t$ bis zur Ordnung $k+1$ äquivalent zu
$\omega_0$. Um das zu sehen definieren wir durch $\alpha =
i_{X_{k+1}}\omega_0$ ein Vektorfeld $X_{k+1}$ und setzen $\phi_t =
\exp(t^{k+1}\Lie_{X_{k+1}})$. Damit folgt
\begin{eqnarray*}
    \phi_t(\omega_0)&=& \omega_0 + t^{k+1} \Lie_{X_{k+1}}\omega_0 + \ldots\\
    &=& \omega_0 + t^{k+1} \dif i_{X_{k+1}} \omega_0 +\ldots\\
    &=& \omega_0 + t^{k+1} \omega_{k+1} + \ldots\\
    &=& \omega_t + t^{k+2}(\ldots).
\end{eqnarray*}
Ist $H_{dR}^2(M) = 0$, so sind also alle Deformationen trivial. Falls
wir nun aber eine präsymplektische Form $\omega_0$ deformieren wollen,
können wir die Gleichung $\alpha = i_X\omega_0$ im Allgemeinen nicht
mehr lösen. Damit lässt sich nicht mehr nur anhand der
deRahm-Kohomologie von $M$ entscheiden, ob es nichttriviale
Deformationen  gibt oder nicht.

\subsection{Poisson-Mannigfaltigkeiten}\label{graphPoisson}
    Sei $L=\op{graph}(\pi)$ für einen Poisson-Bivektor $\pi$ auf $M$.
    Wir können $L$ mit $T^\ast M$ und $L^\ast$ mit $TM$
    identifizieren.  Die Lie-Algebroidstruktur auf $TM$ ist die
    kanonische, während die Lie-Klammer auf $T^\ast M$ die
    Koszul-Klammer
    $$[\alpha,\beta]=\Lie_{\pi(\alpha)}\beta-\Lie_{\pi(\beta)}\alpha -
    \dif(\pi(\alpha,\beta))$$
    und der Anker durch $\pi:T^\ast M\ra TM$
    gegeben ist.  Wir erhalten dadurch eine
    Courant-Algebroid\-struk\-tur 
    auf $T^\ast M \oplus TM$, die aber nicht mit der kanonischen
    Courant-Algebroid\-struk\-tur übereinstimmt. Weiter ist
    $\dL=[\pi,\cdot]$ und für eine $t$-abhängige $2$-Form $\mu_t:L \ra
    L^\ast$ mit $\mu_0=0$ ist die Gleichung $\dL \mu_t +
    \frac{1}{2}[\mu_t,\mu_t]_\ast = 0$ unter den obigen
    Identifikationen äquivalent zu
    $$[\pi + \mu_t,\pi +\mu_t]=0.$$
    Insbesondere erhalten wir, dass
    $\op{graph}(\pi)$ für einen Bivektor $\pi$ genau dann eine
    Dirac-Struktur ist, wenn $[\pi,\pi]= 0$ erfüllt ist.
    
    Die Bedingung für triviale Deformationen lautet jetzt
    $$ \op{graph}(\Pi_t) = \F\phi_t(\op{graph}(\Pi_0))$$
    für einen Diffeomorphismus $\phi_t$ von $M$, wobei jetzt
    $\Pi_t = \pi +\mu_t$ sei. Dies bedeutet aber
    $$\Pi_t=\phi_t^\ast \Pi_0,$$
    so dass wir die bekannte Charakterisierung trivialer
    Deformationen erhalten.

\subsection{Lie-Bialgebroide}

Wir betrachten nun noch denn Fall genauer, dass wir eine Aufspaltung
von $E$ in transversale Dirac-Strukturen $L$ und $L'$ finden können
und $(L,L^\ast)$ damit ein Lie-Bialgebroid ist. Ist $\omega_t =
t\omega_1 +t^2 \omega_2 +\ldots$ eine formale Deformation von
$\omega_0 = 0$, dann lautet die Bedingung, dass
$L_t=\op{graph}(\omega_t)$ eine Dirac-Struktur ist
$$\dL \omega_t +\frac{1}{2}[\omega_t,\omega_t]_\ast = 0.$$
Wir nehmen an,
dass $\omega_t$ diese Gleichung bis zur Ordnung $n$ erfüllt. Dann
folgt
\begin{eqnarray*}
    \dL \omega_t + \frac{1}{2}[\omega_t,\omega_t]_\ast &=&
    t^{n+1}\Big(\dL 
    \omega_{n+1} + \frac{1}{2} \sum_{i=1}^n
    [\omega_{n+1-i},\omega_i]_\ast\Big) + t^{n+2}(\ldots).
\end{eqnarray*}
Um die Deformation bis zur Ordnung $n+1$ fortzusetzen, müssen wir
also $\omega_{n+1}$ so bestimmen, dass  die Gleichung  
$$\dL \omega_{n+1} + \frac{1}{2} \sum_{i=1}^n
[\omega_{n+1-i},\omega_i]_\ast = 0$$
erfüllt ist. Im Allgemeinen wird dies nicht möglich sein, man kann
jedoch zeigen, dass die notwendige Bedingung 
$$\dL \sum_{i=1}^n [\omega_{n+1-i},\omega_i]_\ast = 0$$
immer erfüllt ist. Dazu die folgende Vorüberlegung.
Zunächst gilt für eine beliebige Zweiform $\omega \in  \Omega^2(L)$
aufgrund der Super-Jacobi-Identität, die in diesem Fall ja auch für
$[\dcd]_\ast$ erfüllt ist,  die Gleichung 
$$[\omega,[\omega,\omega]_\ast]_\ast = 0.$$
Definieren wir weiter durch
$$\dif_\omega = \dL + [\omega,\cdot]_\ast$$
eine $\wedge$-Superderivation vom Grad
Eins, dann folgt
\begin{eqnarray*}
    \dif_\omega\big(\dL \omega + \frac{1}{2}[\omega,\omega]_\ast\big)
    &=&  \frac{1}{2}[\dL \omega,\omega]_\ast - \frac{1}{2}[\omega,\dL
    \omega]_\ast + [\omega,\dL \omega]_\ast
    + [\omega,[\omega,\omega]_\ast]_\ast \\[1mm]
    &=& 0.
\end{eqnarray*}
Wenden wir diese Gleichung auf $\omega_t$ an und betrachten die
Ordnung $n+1$, erhalten wir
\begin{eqnarray*}
    \dif_{\omega_t}\big(\dL \omega_t +
    \frac{1}{2}[\omega_t,\omega_t]_\ast\big) &=& \Big(\dL +
    t[\omega_1,\cdot]_\ast  + \ldots\Big)\\
    &&\qquad\Big( t^{n+1}\big(\dL 
    \omega_{n+1} + \frac{1}{2} \sum_{i=1}^n
    [\omega_{n+1-i},\omega_i]_\ast\big) + t^{n+2}(\ldots)\Big)\\
    &=& \frac{1}{2} t^{n+1} \dL
    \sum_{i=1}^n[\omega_{n+1-i},\omega_i]_\ast + t^{n+2}(\ldots) \\
    &=& 0
\end{eqnarray*}
Es gibt also eine Obstruktion für die Fortsetzbarkeit einer formalen
Deformation in der dritten Lie-Algebroid-Kohomologie von $L$. Wir
werden später sehen, dass diese Aussage auch in der allgemeinen
Situation, also wenn $L^\ast$ keine Dirac-Struktur ist, weiterhin
gültig bleibt.

\section{Courant-Algebroide als Lie-quasi-Bialgebroide}

Ist in unserer Aufspaltung $E = L\oplus L'$ das Vektorbündel $L'\cong
L^\ast$ ebenfalls eine Dirac-Struktur, dann wissen wir bereits, dass
$(L,L^\ast)$ ein Lie-Bialgebroid ist. Setzen wir
$$\psi(\alpha,\beta,\gamma) = -\scal{[\alpha,\beta]_\C,\gamma},$$
dann
ist dadurch ein Element $\psi \in \sect(\Wedge^3 L)$ definiert und es
ist $\psi = 0$ genau dann, wenn $L^\ast$ eine Dirac-Struktur ist, und
genau dann erfüllt $[\dcd]_\ast$ die Jacobi-Identität. Man kann
deshalb vermuten, dass sich der Jacobiator von $[\dcd]_\ast$ in Termen
von $\psi$ ausdrücken lässt, und wie wir gleich sehen werden, vermuten
wir richtig.  Wir fassen $\psi$ im weiteren auch als eine Abbildung
$\sect(\Wedge^2 L^\ast)$ $\ra \sect(L)$ auf, und zwar gemäß
$\psi(\alpha,\beta) = -[\alpha,\beta]_L$.
\begin{Lemma}\label{quasiLemma}
    Für das eben definierte $\psi$ gelten die folgenden Gleichungen:
    \begin{enumerate}
    \item $\rho(\psi(\alpha,\beta)) =\rho([\alpha,\beta]_\ast)
        -[\rho(\alpha),\rho(\beta)]_\C $
    \item $\begin{array}[t]{@{}rcl} [[\alpha,\beta]_\ast,\gamma]_\ast
            + [[\beta,\gamma]_\ast,\alpha]_\ast +
            [[\gamma,\alpha]_\ast,\beta]_\ast &=&
            i_{\psi(\alpha,\beta)} \dL \gamma + i_{\psi(\beta,\gamma)}
            \dL
            \alpha + i_{\psi(\gamma,\alpha)} \dL \beta\\[2mm]
            &&\qquad + \dL\bigl(\psi(\alpha,\beta,\gamma)\bigr)
        \end{array}$
    \item $\begin{array}[t]{@{} l }
            \rho(\alpha)\psi(\beta,\gamma,\delta) -
            \rho(\beta)\psi(\alpha,\gamma,\delta) +
            \rho(\gamma)\psi(\alpha,\beta,\delta) -
            \rho(\delta)\psi(\alpha,\beta,\gamma) \\[1.5mm]
            \qquad - \psi([\alpha,\beta]_\ast,\gamma,\delta) +
            \psi([\alpha,\gamma]_\ast,\beta,\delta) -
            \psi([\alpha,\delta]_\ast,\beta,\gamma) \\[1.5mm]
            \qquad- \psi([\beta,\gamma]_\ast,\alpha,\delta) +
            \psi([\beta,\delta]_\ast,\alpha,\gamma) -
            \psi([\gamma,\delta]_\ast, \alpha,\beta)\quad = \quad 0
        \end{array}$
    \end{enumerate}
\end{Lemma}

\begin{proof}
    Die erste Behauptung ergibt sich sofort aus der Definition von
    $\psi$ und der Gleichung $[\rho(\alpha),\rho(\beta)] =
    \rho([\alpha,\beta]_\C)$. Für die zweite Behauptung rechnen wir mit
    Hilfe der Jacobi-Identität für die Courant-Klammer für alle $s \in \sect(L)$
    \begin{eqnarray*}
        \lefteqn{\scal{[\alpha,[\beta,\gamma]_\ast]_\C +
        [\beta,[\gamma,\alpha]_\ast]_\C +
        [\gamma,[\alpha,\beta]_\ast]_\C,s}}\\
    \qquad&=& \scal{[\alpha,[\beta,\gamma]_\ast]_\C-
      [\beta,[\alpha,\gamma]_\ast]_\C -[[\alpha,\beta]_\ast,\gamma]_\C,s}\\
    \qquad&=& \scal{-[\alpha,[\beta,\gamma]_L]_\C +
      [\beta,[\alpha,\gamma]_L]_\C + [[\alpha,\beta]_L,\gamma]_\C,s}\\
    \qquad&=& - \scal{[\alpha,[\beta,\gamma]_L]_\C +
      [\beta,[\gamma,\alpha]_L]_\C + [\gamma,[\alpha,\beta]_L]_\C,s} +
    \rho(s)\scal{[\alpha,\beta]_L,\gamma}\\
    \qquad&=& \scal{i_{[\beta,\gamma]_L} \dL \alpha +
      i_{[\gamma,\alpha]_L} \dL \beta + i_{[\alpha,\beta]_L} \dL
      \gamma + \dL \bigl( \psi(\alpha,\beta,\gamma) \bigr),s}.
  \end{eqnarray*}
  Nun zu der dritte Aussage. Wir rechnen zunächst
  \begin{eqnarray*}
      \rho(\alpha)\psi(\beta,\gamma,\delta) &=&
      -\rho(\alpha)\scal{[\beta,\gamma]_L,\delta}\\ 
      &=& -\scal{[\alpha,[\beta,\gamma]_L]_\C, \delta} -
      \scal{[\beta,\gamma]_L,[\alpha,\delta]_\C} \\
      &=& \psi(\beta,\gamma,[\alpha,\delta]_\ast) -
      \scal{[\alpha,[\beta,\gamma]_L]_\C, \delta} \\
      &=& \psi([\alpha,\delta]_\ast,\beta,\gamma) -
      \scal{[\alpha,[\beta,\gamma]_L]_\C, \delta}.
  \end{eqnarray*}
  Damit folgt unter Verwendung der Jacobi-Identität  
  $$\begin{array}{l} \rho(\alpha)\psi(\beta,\gamma,\delta) -
      \rho(\beta)\psi(\alpha,\gamma,\delta) +
      \rho(\gamma)\psi(\alpha,\beta,\delta) -
      \rho(\delta)\psi(\alpha,\beta,\gamma) \\[1.5mm]
      \qquad\qquad - \psi([\alpha,\delta]_\ast,\beta,\gamma) +
      \psi([\beta,\delta]_\ast,\alpha,\gamma) -
      \psi([\gamma,\delta]_\ast,\alpha,\beta) \\[1.5mm]
      \qquad = -\scal{[\alpha,[\beta,\gamma]_L]_\C,\delta} +
      \scal{[\beta,[\alpha,\gamma]_L]_\C,\delta} -
      \scal{[\gamma,[\alpha,\beta]_L]_\C,\delta} +
      \rho(\delta)\scal{[\alpha,\beta]_L,\gamma}\\[1.5mm]
      \qquad = -\scal{[\alpha,[\beta,\gamma]_L]_\C,\delta} +
      \scal{[\beta,[\alpha,\gamma]_L]_\C,\delta}
      +\scal{[[\alpha,\beta]_L,\gamma]_\C,\delta} \\[1.5mm]
      \qquad = \scal{[\alpha,[\beta,\gamma]_\ast]_\C,\delta} -
      \scal{[\beta,[\alpha,\gamma]_\ast]_\C,\delta}
      -\scal{[[\alpha,\beta]_\ast,\gamma]_\C,\delta} \\[1.5mm]
      \qquad = \psi([\beta,\gamma]_\ast,\alpha,\delta) -
      \psi([\alpha,\gamma]_\ast,\beta,\delta) +
      \psi([\alpha,\beta]_\ast,\gamma,\delta),
  \end{array}$$
  was zu zeigen war.
\end{proof}

Die Struktur, die wir auf dem Paar $(L,L^\ast)$ gefunden haben, erhält
durch die folgende Definition einen Namen.
\begin{Definition}[{\cite{roytenberg:1999a}}]
    Sei $A\ra M$ ein Vektorbündel. Dann heißt $(A,A^\ast)$
    Lie-quasi-Bi\-al\-ge\-broid, wenn folgendes gilt:
    \begin{enumerateR}
    \item $(A,[\dcd],\rho)$ ist ein Lie-Algebroid.
    \item Auf $A^\ast$ ist eine antisymmetrische Verknüpfung
        $[\dcd]_\ast:\sect(A^\ast) \times \sect(A^\ast) \ra
        \sect(A^\ast)$ sowie ein Vektorbündelhomomorphismus
        $\rho_\ast:A \ra TM$ definiert, so dass die Leibniz-Regel
        $$[\alpha,f \beta]_\ast = f [\alpha,\beta]_\ast +
        (\rho_\ast(\alpha)f)\beta$$
        gilt. Die Jacobi-Identität wird jedoch nicht gefordert.
    \item Das Lie-Algebroid-Differential $\dif_A$ von $A$ ist eine
        Superderivation der durch $[\dcd]_\ast$ und $\rho_\ast$
        gegebenen verallgemeinerten Schouten-Nijenhuis-Klammer auf
        $A^\ast$.  
    \item Es gibt ein $\psi \in \sect(\Wedge^3 A)$, so dass die in
        Lemma $\ref{quasiLemma}$ aufgeführten Gleichungen gelten.
    \end{enumerateR}
    Weiter heißt $(A,A^\ast)$ quasi-Lie-Bialgebroid, genau dann, wenn
    $(A^\ast,A)$ ein Lie-quasi-Bialge\-broid ist.
\end{Definition}    

Es gilt nun der folgende Satz:
\begin{Satz}[{\cite{kosmann-schwarzbach:2003a}}]\label{liequasicourant}
    Sei $(A,A^\ast)$ ein Lie-quasi-Bialge\-broid. Dann ist auf
    $E=A\oplus A^\ast$ eine Courant-Algebroid-Struktur gegeben. Dabei
    gilt für $s_1,s_2 \in \sect(A)$ und $\alpha_1, \alpha_2 \in
    \sect(A^\ast)$ 
    \begin{eqnarray*}
    [s_1+\alpha_1,s_2 + \alpha_2]_\C &=& [s_1,s_2] +
    \Lie_{s_1}\alpha_2 - i_{s_2}\dif_{A}\alpha_1\\
    && + [\alpha_1,\alpha_2]_\ast +
    i_{\alpha_1\wedge\alpha_2}\psi  + \Lie^\ast_{\alpha_1}s_2
    - i_{\alpha_2}\dif_{A^\ast} s_1 
    \end{eqnarray*}
    sowie
    $$\rho_E(s_1+\alpha_1)=\rho(s_1) + \rho_\ast(\alpha_1),$$
    wobei
    $\Lie^\ast_\alpha = i_\alpha \dif_{A^\ast} + \dif_{A^\ast}
    i_\alpha $ gilt, und $\dif_{A^\ast}$ wie im Lie-Algebroid-Fall durch
    $[\dcd]_\ast$ und $\rho_\ast$ definiert ist\/\footnote{Da für
      $[\dcd]_\ast$ die Jacobi-Identität nicht zu gelten braucht, ist
      im Allgemeinen $\dif_{A^\ast}^2 \neq 0$.}.
\end{Satz}

Mit dem bisher gezeigten können wir jetzt schließen, dass bereits
jedes Courant-Algebroid, in dem es eine Dirac-Struktur gibt, von
dieser Form ist.
\begin{Satz}
    Sei $E$ ein Courant-Algebroid, $L \subseteq E$ eine
    Dirac-Struktur. Sei weiter $L'$ ein isotropes Komplement zu $L$ in
    $E$. Identifizieren wir $L^\ast$ vermöge der Bilinearform mit
    $L'$, so ist $(L,L^\ast)$ ein Lie-quasi-Bialgebroid und das
    dadurch gemäß Satz \ref{liequasicourant} definierte
    Courant-Algebroid $L\oplus L^\ast$ ist isomorph zu $E$.
\end{Satz}
\begin{proof}
    Mit den Lemmata \ref{dualKlammer}, \ref{compatibility} und
    \ref{quasiLemma} ist klar, dass $(L,L^\ast)$ ein
    Lie-quasi-Bialge\-broid ist. Dass die durch $E$ auf $L\oplus
    L^\ast$ gegebene Courant-Algebroidstruktur mit der durch das
    Lie-quasi-Bialgebroid $(L,L^\ast)$ gegebenen Struktur
    übereinstimmt, ist mit Lemma \ref{courantLemma} eine einfache
    Rechnung.
\end{proof}

\section{Die Rothstein-Klammer}

In der Literatur wird zur Beschreibung von Courant-Algebroiden,
Lie-quasi-Bialgebroiden usw. oft das Konzept der
Supermannigfaltigkeiten verwendet, siehe z.B.
\cite{roytenberg:1999a,roytenberg:2002a,roytenberg:2002b,
  kosmann-schwarzbach:2003a}. Dabei wird die jeweilige
Algebroid-Struktur von $E$ in einer Funktion $\Theta \in C^\infty(T^\ast\Pi
E)$ kodiert, wobei die Gleichung
$$\{\Theta,\Theta\} = 0$$
die erforderlichen Eigenschaften für das
Algebroid garantiert. Dabei ist $\{\dcd\}$ die kanonische
Poisson-Klammer auf der Supermannigfaltigkeit $T^\ast\Pi E$. Die
eigentliche Klammer auf $E$ lässt sich dann als abgeleitete Klammer
aus $\Theta$ rekonstruieren, und auf ähnliche Weise erhält man auch
den Anker. Wir wollen darauf nicht weiter eingehen, sondern verweisen
auf die oben angegebene Literatur. Stattdessen verwenden wir eine dazu
äquivalente Beschreibung ohne die Zuhilfenahme von
Supermannigfaltigkeiten, allerdings zu dem Preis, dass wir einen
Zusammenhang auf der Dirac-Struktur $L$ wählen müssen. Deshalb
zunächst einige Bemerkungen über Vektorbündel und Zusammenhänge.
 
\subsection{Zurückgezogen Zusammenhänge}
Sei $\pi:F \ra M$ ein Vektorbündel, $N$ eine Mannigfaltigkeit und $f:N
\ra M$ eine glatte Abbildung. Mit $C^\infty_f(N,F)$ bezeichnen wir die
Abbildungen von $N$ nach $F$ entlang $f$, also Abbildungen $s:N\ra F$
mit $\pi\circ s = f$. Bezeichnet $f^\sharp F$ das mit $f$
zurückgezogene Bündel, dann ist $\sect(f^\sharp F)$ kanonisch isomorph
zu $C^\infty_f(N,F)$ \cite{michor:2004a}. Sind $a_\alpha \in \sect(F)$
lokale Basisschnitte von $F$, dann sind $f^\sharp a_\alpha \in
\sect(f^\sharp F)$ lokale Basisschnitte von $f^\sharp F$. Denn jede
Abbildung $s \in C^\infty_f(N,F)$ lässt sich ja lokal als $s =
s^\alpha f^\sharp a_\alpha$ mit $s^\alpha \in C^\infty(N)$ schreiben.  Ein
Zusammenhang auf $F$ definiert nun eine kovariante Ableitung
$$f^\sharp\nabla:\mathfrak{X}(N) \times C_f^\infty(N,F) \ra C_f^\infty(N,F),$$
insbesondere gilt für $N=M$ und $f = \op{id}$
$$\nabla:\mathfrak{X}(M) \times \sect(F) \ra \sect(F).$$
Weiter zeigt
sich für Vektorfelder $X\in \mathfrak{X}(N)$ mit $Tf(X) = 0$, dass die
kovariante Ableitung von $s$ nach $X$ unabhängig von dem Zusammenhang
ist. Wir schreiben dann für $\nabla_X s$ einfach $X(s)$.

Die Christoffelsymbole des Zusammenhangs $\nabla$ sind lokal
definierte Funktionen $\Gamma_{i\alpha}^\beta$, die in Koordinaten
$x^i$ von $M$ durch
$$\nabla_{\partdif{x^i}} a_\alpha = \Gamma_{i \alpha}^\beta a_\beta$$
bestimmt sind.  Für $f^\sharp \nabla$ erhält man für Koordinaten $y^j$
von $N$ mit einer kleinen Rechnung die Christoffelsymbole
$$\tilde{\Gamma}_{j\alpha}^\beta = \frac{\partial f^i}{\partial y^j}
f^\ast\Gamma_{i\alpha}^\beta.$$

Der Krümmungstensor des Zusammenhangs $\nabla$ ist für Vektorfelder
$X,Y \in \mathfrak{X}(M)$ und einen Schnitt $s \in \sect(E)$ definiert
als
$$R(X,Y)s = \nabla_X \nabla_Y s - \nabla_Y \nabla_X s - \nabla_{[X,Y]}
s ,$$
für $f^\sharp \nabla$ ergibt sich dann
$$(f^\sharp R)(V,W)\,f^\sharp s =
f^\sharp\(R\big(Tf(V),Tf(W)\big)s\)$$
mit $V,W \in
\mathfrak{X}(N)$.

Im Folgenden betrachten wir den Fall, dass $N = T^\ast M$ und $f=\tau$
die Kotangentialprojektion ist. Auf $T^\ast M$ wollen wir zur
Beschreibung der lokalen Ausdrücke kanonische Koordinaten $(q^i,p_j)$
verwenden, also durch $q^i$ induzierte Vektorbündelkoordinaten. Das
Vorzeichen der kanonischen Poisson-Klammer auf $T^\ast M$
sei so gewählt, dass für $(q^i,p_j)$ die Gleichungen 
$$\{q^i,q^j\} = \{p_i,p_j\} = 0 \qquad \text{und} \qquad \{q^i,p_j\} =
\delta_j^i$$
gelten.  Die lokalen Vektorfelder $\partdif{p_i}$ sind
dann vertikal, d.h. es gilt $T\tau(\partdif{p_i})=0$, und für einen
Schnitt $s \in \sect(f^\ast E)$ ist, wie oben beschrieben, der
Ausdruck
$$\frac{\partial s}{\partial p_i} = \nabla_{\partdif{p_i}} s$$
unabhängig von Zusammenhang.  Weiter gilt
$$\big(\tau^\sharp \nabla)_{\partdif{q^i}} \tau^\sharp a_\alpha =
\tau^\ast(\Gamma_{i \alpha}^\beta)\tau^\sharp a_\beta \qquad \text{sowie} \qquad
\big(\tau^\sharp \nabla)_\partdif{p_j} \tau^\sharp a_\alpha = 0.$$
Ist $E$
ein Lie-Algebroid mit Lie-Klammer $[\dcd]$ und Anker $\rho$, dann
definieren wir die Torsion von $\nabla$ durch
$$T(s_1,s_2) = \nabla_{\rho(s_1)}s_2 - \nabla_{\rho(s_2)}s_1 -
[s_1,s_2],$$
sowie die Torsion von $\tau^\sharp \nabla$ als
$$(\tau^\sharp T)(\tau^\sharp s_1,\tau^\sharp s_2) =
\tau^\sharp(T(s_1,s_2)).$$

\subsection{Die Rothstein-Klammer}

Sei $F\ra N$ ein Vektorbündel über einer symplektischen
Mannigfaltigkeit $(N,\omega)$ und sei $\Lambda$ der zugehörige
Poisson-Tensor. Weiter sei auf $F$ eine Pseudo-Metrik $g$ gegeben,
sowie ein metrischer Zusammenhang $\nabla$ mit Krümmung $R$. Seien
$x^i$ lokale Koordinaten auf $U \subset N$ und seien $s_A$ lokale
Basisschnitte von $F$ auf $U$ und $s^A$ die dualen Basisschnitte von
$F^ \ast$. Die Krümmung ist dann lokal durch
$$
R(\partial_i,\partial_j)s_A = R^B_{Aij} s_B$$
gegeben. Wir
definieren nun einen Schnitt $\hat{R} \in \sect(TN \otimes \Wedge^2
F^\ast \otimes T^\ast N)$ durch
$$\hat{R} = \frac{1}{2} \partial_i \otimes \Lambda^{ij} g_{AB}
R^A_{Cjk} s^B\wedge s^C \otimes \dif x^k.$$
Elemente $(X\otimes \psi
\otimes\alpha) \in \sect(TN \otimes \Wedge^2 F^\ast \otimes T^\ast N)$
können wir als Abbildungen $\sect(TN \otimes\Wedge^\bullet F^\ast) \ra
\sect(TN\otimes \Wedge^{\bullet+2} F^\ast)$ auffassen, wobei
$$(X\otimes \phi \otimes \alpha)(Y\otimes \psi):= \alpha(Y) X\otimes
\phi\wedge \psi.$$
Insbesondere gilt dann
$$\hat{R}(X\otimes \psi) = \frac{1}{2} \partial_i \Lambda^{ij} g_{AB}
R^A_C(\partial_j,X) s^B\wedge s^C\wedge \psi.$$
Durch Verknüpfung der
Abbildungen können wir jetzt $\hat{R}^2$, $\hat{R}^3$ usw. bilden. Da
$\hat{R}$ den Grad des $\Wedge^\bullet F^\ast$-Anteils immer um zwei
erhöht, ist $\hat{R}$ nilpotent und durch 
$$(\op{id}-\hat{R})^{-\frac{1}{2}} = \op{id} + \frac{1}{2}\hat{R} +
\frac{3}{8} \hat{R}^2 +\ldots$$
ist ein Schnitt in $\sect(TM \otimes \Wedge^\bullet F^\ast \otimes
T^\ast M)$ wohldefiniert.
Für $S\in \sect(TN \otimes \Wedge^\bullet F^\ast \otimes T^\ast N)$
sei $S^i_j$ der durch die Gleichung
$$S = \partial_i \otimes S^i_j \otimes\dif x^j$$
lokal definierte Schnitt in $\sect(\Wedge^\bullet F^\ast)$.
Schließlich sollen
$i(s)$ bzw. $j(s)$  das Links-- bzw. Rechtseinsetzen von $s\in
\sect(F)$ in Elemente aus $\sect(\Wedge^\bullet F^\ast)$ bezeichnen.
Damit können wir den folgenden Satz formulieren, für einen Beweis siehe
z.B. \cite{bordemann:2000a}.
\begin{Satz}[Rothstein-Klammer]
    Auf den Schnitten $\sect(\Wedge^\bullet F^\ast)$ ist durch die
    Roth\-stein-Poisson-Klammer, gegeben durch
    $$\roth{\phi,\psi} =
    \Lambda^{ij}\big(1-\hat{R})^{\frac{1}{2}}\big)^k_i\wedge
    \big(1-\hat{R})^{\frac{1}{2}}\big)^l_j \wedge \nabla_{\partial_k}
    \phi \wedge \nabla_{\partial_l} \psi + g^{AB} j(s_A)\phi \wedge
    i(s_B)\psi,$$
    eine Super-Poisson-Klammer erklärt, das heißt, für $\phi \in
    \sect(\Wedge^k F^\ast)$, $\psi \in  \sect( \Wedge^l F^\ast)$ und
    $\eta \in  \sect( \Wedge^\bullet F^\ast) $ gilt
    \begin{enumerate}
    \item $\roth{\phi,\psi} = -(-1)^{kl}\roth{\psi,\phi}$
    \item $\roth{\phi,\psi\wedge \eta} = \roth{\phi,\psi}\wedge \eta +
        (-1)^{kl} \psi\wedge \roth{\phi,\eta}$
    \item $\roth{\phi,\roth{\psi,\eta}} = \roth{\roth{\phi,\psi},\eta}
        + (-1)^{kl} \roth{\psi,\roth{\phi,\eta}}.$
    \end{enumerate}
\end{Satz}

Wir wollen jetzt die Rothstein-Klammer für eine spezielle Situation
bestimmen. Sei $L \ra M$ ein Vektorbündel mit $k$-dimensionaler Faser
über der $n$-dimensionalen Mannigfaltigkeit $M$. Sei auf $L$ ein
Zusammenhang $\nabla$ gegeben mit Krümmung $R$. Seien $q^i$ lokale
Koordinaten von $M$ und seien $a_\alpha$ lokale Basisschnitte von $L$
sowie $a^\alpha$ dazu duale lokale Basisschnitte von $L^\ast$. Auf
$L^\ast \oplus L$ ist dann durch
$$\nabla_X (\alpha,e) = (\nabla_X \alpha,\nabla_X e)$$
ein Zusammenhang
definiert, der bezüglich der kanonischen Bilinearform auf $L^\ast\oplus
L$ metrisch ist. Denn es gilt ja für $X \in \mathfrak{X}(M)$ und
$(\alpha_1,s_1),(\alpha_2,s_2) \in \sect(L^\ast\oplus L)$
\begin{eqnarray*}
    \nabla_X\scal{(\alpha_1,s_1),(\alpha_2,s_2)} &=& \nabla_X
    \(\alpha(s_2)\) +   \nabla_X\(\alpha_2(s_1)\) \\
    &=& \scal{\nabla_X\alpha_1,s_2} + \scal{\alpha_1,\nabla_X s_2} +
    \scal{\nabla_X\alpha_2,s_1} + \scal{\alpha_2,\nabla_X s_1} \\
    &=& \scal{\nabla_X(\alpha_1,s_1),(\alpha_2,s_2)} +
    \scal{(\alpha_1,s_1),\nabla_X(\alpha_2,s_2)}.
\end{eqnarray*}
Sei $\bar{R}$ die Krümmung dieses Zusammenhangs.  Wählen wir
$$\op{f}_1,\ldots,\op{f}_A,\ldots,\op{f}_{2k} = a^1,\ldots
a^k,a_1,\ldots a_k$$
als lokale Trivialisierung von $L^\ast\oplus L$,
dann folgt
$$
\bar{R}^B_{Aij} = \left\{
    \begin{array}{c@{\qquad}l}
        -R^A_{Bij} &  \text{für}\; 1\leq A,B \leq k \\[2mm]
         R^{B-k}_{A-k,ij} &\text{für}\; k+1\leq A,B \leq 2 k \\[2mm]
        0 & \text{sonst}.\\
    \end{array} \right. $$

Wir setzen nun $F = \tau^\sharp(L^\ast\oplus L) = (\tau^\sharp(L
\oplus L^\ast))^\ast$ mit $\tau:T^\ast M
\ra M$. Wie oben beschrieben liefert der Zusammenhang auf $L$ einen
Zusammenhang auf $F$, und sei $\tilde{R}$ dessen Krümmung. Auf $T^\ast
M$ wollen wir kanonische Koordinaten $(q,p)$ verwenden. Der
Poisson-Tensor $\Lambda$ ist dann durch $\Lambda(\dif q^i,\dif p_j) =
\delta^i_j = -\Lambda(\dif p_j,\dif q^i)$ sowie $\Lambda(\dif
q^i,\dif q^j) = \Lambda(\dif p_i,\dif p_j) = 0$ bestimmt.
Beachten wir noch, dass der lokale Ausdruck für den
Krümmungstensor von $\tilde{R}$ bezüglich der $q$-Indizes durch
$\tilde{R}^B_{Aij} = \tau^\ast \bar{R}^B_{Aij}$ gegeben ist und für ein
$p$-Index verschwindet, so erhält man mit einer kleinen Rechnung die
Gleichungen 
\begin{eqnarray*}
    \hat{R}\Big(\partdif{q^i} \otimes \psi\Big) &= &\partdif{p_j}\otimes
    \tau^\ast R^\alpha_{\beta ij} \ta_\alpha \wedge \ta^\beta\wedge
    \psi\\ 
    \hat{R}\Big(\partdif{p_i} \otimes \psi\Big) &= &0,
\end{eqnarray*}
wobei $\psi \in \sect(\Wedge^\bullet\tau^\sharp(L \oplus L^\ast))$ und
$\hat{R}$ die zuvor beschriebene Abbildung zu der Krümmung $\tilde{R}$
ist. Damit folgt $\hat{R}^2=0$ und somit
$$(\op{id}-\hat{R})^{-\frac{1}{2}} = \op{id} + \frac{1}{2} \hat{R}.$$
Mit diesen Vorbereitungen kann man durch eine kleine Rechnung die
Rothstein-Klam\-mer in eine etwas konkretere Form bringen.
\begin{Lemma}[{\cite{eilks:2004a}}]\label{eilks}
Für die eben beschriebene Rothstein-Klammer auf
$\sect(\Wedge^\bullet\tau^\sharp(L\oplus L^\ast))$ gilt 
\begin{eqnarray*}
    \roth{\phi,\psi} &=& \nabla_{\partdif{q^i}}\phi \wedge
    \partdif{p_i}\psi  - \partdif{p_i} \phi
    \wedge \nabla_{\partdif{q^i}} \psi  + \tau^\ast 
    R^\alpha_{\beta ij} \ta_\alpha \wedge \ta^\beta\wedge
    \partdif{p_i}\phi\wedge\partdif{p_j}\psi \\
    && +\; j(\ta_\alpha)\phi \wedge i(\ta^\alpha)\psi +
    j(\ta^\alpha)\phi \wedge i(\ta_\alpha)\psi.
\end{eqnarray*}
Dabei
sind jetzt $\psi$ und $\phi$ Schnitte in $\Wedge^\bullet\tau^\sharp(L
\oplus L^\ast)$, sowie $\ta_\alpha = \tau^\sharp a_\alpha$ die
zurückgeholten Basisschnitte und entsprechend ist $\ta^\alpha =
\tau^\sharp a^\alpha$.
\end{Lemma}
\subsection{Super-Darbouxkoordinaten}

Berechnen wir die Rothstein-Klammern für die
Koordinatenfunktionen $q^i,p_j,\ta^\alpha,\ta_\beta$ von
$\tau^\sharp(L\oplus L^\ast)$, so erhalten wir die Gleichungen
$$\begin{array}{rclrclrcl} \roth{q^i,q^j} &=& 0 & \roth{q^i,p_j}& =&
    \delta^i_j & \roth{p_i,p_j}&
    =&\tau^\ast R^\alpha_{\beta ij}\, \ta_\alpha\wedge\ta^\beta\\[2mm]
    \roth{q^i,\ta_\alpha}& =& 0 & \roth{q^i,\ta^\alpha}& =& 0\\[2mm]
    \roth{p_i,\ta_\alpha}& =& -\tau^\ast\Gamma_{i\alpha}^\beta\,
    \ta_\beta \quad& \roth{p_i,\ta^\alpha}& =&
    \tau^\ast\Gamma_{i\beta}^{\alpha}\,\ta^\alpha \\[2mm]
    \roth{\ta_\alpha,\ta_\beta}& =& 0 &
    \roth{\ta^\alpha,\ta^\beta}&=&0 & \roth{\ta_\alpha,\ta^\beta}& =&
    \delta_\alpha^\beta,
\end{array}
$$
und mit Hilfe der Leibniz-Regel ist die Rothstein-Klammer damit für
alle Schnitte
$$\Psi \in \sect(\Wedge^\bullet \tau^\sharp(L\oplus L^\ast))$$
bekannt.
\begin{prop}
Definieren wir
$$r_i = p_i - \tau^\ast\Gamma_{i\alpha}^\beta\, \ta^\alpha \wedge
\ta_\beta,$$
dann werden durch $q^i,r_j,\ta^\alpha,\ta_\beta$ alle in
den Impulsen linearen Schnitte in $\sect(\Wedge^\bullet
\tau^\sharp(L\oplus L^\ast))$ erzeugt, und es gelten die Gleichungen
$$\roth{q^i,r_j} = \delta^i_j \qquad \text{und} \qquad
\roth{\ta^\alpha,\ta_\beta} = \delta^\alpha_\beta,$$
sowie
\begin{eqnarray*}
    &&\lefteqn{\roth{q^i,q^j} = \roth{q^i,\ta^\alpha} =
      \roth{q^i,\ta_\beta} 
      =\roth{r_i,r_j}}\\[1mm] &=&  \roth{r_i,\ta^\alpha} =
    \roth{r_i,\ta_\beta} = 
    \roth{\ta^\alpha,\ta^\beta} = \roth{\ta_\alpha,\ta_\beta} = 0.
\end{eqnarray*}
\end{prop}
\begin{proof}
    Der Beweis erfolgt durch direktes Nachrechnen. Zunächst ist
    $$\roth{q^i,\tau^\ast \Gamma_{j
        \alpha}^\beta\,\ta^\alpha\wedge\ta_\beta} = 0$$
    und damit
    $$\roth{q^i,r_j}  = \delta^i_j.$$
    Weiter haben wir
    \begin{eqnarray*}
        \roth{p_i,-\tau^\ast\Gamma_{j\alpha}^\beta\, \ta^\alpha 
          \wedge \ta_\beta}
        &=&
        \nabla_{\partdif{q^i}}(\tau^\ast\Gamma_{j\alpha}^\beta\,
        \ta^\alpha 
        \wedge 
        \ta_\beta)\\
        &=&   \tau^\ast\Bigg(\frac{\partial\Gamma_{j\alpha}^\beta}
        {\partial q^i} -\Gamma_{i\alpha}^{\gamma}
        \Gamma_{j\gamma}^\beta + \Gamma_{j\alpha}^{\gamma}
        \Gamma_{i\gamma}^\beta\Bigg)\ta^\alpha\wedge\ta_\beta,
    \end{eqnarray*}
    und damit wegen der Superantisymmetrie 
    $$
    \roth{-\tau^\ast\Gamma_{i\alpha}^\beta\, \ta^\alpha \wedge
      \ta_\beta,p_j}
    =-\tau^\ast\Bigg(\frac{\partial\Gamma_{i\alpha}^\beta} {\partial
      q^j} -\Gamma_{j\alpha}^{\gamma} \Gamma_{i\gamma}^\beta +
    \Gamma_{i\alpha}^{\gamma}
    \Gamma_{j\gamma}^\beta\Bigg)\ta^\alpha\wedge\ta_\beta.
    $$
    Jetzt brauchen wir noch
    \begin{eqnarray*}
        \roth{\tau^\ast\Gamma_{i\alpha}^\beta\, \ta^\alpha 
          \wedge \ta_\beta,\tau^\ast\Gamma_{j\gamma}^\delta\, \ta^\gamma
          \wedge \ta_\delta}  &=& \tau^\ast\big(\Gamma_{i\alpha}^\beta
        \Gamma_{j\gamma}^\delta\big) (\delta^\gamma_\beta \ta^\alpha
        \wedge \ta_\delta - \delta^\alpha_\delta \ta^\gamma\wedge
        \ta_\beta)\\
        &=& \tau^\ast\big(\Gamma_{i\alpha}^{\gamma}
        \Gamma_{j\gamma}^\beta - \Gamma_{j\alpha}^{\gamma}
        \Gamma_{i\gamma}^\beta\big)\ta^\alpha\wedge\ta_\beta,
    \end{eqnarray*}
    und wir erhalten insgesamt 
    \begin{eqnarray*}
        \roth{r_i,r_j}
        &=&\roth{p_i,p_j} +
        \tau^\ast\Bigg(\frac{\partial\Gamma_{j\alpha}^\beta}  
        {\partial q^i} -\frac{\partial\Gamma_{i\alpha}^\beta} {\partial
          q^j} + \Gamma_{j\alpha}^{\gamma}
        \Gamma_{i\gamma}^\beta -\Gamma_{i\alpha}^{\gamma}
        \Gamma_{j\gamma}^\beta\Bigg)\ta^\alpha\wedge \ta_\beta \\
        &=&\tau^\ast R^\alpha_{\beta ij}\, \ta_\alpha\wedge\ta^\beta +
        \tau^\ast R^\beta_{\alpha ij}\ta^\alpha\wedge \ta_\beta\\
        &=&0.
    \end{eqnarray*}
    Schließlich gilt
    \begin{eqnarray*}
        \roth{r_i,\ta_\gamma} &=& \roth{p_i,\ta_\gamma} -
        \roth{\tau^\ast\Gamma_{i \alpha}^\beta \,
          \ta^\alpha\wedge\ta_\beta} \\ 
        &=& -\tau^\ast\Gamma_{i\gamma}^\beta\, \ta_\beta +
        \tau^\ast\Gamma_{i\gamma}^\beta\, \ta_\beta \\
        &=&0
    \end{eqnarray*}
    und genauso
    $$  \roth{r_i,\ta^\gamma} = 0.$$
\end{proof}
\begin{Bemerkung}
    Man beachte, dass die $r_i$'s im Gegensatz zu den $p_i$'s keine
    Koordinaten auf $T^\ast M$ mehr sind, da erstere im Allgemeinen
    Anteile in $\sect(\Wedge^2 \tau^\sharp(L \oplus L^\ast)$ habe.
\end{Bemerkung}

\section{Die Courant-Klammer als abgeleitete Klammer}

Mit den Vorbereitungen im letzten Abschnitt sind wir jetzt in der Lage, die
Courant-Klammer auf $L\oplus L^\ast$ mit Hilfe der Rothstein-Klammer
als abgeleitete Klammer schreiben zu können. Im Gegensatz zu der
Arbeit von Roytenberg \cite{roytenberg:1999a} verwenden wir jedoch
nicht die Theorie der Supermannigfaltigkeiten, sondern greifen ausschließlich
auf Mittel  aus der  "`konventionellen"' Differentialgeometrie zurück.

\subsection{Die BRST-Ladung}

Sei jetzt wieder $E=L\oplus L^\ast$ ein Courant-Algebroid mit
Dirac-Struktur $L$.  Koordinaten von $M$ bezeichnen wir mit
$q^1,\ldots,q^n$ und lokale Basisschnitte von $L$ mit $a_1,\ldots,a_k$
sowie lokale Basisschnitte von $L^\ast$ mit $a^1,\ldots,a^k$. Weiter
definieren wir die lokalen Funktionen
$$c_{\alpha\beta}^\gamma = \scal{[a_\alpha, a_\beta]_\C,a^\gamma}$$
und
$$\bar{c}^{\alpha\beta}_\gamma =
\scal{[a^\alpha,a^\beta]_\C,a_\gamma}.$$
Sei $\tau:T^\ast M \ra M$
wieder die Kotangentialprojektion.  Um die Rothstein-Klammer auf der
Algebra $\sect(\Wedge^\bullet \tau^\sharp(L\oplus L^\ast))$ bilden zu
können, müssen wir noch einen Zusammenhang $\nabla$ auf $L$ wählen.
Dies stellt jedoch kein Problem dar, da die für uns wichtigen Aussagen
später unabhängig von dieser Wahl sein werden. Im Folgenden werden
wir, um Schreibarbeit zu sparen, die auf $M$ lokal definierten
Funktionen wie $\Gamma_{i\alpha}^\beta$, $c_{\alpha\beta}^\gamma$ usw.
und die entsprechenden auf $T^\ast M$ zurückgezogene Funktion mit dem
gleichen Symbol bezeichnen.

Sei $T$ die Torsion auf $L$, also
$$T_{\alpha\beta}^\gamma = \rho(a_\alpha)^i\Gamma_{i \beta}^\gamma-
\rho(a_\beta)^i \Gamma_{i \alpha}^\gamma - c_{\alpha\beta}^\gamma$$
und
$\bar{T}$ die Torsion auf $L^\ast$,
$$\bar{T}^{\alpha\beta}_\gamma = \rho(a^\beta)^i\Gamma^{
  \alpha}_{i\gamma}- \rho(a^\alpha)^i \Gamma_{i \gamma}^\beta -
\bar{c}^{\alpha\beta}_\gamma$$
Weiter bezeichnen wir mit
$\mathcal{J}:\mathfrak{X}(M) \ra C^\infty(M)$ die Abbildung, die einem
Vektorfeld die dadurch gegebene lineare Funktion auf $T^\ast M$
zuordnet, also $\mathcal{J}(X)(\alpha) = \scal{X|_m,\alpha} =
\alpha(X_m)$ für $\alpha \in T^\ast M|_m$.
\begin{Satz}\label{derived}
    Durch die Definitionen
    \begin{eqnarray*}
        \mu &=& -\mathcal{J}(\rho(a_\alpha))\ta^\alpha + \frac{1}{2}
        T_{\alpha\beta}^\gamma \ta^\alpha\wedge\ta^\beta\wedge\ta_\gamma\\
        &=& -r_i\, \rho(a_\alpha)^i \ta^\alpha - \frac{1}{2}
        c_{\alpha\beta}^\gamma \ta^\alpha\wedge\ta^\beta\wedge\ta_\gamma
    \end{eqnarray*}
    und
    \begin{eqnarray*}
        \gamma &=& -\mathcal{J}(\rho(a^\alpha))\ta_\alpha + \frac{1}{2}
        \bar{T}^{\alpha\beta}_\gamma
        \ta_\alpha\wedge\ta_\beta\wedge\ta^\gamma\\
        &=& -r_i\, \rho(a^\alpha)^i \ta_\alpha - \frac{1}{2}
        \bar{c}^{\alpha\beta}_\gamma\ta_\alpha\wedge
        \ta_\beta\wedge\ta^\gamma\\
        \end{eqnarray*}
    sind zwei globale Schnitte  in $\Wedge^\bullet \tau^\sharp(L\oplus L^\ast)$
    gegeben. Weiter sei 
    $$\psi(\alpha_1,\alpha_2,\alpha_3) =
    -\scal{[\alpha_1,\alpha_2]_\C,\alpha_3},$$
    wobei wir $\psi$ als Schnitt in $\Wedge^3\tau^\sharp( L\oplus
    L^\ast)$ auffassen. Lokal ist $\psi$ also von der Form
    $$\psi = \psi^{\alpha\beta\gamma} \ta_\alpha\wedge \ta_\beta
    \wedge \ta_\gamma.$$
    \begin{enumerateR}
    \item Für alle $s_1,s_2 \in \sect(\Wedge^\bullet L)$ gilt
        $$\roth{\roth{\tilde{s}_1,\mu},\tilde{s}_2} =
        \tau^\sharp[s_1,s_2],$$
        und für alle $\alpha \in
        \sect(\Wedge^\bullet L^\ast)$ ist
        $$\roth{\mu,\tilde{\alpha}} = \tau^\sharp \dif_L \alpha .$$
    \item
        Für alle $\alpha,\beta \in \sect(\Wedge^\bullet L^\ast)$ gilt
        $$\roth{\roth{\tilde{\alpha},\gamma},\tilde{\beta}} =
        \tau^\sharp[\alpha,\beta]_\ast,$$ 
        und für alle $s_1 \in \sect(\Wedge^\bullet L)$ ist
        $$\roth{\gamma,\tilde{s}_1} = \tau^\sharp \dif_{L^{\!\ast}}\, s_1.$$
    \item
        Sei $\Theta = \mu +\gamma+\psi$. Dann gilt für alle $e_1,e_2
        \in \sect(E)$
        $$\roth{\roth{\te_1,\Theta},\te_2} = \tau^\sharp[e_1,e_2]_\C$$
        sowie für $f\in C^\infty(M)$
        $$\roth{\Theta,\tau^\ast f} = \tau^\sharp \D f,$$
        beziehungsweise
        $$\roth{\roth{e_1,\Theta},\tau^\ast f}=\tau^\ast(\rho(e_1)f).$$
    \end{enumerateR}
\end{Satz}

\begin{proof}
    Zunächst rechnet man nach, dass die beiden für $\mu$ angegebenen
    Ausdrücke tatsächlich übereinstimmen. Mit dem
    Transformationsverhalten für die  Torsion $T_{\alpha\beta}^\gamma$
    folgt weiter, dass $\mu$ als globaler Schnitt wohldefiniert
    ist. Für $\gamma$ gelten natürlich die analogen Aussagen. 
    \begin{enumerateR}
    \item Wenn wir in unseren Super-Darbouxkoordinaten rechnen,
        erhalten wir leicht folgende Gleichungen, wobei $f,g \in
        C^\infty(M)$
        \begin{eqnarray*}
        \roth{\tau^\ast f,\tau^\ast g} &=&0,\\
        \roth{\roth{\ta_\nu,\mu},\ta_\kappa} &=& c_{\nu\kappa}^\gamma
        \ta_\gamma = \tau^\sharp[a_\nu,a_\kappa],\\
        \roth{\roth{\ta_\nu,\mu},\tau^\ast f}&=&
        \tau^\ast(\rho(\ta_\nu) f) = \tau^\ast[a_\nu,f].
        \end{eqnarray*}
        Damit sieht
        man, dass für $s_1 \in\sect(\Wedge^k L),s_2 \in
        \sect(\Wedge^l L)$ durch
        $$[\tilde{s}_1,\tilde{s}_2]_\mu =
        \roth{\roth{\tilde{s}_1,\mu},\tilde{s}_2} $$
        eine Abbildung
        $$[\dcd]_\mu :\tau^\sharp(\sect(\Wedge^\bullet L)) \times
        \tau^\sharp(\sect(\Wedge^\bullet L)) \ra
        \tau^\sharp(\sect\Wedge^\bullet L))  $$ 
        definiert ist. Weiter gilt
        $\roth{\tilde{s}_1,\tilde{s}_2} = 0$, und mit der
        Jacobi-Identität für die Rothstein-Klammer folgt
        $$\roth{\roth{\tilde{s}_1,\mu},\tilde{s}_2} =
        -(-1)^{(k-1)(l-1)}\roth{\roth{\tilde{s}_2,\mu},\tilde{s}_1},$$
        also die Superantisymmetrie für $[\dcd]_\mu$.
        Mit Hilfe der Leibniz-Regel folgt jetzt 
        $$[\tilde{s}_1,\tilde{s}_2]_\mu = \tau^\sharp[s_1,s_2]$$
        für alle $s_1,s_2 \in\sect(\Wedge^\bullet L)$ und damit der
        erste Teil von \textit{i.}).

        Weiter rechnen wir leicht nach, dass
        $$\roth{\mu,\tau^\ast f} = \rho(a_\alpha)f\;\ta^\alpha =
        \tau^\sharp(\dif_L f)$$
        sowie 
        $$\roth{\mu,\ta^\alpha} = -\frac{1}{2}c_{\beta\gamma}^\alpha
        \ta^\beta\wedge\ta^\gamma  = \tau^\sharp \dif_L a^\alpha$$
        womit \textit{i.}) gezeigt ist.

    \item Analog zu \textit{i.})
        
    \item Seien $s_1,s_2\in\sect(\Wedge^k L)$ und $\eta_1,\eta_2 \in
        \sect(\Wedge^l L^\ast)$. Dann gilt
        $$\begin{array}{ccc@{\qquad\quad}ccc}
            \roth{\ta_\nu,\tilde{\eta}_1} &=&
            \tau^\sharp(i(a_\nu)\eta_1) &
            \roth{\tilde{\eta}_1,\ta_\nu} &=&
            \tau^\sharp(j(a_\nu)\eta_1)
            = -(-1)^l \tau^\sharp(i(a_\nu)\eta_1) \\[2mm]
            \roth{\ta^\nu,\tilde{s}_1} &=& \tau^\sharp(i(a^\nu)s_1) &
            \roth{\tilde{s}_1,\ta^\nu} &=& \tau^\sharp(j(a^\nu)s_1) =
            -(-1)^k \tau^\sharp(i(a^\nu)s_1)
        \end{array}$$
        
       Damit folgt insbesondere
       $$\roth{\roth{\teta_1,\psi},\teta_2} =
       \tau^\sharp[\eta_1,\eta_2]_L.$$
       Mit dieser Vorbereitung kann die Behauptung jetzt leicht
       nachgerechnet werden, z.B. gilt
       \begin{eqnarray*}
           \roth{\roth{\tilde{s}_1,\mu},\teta_2} &=&
           \roth{\mu,\roth{\tilde{s}_1,\teta_2}} +
           \roth{\tilde{s}_1,\roth{\mu,\teta_2}}\\
           &=& \tau^\sharp(\dL i_{s_1} \eta_2 + i_{s_1} \dL
           \eta_2) = \tau^\sharp\Lie_{s_1} \eta_2
       \end{eqnarray*}
       sowie
       $$
       \roth{\roth{\eta_1,\mu},\tilde{s}_2} = \roth{\tau^\sharp \dL
         \teta_1,\tilde{s}_2} = -\tau^\sharp(i_{s_2} \dL \eta_1).$$
       Rechnet man die restlichen Terme in
       $$\roth{\roth{\tilde{s}_1+\teta_1,\Theta},\tilde{s}_2+\teta_2}$$
       auf ähnliche Weise aus, erhält man insgesamt die in Satz
       \ref{liequasicourant} angegebene Gleichung für die Courant-Klammer.
       Außerdem gilt
       $$\roth{\Theta,\tau^\ast f}= \tau^\sharp(\dL f + \dif_{L^\ast}
       f) = \tau^\sharp\D f,$$
       bzw. mit $e \in \sect(L\oplus L^\ast)$
       $$\roth{\roth{\Theta,\te},f} = \roth{\Theta,\roth{\te,f}} +
       \roth{\te,\roth{\Theta,f}} = \tau^\ast(\rho(e)f).$$
   \end{enumerateR}
\end{proof}
\begin{Bemerkungen}
    \begin{enumerateR}
    \item In Analogie zur BRST-Quantisierung nennen wir $\Theta$ auch
        BRST-Ladung, vgl.  z.B.
        \cite{bordemann.herbig.waldmann:2000a,bordemann:2000a,eilks:2004a}.
    \item Weiter beachte man die formale Übereinstimmung unserer
        Ausdrücke für $\mu$ und $\gamma$, notiert mit Hilfe der
        Super-Darbouxkoordinaten, mit den in \cite{roytenberg:1999a}
        angegebenen Ausdrücken.
    \end{enumerateR}
\end{Bemerkungen}

Da die Abbildungen $\tau^\ast$ bzw. $\tau^\sharp$ injektiv sind,
können wir Elemente in $\sect(\Wedge^\bullet(L\oplus L^\ast))$ mit
ihren Bildern in $\sect(\Wedge^\bullet \tau^\sharp(L \oplus L^\ast))$
identifizieren.  Wir werden deshalb beispielsweise für $\tilde{s} =
\tau^\sharp s$ im folgenden einfach wieder $s$ schreiben. Da $\mu$ und
$\gamma$ die einzigen Elemente in $\sect(\Wedge^\bullet \tau^\sharp(L
\oplus L^\ast))$ sind, die nicht im Bild von $\tau^\sharp$ liegen und
wir im weiteren betrachten werden, sollte dadurch keine Verwirrung
entstehen.

\begin{prop}\label{Theta}
Für das oben definierte $\Theta$  gilt $\roth{\Theta,\Theta} = 0$,
oder äquivalent dazu
\begin{eqnarray*}
\roth{\mu,\mu} &=& 0\\
\frac{1}{2} \roth{\gamma,\gamma} + \roth{\mu,\psi}& =& 0\\
\roth{\mu,\gamma} &=&0\\
\roth{\gamma,\psi} &=& 0.
\end{eqnarray*}
\end{prop}

\begin{proof}
    Seien $e_1,e_2,e_3 \in \sect(E)$ und $f,g \in C^\infty(M)$.
    Mit der Jacobi-Identität für die Rothstein-Klammer rechnen wir
    zunächst, dass
    \begin{eqnarray*}
        \tau^\sharp([[e_1,e_2]_\C,e_3]_\C) &=&
        \roth{\roth{\roth{\roth{e_1,\Theta},e_2},\Theta},e_3}\\
        &=&  \roth{\roth{\roth{e_1,\Theta},\roth{e_2,\Theta}},e_3} -
        \roth{\roth{\roth{\roth{e_1,\Theta},\Theta},e_2},e_3} \\
        &=& \roth{\roth{\roth{e_1,\Theta},e_3},\roth{e_2,\Theta}} +
        \roth{\roth{e_1,\Theta},\roth{\roth{e_2,\Theta},e_3}}\\
        &&\qquad\quad +
        \frac{1}{2}\roth{\roth{\roth{\roth{\Theta,\Theta},e_1},e_2},e_3}\\
        &=& \tau^\sharp([[e_1,e_3]_\C,e_2]_\C + [e_1,[e_2,e_3]_\C]_\C)
        +\frac{1}{2}\roth{\roth{\roth{\roth{\Theta,\Theta},e_1},e_2},e_3}.
    \end{eqnarray*}
    Aufgrund der Jacobi-Identität für die Courant-Klammer gilt damit
    die Gleichung
    $$\roth{\roth{\roth{\roth{\Theta,\Theta},e_1},e_2},e_3} = 0.$$
    Die
    Aussage $\rho([e_1,e_2]_\C) = [\rho(e_1),\rho(e_2)]$ übersetzt sich
    auf ähnliche Weise zu
    $$\roth{\roth{\roth{\roth{\Theta,\Theta},e_1},e_2},f} = 0,$$
    und
    schließlich liefert $\scal{\D f,\D g} = 0$ mit
    \begin{eqnarray*}
        \tau^\sharp\scal{\D f,\D g} &=&
        \roth{\roth{\Theta,f},\roth{\Theta,g}}\\
        &=& \roth{\roth{\roth{\Theta,f},\Theta},g} -
        \roth{\Theta,\roth{\roth{\Theta,f},g}} \\
        &=& \frac{1}{2}\roth{\roth{\roth{\Theta,\Theta},f},g}
    \end{eqnarray*}
    die Gleichung
    $$\roth{\roth{\roth{\Theta,\Theta},f},g} = 0.$$
    Wenn wir uns jetzt
    aber die Definition von $\mu$, $\gamma$ und $\psi$ genau
    anschauen, dann erkennt man, dass $\roth{\Theta,\Theta}$ nur Terme
    der folgenden Form enthält:
    \begin{enumerate}
    \item Zurückgezogenen Schnitte in
        $\Wedge^4\tau^\sharp(E\oplus E^\ast)$.
    \item Produkte von einem $r_i$ mit einem zurückgezogenen
        Schnitt in   $\Wedge^2\tau^\sharp(E\oplus E^\ast)$.
    \item Produkte von $r_i r_j$ mit einer zurückgezogenen Funktion.
    \end{enumerate}
    Durch die Gültigkeit der obigen Gleichungen für beliebige Schnitte
    in $E$ und Funktionen auf $M$ kann man folgern, dass keine der
    drei genannten Typen in $\roth{\Theta,\Theta}$ vorkommen kann und
    somit $\roth{\Theta,\Theta} = 0$ sein muss. Die Gleichungen für
    $\mu$, $\gamma$ und $\psi$ ergeben sich durch Sortieren nach dem
    Grad, wobei wir z.B. den Geistgrad $\op{gh}$ verwenden können, der
    für ein Element in $\phi \in \sect(\tau^\ast(\Wedge^k E \otimes
    \Wedge^l)) \subset \sect(\Wedge^{k+l} \tau^\sharp(E\oplus
    E^\ast))$ durch $\op{gh} \phi = (l-k) \phi$ gegeben ist. Da
    $\op{gh}$ eine Derivation der Rothstein-Klammer ist
    \cite{eilks:2004a}, kann damit die Behauptung gefolgert werden.
\end{proof}

\begin{Bemerkung}
    Die Gleichung $\roth{\mu,\mu}=0$ ist die Jacobi-Identität für die
    Klammer auf $L$, wohingegen $\frac{1}{2}\roth{\gamma,\gamma} +
    \roth{\mu,\psi} = 0$ die zweite Gleichung in Lemma
    \ref{quasiLemma} liefert. $\roth{\mu,\gamma}=0$ bedeutet die
    Verträglichkeit von $\dL$ mit der Klammer $[\dcd]_\ast$, und
    $\roth{\gamma,\psi} = 0$ ist die dritte Gleichung von Lemma
    \ref{quasiLemma}. Wir erhalten jetzt also auf einfache Weise die
    Ergebnisse wieder, die wir zuvor mühsam nachgerechnet hatten.
\end{Bemerkung}

\subsection{Proto-Bialgebroide}

Wir haben gesehen, wie wir einer Courant-Algebroid-Struktur auf
$L\oplus L^\ast$ ein Element $\Theta \in \sect(\Wedge^\bullet
\tau^\sharp(L\oplus L^\ast))$ mit $\roth{\Theta,\Theta}=0$ zuordnen
können, so dass sich die Courant-Klammer mit Hilfe von $\Theta$ und
der Rothstein-Klammer als abgeleitete Klammer schreiben lässt.
Umgekehrt kann man auch durch die Vorgabe eines geeigneten $\Theta \in
\sect(\Wedge^\bullet \tau^\sharp(L\oplus L^\ast))$ mit
$\roth{\Theta,\Theta} = 0$ das Bündel $L\oplus L^\ast$ zu einem
Courant-Algebroid machen. Dabei muss $\Theta$ so gewählt werden, dass
die Ausdrücke $$\roth{\roth{\te_1,\Theta},\te_2}$$
sowie 
$$\roth{\Theta,\tau^\ast f}$$
für zwei zurückgezogene Schnitte $\te_1$
und $\te_2$ sowie eine Funktion $f\in C^\infty(M)$ wieder einen
zurückgezogenen Schnitt von $L\oplus L^\ast$ definieren, also in
$\tau^\sharp(\sect(L\oplus L^\ast))$ liegen. Damit dürfen die beiden
oben genannten Ausdrücke nicht
von den Impulsen $p$ abhängen, und mit einem Blick auf die in Lemma
\ref{eilks} angegebene Form der Rothstein-Klammer folgt, dass $\Theta$
höchstens quadratisch in den Impulsen $p$ sein darf. Um die erste
Bedingung weiter zu untersuchen, bezeichnen wir mit $\Pol$ die in den
Impulsen $p$ polynomialen Elemente aus
$\sect(\Wedge^\bullet\tau^\sharp(L\oplus L^\ast))$ und definieren
durch
$$\epsilon =  p_i\frac{\partial}{\partial p_i} + \ta_\alpha\wedge
i(\ta^\alpha)$$
und
$$\lambda = p_i\frac{\partial}{\partial p_i} + \ta^\alpha\wedge
i(\ta_\alpha)$$
eine Bigradierung auf $\Pol$. Ein Element $\Psi \in \Pol$ hat den Bigrad
$(k,\ell)$, wenn die Gleichungen $\epsilon \Psi = k \Psi$ und $\lambda
\Psi = \ell\Psi$ erfüllt sind. Wir schreiben dann $\Psi \in \Pol^{k,\ell}$.
Weiter sei der Totalgrad   durch die
Summe
$$\chi = \epsilon +\lambda$$
gegeben.  

Betrachtet man wieder die Formel für die Rothstein-Klammer, so erkennt
man, dass für $\Phi, \Psi \in \sect(\Wedge^\bullet \tau^\sharp(L
\oplus L^\ast))$ mit $\chi(\Phi) = k \Phi$ und $\chi(\Psi) = \ell
\Psi$ die Gleichung
$$\chi(\roth{\Phi,\Psi} = (k+\ell -2)\roth{\Phi,\Psi}$$
folgt, d.h. die Rothstein-Klammer ist vom Grad $-2$ für die $\chi$-Gradierung.
Es gilt nun der folgende
\begin{Satz}
    Sei $\Theta \in \sect(\Wedge^\bullet \tau^\sharp(L\oplus
    L^\ast))$ vom $\chi$-Grad drei, d.h. es gilt $\chi(\Theta) = 3
    \Theta$. Dann ist durch
    $$\tau^\sharp[e_1,e_2]_\Theta = \roth{\roth{\te_1,\Theta},\te_2}$$ eine
    Klammer auf $\sect(L \oplus L^\ast)$ und durch
    $$\tau^\sharp( \rho(e_1)f) = \roth{\roth{\te_1,\Theta},\tau^\ast f}$$
    ein
    Vektorbündelhomomorphismus $\rho: L\oplus L^\ast \ra TM$
    definiert. Ist außerdem $\roth{\Theta,\Theta} = 0$, so wird
    $L\oplus L^\ast$ durch diese Konstruktion zusammen mit der
    kanonischen symmetrischen  Bilinearform zu einem
    Courant-Algebroid.
\end{Satz} 
Die zweite Aussage folgt dabei mit Rechnungen wie beim Beweis zu Lemma
\ref{Theta}. Man vergleiche auch die entsprechenden Rechnungen mit
Hilfe von Supermannigfaltigkeiten in
\cite{roytenberg:1999a,kosmann-schwarzbach:2003a} sowie
allgemeine  Betrachtungen zu abgeleiteten Klammern in
\cite{kosmann-schwarzbach:2003b}.

Ein Element $\Theta$ mit Totalgrad drei ist von der Form 
$$\Theta = \phi + \mu + \gamma + \psi$$
mit $\phi \in \Pol^{0,3} \cong
\sect(\Wedge^3 L^\ast)$, $\mu = \Pol^{1,2}$, $\gamma \in \Pol^{2,1}$
und $\psi \in \Pol^{3,0} \cong \sect(\Wedge^3 L^\ast)$.  Durch
Sortieren nach den Graden erhält man das folgende, zu
$\roth{\Theta,\Theta} = 0$ äquivalente, System von Gleichungen:
\begin{eqnarray*}
\tfrac{1}{2} \roth{\mu,\mu} + \roth{\gamma,\phi} &=& 0\\
\tfrac{1}{2} \roth{\gamma,\gamma} + \roth{\mu,\psi}& =& 0\\
\roth{\mu,\gamma} + \roth{\phi,\psi}&=&0\\
\roth{\mu,\phi} &=& 0\\
\roth{\gamma,\psi} &=& 0.
\end{eqnarray*}
Wie in Satz \ref{derived} können wir durch $\mu$ und $\gamma$
abgeleitete Klammern auf $L$ bzw. $L^\ast$ sowie die entsprechenden
Anker definieren. Die Jacobi-Identität auf $L$ ist dann äquivalent zu
$\phi = 0$. In diesem Fall wird $(L,L^\ast)$ zu einem
Lie-quasi-Bialgebroid, wie in dem von uns bisher betrachteten Fall.
Analog folgt die Jacobi-Identität auf $L^\ast$ genau dann, wenn $\psi
= 0$ gilt, und $(L,L^\ast)$ ist dann ein Quasi-Lie-Bialgebroid. Im
allgemeinen Fall, wenn also weder auf $L$ noch auf $L^\ast$ die
Jacobi-Identität zu gelten braucht, nennt man $(L,L^\ast)$
Proto-Bialgebroid, vgl. \cite{kosmann-schwarzbach:2003a}.

\begin{Bemerkung}
    Die glatte Deformation einer Dirac-Struktur $L$ kann nach Lemma
    \ref{smoothDiracDef} als triviale Deformation der
    Courant-Algebroidstruktur aufgefasst werden, so dass $L$ auch für
    das deformierte Algebroid eine Dirac-Struktur bleibt. Dies
    übersetzt sich zu einer trivialen Deformation $\Theta_t$ von
    $\Theta = \mu +\gamma + \psi$, so dass der $\Pol^{0,3}$-Anteil von
    $\Theta_t$ für alle $t$ Null ist. Das kann man auch auffassen als
    die Deformation des Lie-quasi-Algebroids $(L,L^\ast)$ derart, dass
    die auf $L\oplus L^\ast$ definierten Courant-Algebroidstrukturen
    isomorph sind. Dies ist der Ausgangspunkt der Betrachtungen in
    \cite{roytenberg:2002a}, was auf die gleiche Bedingung führt, die
    wir in Satz \ref{CourantAlgDef} erhalten werden, nur formuliert in
    der Sprache der Supermannigfaltigkeiten.
\end{Bemerkung}

\section{Obstruktion für die Fortsetzbarkeit von Deformationen}

Der Hauptgrund für die Einführung der Rothstein-Klammer ist, dass wir jetzt
auch den kubischen Term in der Bedingung für die Deformation aus Lemma
\ref{diracdeformation} systematisch behandeln können.
\begin{Lemma}
    Seien $\omega_1,\omega_2,\omega_3 \in \sect(\Wedge^2 L^\ast)$ und
    $s_1,s_2,s_3 \in \sect(L)$. Dann gilt

    $$\roth{s_3,\roth{s_2,\roth{s_1,{
            \roth{\roth{\roth{\psi,\omega_1},\omega_2},\omega_3}}}}} =
    -\sum_{\pi\in \mathcal{S}_3}
    \psi\(\omega_1(s_{\pi(1)}),\omega_2(s_{\pi(2)}),
    \omega_3(s_{\pi(3)})\)$$
    Insbesondere ist also
    $$\frac{1}{6}\roth{\roth{\roth{\psi,\omega},\omega},\omega}
    = -\wedge^3\omega\; \psi = T_\omega,$$
    wobei  $T_\omega(s_1,s_2,s_3) =
    \scal{[\omega(s_1),\omega(s_2)]_\C,\omega(s_3)}$ gilt. 
\end{Lemma}

\begin{proof}
    Die Behauptung rechnet man mit Hilfe der Jacobi-Identität nach,
    wobei man beachtet, dass Terme der Form
    $$\roth{\roth{\roth{\psi,\omega},\omega},\omega_1(s_1,s_2)}$$
    usw. verschwinden.
\end{proof}

Um Schreibarbeit zu sparen, formulieren wir folgendes 
\begin{Lemma}
Durch die Forderung
$$\tau^\sharp[\eta_1,\eta_2,\eta_3]_\psi =
\roth{\roth{\roth{\psi,\eta_1},\eta_2},\eta_3} $$
ist eine super-antisymmetrische Abbildung 
$$[\dcd,\cdot\;]_\psi:\sect(\Wedge^k L^\ast) \times
\sect(\Wedge^\ell L^\ast) \times \sect(\Wedge^m L^\ast) \ra
\sect(\Wedge^{k+\ell+m-3} L^\ast)$$
definiert.
\end{Lemma}
\begin{proof}
    Da $\psi$ ein zurückgezogener Schnitt ist und damit auch die
    rechte Seite der Definitionsgleichung einen zurückgezogenen
    Schnitt ergibt, folgt zusammen mit der Injektivität von
    $\tau^\sharp$, dass $[\dcd,\cdot\;]_\psi$ wohldefiniert ist. Die
    Antisymmetrie erhält man aus der Super-Jacobi-Identität für die
    Rothstein-Klammer, wobei man beachtet, dass Terme der Form
    $\roth{\eta_1,\eta_2}$ usw. verschwinden.  Die Aussage über die Grade
    folgt, da $\psi$ den Totalgrad $3$ und die Rothstein-Klammer den
    den Grad $-2$ hat.
\end{proof}
Mit dieser Definition ist jetzt also $T_\omega =
\tfrac{1}{6}[\omega,\omega,\omega]_\psi$. Damit haben wir 
folgende Umformulierung von Satz \ref{diracdeformation}.
\begin{Satz}\label{CourantAlgDef}
    Sei $E$ ein Courant-Algebroid mit Dirac-Struktur $L$ und sei $L'$
    ein isotropes Komplement zu $L$. Identifizieren wir $E$ mit
    $L\oplus L^\ast$, dann ist $\op{graph}(\omega)$ für eine Zweiform
    $\omega$ auf $L$ genau dann eine Dirac-Struktur, wenn die
    Gleichung
    $$\roth{\mu,\omega} +
    \frac{1}{2}\roth{\roth{\omega,\gamma},\omega} + \frac{1}{6}
    \roth{\roth{\roth{\psi,\omega},\omega},\omega} = 0,$$
    beziehungsweise die äquivalente Gleichung
    $$\dL \omega + \frac{1}{2}[\omega,\omega]_\ast +
    \frac{1}{6}[\omega,\omega,\omega]_\psi = 0$$
    erfüllt ist.
\end{Satz}

\begin{Lemma}\label{universalid}
    Sei für ein $\omega \in \sect(\Wedge^2 L^\ast)$ durch 
    \begin{eqnarray*}
        \dif_\omega &=& \roth{\mu,\,\cdot\,} +
        \roth{\roth{\omega,\gamma},\,\cdot\,} +
        \frac{1}{2}\roth{\roth{\roth{\psi,\omega},\omega},\,\cdot\,}\\
        &=& \dL + [\omega,\,\cdot\,]_\ast +
        \frac{1}{2}[\omega,\omega,\,\cdot\,]_\psi 
        \end{eqnarray*}
        eine Abbildung $\dif_\omega:\sect(\Wedge^\bullet L^\ast) \ra
        \sect(\Wedge^\bullet L^\ast) $ definiert. Dann ist
        $\dif_\omega$ eine Superderivation bezüglich des Dachprodukts
        vom Grad eins, und es gilt die Identität
        $$\dif_\omega\Big( \roth{\mu,\omega} +
        \frac{1}{2}\roth{\roth{\omega,\gamma},\omega} + \frac{1}{6}
        \roth{\roth{\roth{\psi,\omega},\omega},\omega}\Big) = 0$$
        bzw.
        $$\dif_\omega\Big( \dL \omega +
        \frac{1}{2}[\omega,\omega]_\ast +
        \frac{1}{6}[\omega,\omega,\omega]_\psi\Big) = 0.$$
\end{Lemma}

\begin{proof}
    Das durch $\dif_\omega$ eine Superderivation vom Grad eins gegeben
    ist, folgt leicht aus den Eigenschaften der Rothstein-Klammer.
    Die zu zeigende Gleichung schreiben wir zunächst einmal aus, wobei
    wir bereits die Identität $\roth{\mu,\psi} +
    \frac{1}{2}\roth{\gamma,\gamma} = 0$ verwenden.
    \begin{eqnarray*}
        \lefteqn{\dif_\omega\Big( \roth{\mu,\omega} +
          \frac{1}{2}\roth{\roth{\omega,\gamma},\omega} + \frac{1}{6}
          \roth{\roth{\roth{\psi,\omega},\omega},\omega}\Big)} \qquad&& \\
        &=& \roth{\roth{\roth{\mu,\omega},\gamma},\omega} -
        \frac{1}{2}
        \roth{\roth{\roth{\psi,\omega},\omega},\roth{\mu,\omega}}\\
        &&
        -\frac{1}{12}\roth{\roth{\roth{\roth{\gamma,\gamma},\omega},
            \omega},\omega}  + 
        \roth{\roth{\omega,\gamma},\roth{\mu,\omega}}\\
        &&+ \frac{1}{2}
        \roth{\roth{\omega,\gamma},\roth{\roth{\omega,\gamma},\omega}}
        + \frac{1}{6}
        \roth{\roth{\omega,\gamma},\roth{\roth{\roth{\psi,\omega},\omega},
              \omega}}\\
        &&
        +\frac{1}{2}\roth{\roth{\roth{\psi,\omega},\omega},\roth{\mu,\omega}}  
        +\frac{1}{4}\roth{\roth{\roth{\psi,\omega},\omega},
          \roth{\roth{\omega,\gamma},\omega}} \\  
        &&
        +\frac{1}{12}\roth{\roth{\roth{\psi,\omega},\omega},
          \roth{\roth{\roth{\psi,\omega},\omega},\omega}}
    \end{eqnarray*}
    Die Behauptung ergibt sich jetzt mit den folgenden Gleichungen,
    die man mit Hilfe der Super-Jacobi-Identität nachrechnet.
    \begin{enumerateR}
    \item $\roth{\roth{\roth{\mu,\omega},\gamma},\omega} = -
        \roth{\roth{\omega,\gamma},\roth{\mu,\omega}} $
     
    \item $\roth{\roth{\roth{\roth{\gamma,\gamma},\omega},
            \omega},\omega} = 6
        \roth{\roth{\omega,\gamma},\roth{\roth{\omega,\gamma},\omega}}
        $
        
    \item $\begin{array}[t]{@{}rcl} 0 &=&
            \roth{\roth{\roth{\roth{\roth{\gamma,\psi},\omega},
                  \omega}, \omega},\omega} \\[2mm]
            &=& 6 \roth{\roth{\roth{\psi,\omega},\omega},
              \roth{\roth{\omega,\gamma},\omega}} + 4
            \roth{\roth{\roth{\psi,\omega},\omega},
              \roth{\roth{\omega,\gamma},\omega}}
        \end{array}$
    \item $\begin{array}[t]{@{}rcl} 0 &=&
            \roth{\roth{\roth{\roth{\roth{\roth{\psi,\psi},\omega},
                    \omega},\omega},\omega},\omega} \\[2mm]
            &=&
            20\roth{\roth{\roth{\psi,\omega},\omega},
              \roth{\roth{\roth{\psi,\omega},\omega},\omega}}
        \end{array}$
    \end{enumerateR}
\end{proof}

Mit diesen Vorbereitungen können wir jetzt die Obstruktion für
die Fortsetzbarkeit von Deformationen bestimmen. Der folgende Satz
liefert das zentrale Ergebnis dieser Arbeit.
\begin{Satz}
    Sei $E = L\oplus L^\ast$ ein Courant-Algebroid mit Dirac-Struktur
    $L$ und sei durch $\omega_t = t \omega_1 + t^2\omega_2 + \ldots +
    t^N \omega_N \in \sect(\Wedge^2 L^\ast)[[t]] $ eine formale
    Deformation von $L$ der Ordnung $N$ gegeben, d.h. die Gleichung
    $$\dL \omega_t + \frac{1}{2}[\omega_t,\omega_t]_\ast
    +\frac{1}{6}[\omega_t,\omega_t,\omega_t]_\psi= 0$$
    ist bis zur
    Ordnung $N$ erfüllt. Dann ist
    $$R_{N+1} =
    -\frac{1}{2}\sum_{i=1}^{N}[\omega_i,\omega_{N+1-i}]_\ast 
    -\frac{1}{6}\sum_{i+j+k=N+1} [\omega_i,\omega_j,\omega_k]_\psi$$
    geschlossen bezüglich $\dL$, und die Deformation $\omega_t$ lässt
    sich genau dann bis zur Ordnung $N+1$ fortsetzen, wenn $R_{N+1}$ exakt
    ist. Die Obstruktion für die Fortsetzbarkeit einer Deformation
    liegt also in der dritten Lie-Algebroidkohomologie der
    Dirac-Struktur $L$.
\end{Satz}

\begin{proof}
    Sei $\omega_{N+1} \in \sect(\Wedge^2 L^\ast)$ und $\omega'_t =
    \omega_t + 
    t^{N+1}\omega_{N+1}$. Wir rechnen zunächst die $(N+1)$--te Ordnung
    der Deformationsbedingung aus.
    \begin{eqnarray*}
        \lefteqn{\dL \omega'_t + \frac{1}{2}[\omega'_t,\omega'_t]_\ast
          +\frac{1}{6}[\omega'_t,\omega'_t,\omega'_t]_\psi}&&\\
        \qquad&=& t^{N+1}\bigg( \dL \omega_{N+1} +
        \frac{1}{2}\sum_{i=1}^{N}[\omega_i,\omega_{N+1-i}]_\ast + 
        \frac{1}{6}\sum_{i+j+k=N+1}
        [\omega_i,\omega_j,\omega_k]_\psi\bigg) + t^{N+2} \ldots\\
    \end{eqnarray*}
        
    Wir können die Deformation also genau dann fortsetzen, wenn wir
    ein $\omega_{N+1}$ mit
    $$\dL \omega_{N+1} = R_{N+1}$$
    finden. Dass $R_{N+1}$ geschlossen ist, ergibt sich mit Lemma
    \ref{universalid} durch Anwenden von $\dif_{\omega'_t}$ auf die
    Deformationsgleichung.
    \begin{eqnarray*}
        0 &=& \dif_{\omega'_{t}} \Big(\dL \omega'_t +
          \frac{1}{2}[\omega'_t,\omega'_t ]_\ast 
          +\frac{1}{6}[\omega'_t,\omega'_t,\omega'_t]_\psi \Big)\\
        &=& t^{N+1}\dL\bigg(
        \frac{1}{2}\sum_{i=1}^{l}[\omega_i,\omega_{N+1-i}]_\ast + 
        \frac{1}{6}\sum_{i+j+k=N+1}
        [\omega_i,\omega_j,\omega_k]_\psi\bigg) + t^{N+2}\ldots
    \end{eqnarray*}
    Da der Ausdruck in jeder Ordnung von $t$ einzeln Null sein muss,
    folgt damit die Behauptung.
\end{proof} 

\begin{Bemerkung}
    Wir erhalten damit also die bekannten Ergebnisse aus der
    Deformationstheorie von Poisson-Tensoren oder präsymplektischen
    Formen wieder, vgl. die Abschnitte \ref{graphPreSymp} und
    \ref{graphPoisson}.  Weiter sehen wir, dass die Obstruktion für
    die Fortsetzbarkeit, die ja durch die dritte
    Lie-Algebroid\-koho\-mo\-lo\-gie von $L$ gegeben ist, unabhängig
    von der Wahl des zu $L$ komplementären Bündels $L'$ und des
    Zusammenhangs auf $L$ ist. Schließlich sei noch erwähnt, dass die
    Klassifikation der Äquivalenz von Deformationen in diesem Fall im
    Allgemeinen nicht durch die zweite Lie-Algebroidkohomologie von
    $L$ gegeben ist, vgl. auch die Folgerungen \ref{starr} und
    \ref{starrAlg}.  Zwar ist das für Poisson-Mannigfaltigkeiten noch
    richtig, für präsymplektischen Mannigfaltigkeiten haben wir jedoch
    in Abschnitt \ref{graphPreSymp} gesehen, dass das Verschwinden der
    zweiten deRahm-Kohomologie für die Starrheit noch nicht
    hinreichend ist.
\end{Bemerkung}

 


\begin{appendix}

\chapter[Lie-Algebroide]{Lie-Algebroide, Lie-Bialgebroide und die
  Schouten-Nijenhuis-Klammer}

\ihead{\leftmark}
\ohead[]{\pagemark}
Wir wollen hier einige Definitionen und grundlegende Aussagen zu
Lie-Algebroiden und verwandten Themen aufführen. Für  ausführlichere
Darstellungen siehe z.B. \cite{marle:2002,cannasdasilva.weinstein:1999a}.  
\begin{ADefinition}[\cite{cannasdasilva.weinstein:1999a}]\label{defLieAlg}
    Ein Lie-Algebroid ist ein Vektorbündel $E \ra M$ über einer
    Mannigfaltigkeit $M$ zusammen mit einer Lieklammer
    $[\cdot\,,\cdot\,]$ auf den Schnitten $\sect(E)$ von $E$, sowie
    einem Vektorbündelhomomorphismus $\rho:E\ra TM$, genannt Anker, so
    dass folgende Leibnizregel gilt:
    $$[e_1,f e_2] = f [e_1,e_2] + \(\rho(e_1)f\) e_2 \qquad \mbox{für alle}\;
    e_1,e_2 \in  \sect(E), f\in C^\infty(M).$$
\end{ADefinition}
\begin{ABemerkung}
    Man kann zeigen \cite{kosmann-schwarzbach.magri:1990a}, dass für
    jedes Lie-Algebroid der Anker ein Liealgebrenhomomorphismus
    $\rho:\sect(E) \ra \mathfrak{X}(M) = \sect(TM)$ ist,
    $$\rho([e_1,e_2]) = [\rho(e_1),\rho(e_1)] \qquad \mbox{für alle}\;
    e_1,e_2 \in  \sect(E).$$ 
\end{ABemerkung}

\begin{ABeispiele}
\begin{enumerateR}
\item Das Standardbeispiel für ein Lie-Algebroid ist das
    Tangentialbündel $TM$ einer Mannigfaltigkeit $M$ mit der
    Lie-Klammer von Vektorfeldern und der Identität als Anker. Ebenso
    wird jedes integrable Unterbündel von $TM$ ein Lie-Algebroid.
\item Ist der Anker $\rho = 0$, dann ist $E$ ein Bündel von
    Lie-Algebren. Insbesondere ist jede Lie-Algebra $\g$ mit dem Anker
    $\rho = 0$ ein Lie-Algebroid über einem Punkt.
 \end{enumerateR}
\end{ABeispiele}   

\begin{ADefinition}\label{defLieAlgdif}
    Sei $E$ ein Lie-Algebroid. Dann definieren wir die Abbildung 
    $$\dE:\sect(\Wedge^k E^\ast) \ra \sect(\Wedge^{k+1} E)$$
    durch die Forderungen 
    $$\dE f(e_1) = \rho(e_1)f $$
    und
    \begin{eqnarray*}
        \dE \omega(e_1,\ldots,e_{k+1}) &=& \sum_{i=1}^{k+1}
        (-1)^{i+1}\rho(e_i)\omega(e_1,\elide{i},e_{k+1}) \\
        &&\qquad -
        \sum_{i<j}(-1)^{i+j}\omega([e_i,e_j],e_1,\elidetwo{i}{j},e_{k+1}) 
    \end{eqnarray*}
    für alle $f\in C^\infty(M)$ und $e_1,\ldots,e_{k+1}\in \sect(E)$.
    Wir nennen $\dif_E$ das Lie-Algebroiddifferen\-tial von $E$.
\end{ADefinition}
Man überzeugt sich leicht davon, dass $\dE$ eine Superderivation
bezüglich des Dachprodukts ist. Für den Fall $E = TM$ ist $\dE$ das
gewöhnliche deRahm-Differential, und es gilt dann $\dif_E^2 = 0$. Mit Hilfe
der Jacobi-Identität für $[\dcd]$ lässt sich diese Aussage jedoch
auch allgemeiner zeigen.
\begin{ALemma}
    Für das Lie-Algebroiddifferential $\dif_E$ gilt $\dif_E^2 = 0$.
\end{ALemma}

Wir können aber auch umgekehrt vorgehen und mit Hilfe eines
Differentials auf der Grassmannalgebra $\sect(\Wedge^\bullet E^\ast)$
das Vektorbündel $E$ zu einem Lie-Algebroid machen.
\begin{ASatz}
    Sei  $E$ ein Vektorbündel und 
    $$\dE:\sect(\Wedge^k E^\ast) \ra \sect(\Wedge^{k+1} E)$$
    eine
    Superderivation von Grad eins. Dann ist durch die Definitionen
    $$\rho(e)f = \dE f(e)\qquad \forall f\in C^\infty(M),$$
    $$\alpha([e_1,e_2]) = \rho(e_1)(\alpha(e_2))-\rho(e_2)(\alpha(e_1)) -
    \dE\alpha(e_1,e_2) \qquad \forall \alpha\in \sect(L^\ast)$$
    eine $\R$-bilineare, antisymmetrische Verknüpfung auf $\sect(E)$
    gegeben, die die Leibniz-Regel
    $$[e_1,f e_2] = f[e_1,e_2] +\rho(e_1)f\, e_2$$
    erfüllt. Genau dann erfüllt $[\dcd]$ die Jacobi-Identität und 
    $(E,[\dcd],\rho)$ ist ein Lie-Algebroid, wenn $\dif_E^2 = 0$ gilt.
\end{ASatz}

Die Lieklammer auf den Schnitten $\sect(E)$ kann mit Hilfe des Ankers
zu einer Super-Lie\-klam\-mer auf ganz $\sect(\Wedge^\bullet E)$ erweitert
werden.
\begin{ADefinition}\label{defLieAb}
     Sei $(E,[\dcd],\rho)$ ein Lie-Algebroid und sei $P\in
    \sect(\Wedge^k E)$.
    \begin{enumerateR}
    \item Die Superderivation 
        $$i_P:\sect(\Wedge^\bullet E^\ast) \ra \sect(\Wedge^{\bullet
          - k} E)$$ 
        ist durch die Forderungen
        $$i_f \omega = f\omega \qquad \text{für} \quad f\in
        C^\infty(M)$$
        und
        $$i_{e_1\wedge\ldots\wedge e_k} = i_{e_1}\ldots i_{e_k}$$
        für $e_1,\ldots,e_k \in \sect(E)$ definiert.
    \item Die Lieableitung 
        $$\Lie_P:\sect(\Wedge^\bullet E^\ast) \ra
        \sect(\Wedge^{\bullet-k+1} 
        E^\ast)$$ 
        ist durch
        $$\Lie_P = [i_P,\dE] = i_P\dE - (-1)^k \dE i_P$$
        gegeben.
    \end{enumerateR}
\end{ADefinition}

\begin{ADefinition}\label{defSchouten}
    Durch die Forderung 
    $$i_{[P,Q]} = [\Lie_P,i_Q] = \Lie_P i_Q - (-1)^{(k-1)(\ell-1)}i_Q
    \Lie_P $$
    für $P \in \sect(\Wedge^k E)$ und $ Q\in
    \sect(\Wedge^\ell E)$ wird die Lie-Algebroidklammer auf ganz
    $\sect(\Wedge^\bullet E)$ fortgesetzt.  Diese so erhaltene Klammer
    heißt Schouten-Nijenhuis-Klammer oder auch kurz Schouten-Klammer.
    Zur Wohldefiniertheit dieser Abbildung siehe z.B.
    \cite{marle:2002}.
\end{ADefinition}

\begin{ASatz}
    Die Algebra $\(\sect(\Wedge^\bullet E),[\dcd],\wedge\)$ ist eine
    Gerstenhaber-Algebra, dass heißt für $P \in \sect(\Wedge^k E),Q \in
    \sect(\Wedge^\ell E)$ und $R\in \sect(\Wedge^\bullet E)$ gilt
    die Super-Antisymmetrie
    $$[P,Q] = -(-1)^{(k-1)(\ell -1)} [Q,P],$$
    die Super-Jacobi-Identität
    $$[P,[Q,R]] = [[P,Q],R] + (-1)^{(k-1)(l-1)}[Q,[P,R]]$$
    sowie die Super-Leibniz-Regel
    $$[P,Q\wedge R] = [P,Q]\wedge R + (-1)^{(k-1)\ell} Q\wedge
    [P,R].$$
\end{ASatz}

Insbesondere wird mit der Schouten-Nijenhuis-Klammer  die Algebra
der Multivektorfelder $\mathfrak{X}^\bullet(M) = \sect(\Wedge^\bullet
TM)$ auf kanonische Weise zu einer Gerstenhaber-Algebra.
    
Für Funktionen $f,g \in C^\infty(M)$ und faktorisierende Schnitte $P =
X_1\wedge\ldots\wedge X_k$, $Q = Y_1\wedge\ldots\wedge
Y_\ell$ erhalten wir für die Schouten-Nijenhuis-Klammer die
Gleichungen
\begin{eqnarray*}
    [f,g] &=& 0,\\
    {[f,P]} &=& -i_{\dE f} P = \sum_{i=1}^k (-1)^i X_i(f)
    X_1\wedge\elide{i}\wedge X_k,\\
    {[P,Q]} &=& \sum_{i=1}^k \sum_{j=1}^\ell [X_i,Y_j]
    X_1\wedge\elide{i}\wedge X_k \wedge Y_1\wedge\elide{j}\wedge
    Y_\ell.
\end{eqnarray*}
Umgekehrt hätten wir die Schouten-Nijenhuis-Klammer auch durch diese
Gleichungen definieren können, siehe z.B. \cite{waldmann:2004a}.

\begin{ADefinition}[\cite{mackenzie.xu:1994a}]\label{defLieBiAlg}
    Ein Lie-Bialgebroid ist ein Paar $(E,E^\ast)$ von dualen
    Vektorbündeln, wobei $(E,[\dcd],\rho)$ und
    $(E^\ast,[\dcd]_\ast,\rho_\ast)$ Lie-Algebroide sind, so dass das
    Lie-Algebroiddiffer\-en\-tial $\dif_{E^\ast}$ von $E^\ast$ eine
    Superderivation der Schouten-Nijenhuis-Klammer auf $E$ ist, d.h.
    für alle $P \in \sect(\Wedge^k E)$ und $Q \in \sect(\Wedge^\ell
    E)$ gilt
    $$ \dif_{E^\ast} [P,Q] = [\dif_{E^\ast} P,Q] +
    (-1)^{k-1}[P,\dif_{E^\ast} Q].$$
\end{ADefinition}
Das folgende Lemma zeigt, dass die Definition nur scheinbar
asymmetrisch in $E$ und $E^\ast$ ist.
\begin{ALemma}[\cite{kosmann-schwarzbach:1995a})]
    $(E,E^\ast)$ ist genau dann ein Lie-Bialgebroid, wenn
    $(E^\ast,E)$ ein Lie-Bialgebroid ist.
\end{ALemma}

Für ein Lie-Bialgebroid $(E,E^\ast)$ ist also auch das
Lie-Algebroiddifferential $\dE$ von $E$ eine
Superderivation der Schouten-Nijenhuis-Klammer auf $E^\ast$.

\chapter{Formale Reihen}\label{formalSeries}
 
Wir geben noch eine sehr kurze Darstellung über die Grundlagen der
Theorie formaler Potenzreihen. Für die Beweise der folgenden
Aussagen sowie weiterführende Informationen siehe z.B.
\cite{niven:1969a,waldmann:1999a,waldmann:2004a}.

Sei $V$ ein Vektorraum, oder allgemeiner ein Modul über einem Ring
$\op{R}$, dann bezeichnen wir die Menge aller Folgen $(v_n)_{n\in
  \mathbb{N}}$ in $V$ mit
$$V[[t]] = \prod_{k\in \mathbb{N}} V_k \qquad \text{mit}\quad V_k =
V\;\; \forall\; k $$
und schreiben Elemente in $V[[t]]$ als formale
Potenzreihen in dem formalen Parameter $t$,
$$v_t = \sum_{k=0}^\infty t^k v_k\qquad \text{mit}\quad v_k \in V.$$
Es
sei betont, dass wir hier keinerlei Konvergenz der unendlichen Reihe
fordern. Durch gliedweise Multiplikation mit Elementen aus $\op{R}$
sowie gliedweise Addition wird $V[[t]]$ auf kanonische Weise zu einem
$\op{R}$-Modul. Durch die Definition
$$\Bigg(\sum_{i=0}^\infty t^i r_i\Bigg) \Bigg(\sum_{j=0}^\infty t^j
v_j\Bigg) = \sum_{k=0}^\infty t^k \sum_{\ell=0}^k r_\ell v_{k-\ell}$$
wird $V[[t]]$ sogar zu einem $R[[t]]$-Modul.

Ist  $V$ eine Algebra, so wird auch $V[[t]]$ durch die Forderung
$$\Bigg(\sum_{i=0}^\infty t^i v_i\Bigg) \Bigg(\sum_{j=0}^\infty t^j
w_j\Bigg) = \sum_{k=0}^\infty t^k \sum_{\ell=0}^k v_\ell w_{k-\ell}$$
zu einer Algebra. Dies erklärt auch die Schreibweise als formale
Reihen.

Ist $v_t = 1 + t w_t$ eine formale Reihe mit einer beliebigen formalen
Reihe $w_t \in V[[t]]$ für eine assoziative Algebra $V$ mit Eins, so
ist durch
$$v_t^{-1} = \sum_{k=0}^\infty(-t w_t)^k$$
das Inverse von $v_t$
erklärt, wie man durch direktes Nachrechnen zeigt.

Im folgenden
wollen wir zusätzlich annehmen, dass der Ring $\op{R}$ die rationalen
Zahlen enthält, $\mathbb{Q} \subseteq \op{R}$.  Dann können wir die
Exponential- und Logarithmusfunktionen von formalen Reihen bilden,
indem wir
$$\exp(t w_t) = \sum_{k=0}^\infty \tfrac{1}{k!}(t w_t)^k$$
und 
$$\ln(1 + t w_t) = \sum_{k=1}^\infty \tfrac{(-1)^{k-1}}{k}(t w_t)^k$$
definieren. Es gelten dann Rechenregel analog zu den für $\exp$ und
$\ln$ bekannten Regel. Insbesondere ist $\exp$ auch hier die
Umkehrfunktion zu $\ln$. Wir können deshalb eine formale Reihe $v_t =
1+t w_t$, die in unterster Ordnung mit der Eins beginnt, immer als $\exp(t
u_t)$ schreiben, wobei dann $u_t = \ln(1+t w_t)$ gilt.

Weiter definieren wir das Anwenden einer formalen Reihe
$$\phi_t=\sum_{i=0}^\infty t^i \phi_i \in \op{Hom}(V,W)[[t]]$$
von
Abbildungen $\phi_i:V \ra W$ auf eine formale Reihe $\sum_{j=0}^\infty
t^j v_j \in V[[t]]$ durch
$$\Bigg(\sum_{i=0}^\infty t^i \phi_i\Bigg) \Bigg(\sum_{j=0}^\infty t^j
v_j \Bigg) = \sum_{k=0}^\infty t^k \sum_{\ell=0}^j
\phi_\ell(v_{k-\ell})$$
und können $\phi$ damit als ein Element in
$\op{Hom}_{\op{R}[[t]]}\big(V[[t]],W[[t]]\big)$ auffassen.
Tatsächlich sind sogar alle $\op{R}[[t]]$-linearen Abbildungen von
dieser Form, d.h. es gilt
$$\op{Hom}(V,W)[[t]] \cong
\op{Hom}_{\op{R}[[t]]}\big(V[[t]],W[[t]]\big).$$

\end{appendix}



\bibliographystyle{plain}
\bibliography{Diplom,Diplomproceeding}

\newpage
\pagestyle{empty}
\section*{Danksagung}

Ich möchte mich an dieser Stelle bei allen bedanken, die mich bei
dieser Arbeit unterstützt haben; zuallererst bei PD Dr. Stefan
Waldmann für eine intensive und verständnisvolle Betreuung; aber
überhaupt bei allen Mitarbeitern des achten Stocks, insbesondere bei
Dr. Nikolai Neumaier für sorgfältiges Korrekturlesen, sowie bei
Michael Carl und Carsten Eilks für hilfreiche Kritik und  Diskussionen.


\end{document}